\documentclass[12pt,article,dvips]{article}
\usepackage{a4p}
\usepackage{appendix}
\usepackage{cite}
\usepackage{mcite}
\usepackage{graphicx}
\usepackage{subfigure}
\usepackage{float}
\usepackage{atlasphysics}
\usepackage{atlas_title,ifthen}
\usepackage{drafthead}
\usepackage{mathptmx}
\usepackage{amsmath}
\usepackage{helvet}
\usepackage[section]{placeins}
\usepackage{rotating}
\usepackage{Zprime}
\usepackage{array} 
\newlength{\capindent}
\newlength{\capwidth}
\newlength{\figwidth}
\newcommand{\icaption}[2][!*!,!]{\hspace*{\capindent}%
  \begin{minipage}{\capwidth}
    \ifthenelse{\equal{#1}{!*!,!}}%
      {\caption{#2}}%
      {\caption[#1]{#2}}
      \vspace*{3mm}
  \end{minipage}}

\begin{document}
\begin{titlepage}
\vspace*{-6mm}
\begin{center}
\end{center}
\vspace*{-0.5cm}
\title{Heavy Neutral Gauge Bosons at the LHC\\
\vspace{0.2cm}
in an Extended MSSM}
\vspace*{-2.cm}
\author{\bf Gennaro Corcella$^1$,
\ \ \bf Simonetta Gentile$^2$ }
\begin{center} 
$^1${\sl INFN, Laboratori Nazionali di Frascati}, \\
{\sl Via E.~Fermi 40, I-00044, Frascati,
Italy}\\
\vspace{0.5cm}

\it $^2${\sl Dipartimento di Fisica,  Universit\`a di Roma 'La Sapienza'\\ 
and INFN, Sezione di Roma,}\\
{\sl Piazzale A.~Moro 2, I-00185, Roma, Italy}
\end{center}

\vspace*{-1cm}
\begin{abstract}
Searching for heavy neutral gauge bosons \Zprime, predicted in extensions of
the Standard Model based on a \Uprime\  gauge symmetry, 
is one of the challenging objectives of the experiments carried out at 
the Large Hadron Collider. 
In this paper, we study \Zprime\  phenomenology at hadron colliders 
according to several  \Uprime-based models and in the Sequential Standard Model.
In particular, possible \Zprime\  decays into supersymmetric particles
are included, in addition 
to the Standard Model modes so far investigated.  
We point out the impact of the \Uprime\ group on the MSSM spectrum and,
for a better understanding, we consider 
a few benchmarks points in the parameter space.
We account for the D-term contribution,
due to the breaking of \Uprime, to slepton and squark masses
and investigate its effect on $Z'$ decays into sfermions.
Results on branching ratios and cross sections are
presented, as a function of the MSSM and \Uprime\ parameters,
which are varied within suitable ranges.
We pay special attention 
to final states with leptons and missing energy and make predictions
on the number of events with sparticle production in
$Z'$ decays, for a few values of 
integrated luminosity and centre-of-mass energy of the LHC.

\vspace{0.3cm}
\noindent
{\em Keywords: Physics Beyond the Standard Model; Collider
Phenomenology;\\ Supersymmetry;
Heavy Gauge Bosons; Grand Unification Theories.}
\end{abstract}
\end{titlepage}
%
%
\section{Introduction}\label{sec:Intro}

The Standard Model (SM) of 
the strong and electroweak interactions has been so far successfully tested at several machines, such 
the LEP and Tevatron accelerators and 
has been lately confirmed by the data collected by the Large Hadron Collider (LHC).
New physics models have nonetheless been proposed to solve the drawbacks of the SM,
namely the hierarchy problem, the Dark Matter observation 
or the still undetected Higgs boson, responsible for the mass
generation.  
The large amount of 
data collected at the centre-of-mass energy of 7 TeV
at the LHC opens a window to extensively search 
for new physics.
The further increase to 8 and ultimately 14 TeV, 
as well as
higher integrated luminosities, will extend  this investigation in the near future.
 
The simplest possible extension of the SM consists in   
a gauge group of larger rank involving 
the introduction of one extra \Uprime\  factor, inspired by Grand Unification
Theories (GUTs), which
leads to the prediction of a new neutral gauge boson $Z'$.
The phenomenology of the $Z'$  has been studied 
from a theoretical viewpoint 
(see, e.g., the reviews \cite{Langacker:2008yv,rizzo} or the
more recent work in Refs.~\cite{zwirner1,zwirner2}), 
whereas searches for new heavy gauge bosons have been carried out at the 
Tevatron by the CDF \cite{cdfzprime} and D0 \cite{d0zprime} 
Collaborations and at the 
LHC by ATLAS \cite{atlas} and CMS \cite{cms}.
Besides the $Z'$  bosons yielded by the extra \Uprime\  group, the 
analyses have also investigated 
the so-called Sequential Standard Model ($Z'_{\rm SSM}$),
i.e. a $Z'$ with the same couplings to fermions and
gauge bosons as the $Z$ of the SM. The Sequential Standard Model does not have
theoretical bases like the \Uprime\ models,
but it is used as a benchmark,
since, as will be seen later on, the production cross section
is just function of the $Z'$ mass and there is no dependence on other
parameters.

The Tevatron analyses searched 
for high-mass dielectron resonances in $p\bar p$ collisions at
1.96 TeV and set a lower $Z'$  mass limit of about 1023 (D0) and 963 (CDF) GeV
for the $Z'_{\rm SSM}$.
The LHC experiments investigated the 
production of both dielectrons and dimuons at large invariant masses and
several models of $Z'$  production, i.e. different \Uprime\  gauge groups.
The CMS Collaboration, by using event samples corresponding to an integrated luminosity
of $1.1$~fb$^{-1}$,
excluded a $Z'$  with SM-like couplings and mass below 2.32 TeV, 
a GUT-inspired $Z'$  below 1.49-1.69 TeV and a Kaluza--Klein graviton
in extra-dimension models \cite{randall}
below 0.71-1.63 TeV.
The ATLAS Collaboration analyzed 5~fb$^{-1}$ of data and obtained
a bound of 2.21 TeV for the SM-like case, in
the range 1.76-1.96 TeV for the \Uprime\ scenarios and about
0.91-2.16 TeV for the Randall--Sundrum gravitons \footnote{The exclusion 
ranges depend on the specific \Uprime\ model and, for the graviton searches,
on the coupling value.}.

All such analyses, and therefore the obtained exclusion limits,
crucially rely on the assumption that 
the $Z'$  decays into Standard Model particles, with 
branching ratios depending on its
mass and, in the GUT-driven case, 
on the parameters characterizing the specific \Uprime\  model:
such a choice is dictated by the sake of minimizing the parameters 
ruling the $Z'$ 
phenomenology.
As a matter of fact, in the perspective of searching for new physics at the LHC, 
there is no actual reason to exclude $Z'$ decays into
channels beyond the SM, such as its supersymmetry.
In fact, new physics contributions to the
$Z'$ width will significantly decrease the
branching ratios into SM particles, and therefore the mass limits quoted by the experiments
may have to be revisited. Furthermore, $Z'$ decays into supersymmetric particles, if existing, 
represent an excellent tool to
investigate the electroweak interactions at the LHC in a phase-space corner that cannot
be explored by employing the usual techniques. Therefore, the possible discovery of supersymmetry in
$Z'$ processes would open the road to additional investigations,
since one would need to formulate a scenario accommodating
both sparticles and heavy gauge bosons.

The scope of this paper is indeed the investigation 
of the phenomenology of $Z'$  bosons at the LHC, 
assuming that they can decay into both SM and supersymmetric particles. As for supersymmetry,
we shall refer to the Minimal Supersymmetric Standard Model (MSSM)
\cite{Haber:1984rc,Barbieri:1982eh} and 
study the dependence on the MSSM parameters.
A pioneering study of supersymmetric contributions to $Z'$ decays was 
carried out in
\cite{Gherghetta:1996yr}, wherein
the partial widths in all SM and MSSM channels were derived analytically,
and the branching ratios computed for a few \Uprime\  scenarios.
However, the numerical analysis was performed for a mass \MZprime=700~GeV,
presently ruled out by the late experimental measurements, and only for one point of the
supersymmetric phase space. Therefore, no firm 
conclusion could be drawn about the feasibility to
search for the $Z'$  within supersymmetry at the LHC.
This issue was tackled again more recently.
Ref.~\cite{kang}  
studied how the $Z'$ mass exclusion limits change once sparticle 
and exotic decay modes
are included, for many \Uprime\ models and varying the
supersymmetric particle masses from 0 to 2.5 TeV.
The Higgs and neutralino sectors in extensions of the MSSM, including
GUT-inspired \Uprime\ models, were thoroughly debated in 
\cite{barg1} and \cite{barg2}, respectively.
Ref.~\cite{baum} considered the \Uprime$_{\rm B-L}$ gauge group,
B and L being the baryon and lepton numbers, and focused on
the decay of the $Z'$ into charged-slepton pairs for a few points
in the MSSM phase space and various values of $Z'$ and slepton masses.
Ref.~\cite{chang} investigated all possible decays of the $Z'$ 
in the SM and MSSM, and several \Uprime\ models, for two sets of supersymmetric
parameters and a $Z'$  mass in the 1-2 TeV range.

In the following, we shall extend the above work in several aspects.
Special attention will be paid to the
MSSM spectrum after the addition of the \Uprime\ gauge symmetry.
Squark and slepton masses will be parametrized
as the sum of a soft mass and the so-called D- and F-terms \cite{Martin:1997ns}. 
In particular, accounting for the D-term has an impact on the sfermion masses,
which get an extra contribution driven by the \Uprime\  group.
Higgs, chargino and neutralino masses will be determined
by diagonalizing the corresponding
mass matrices. 
A detailed study will be thereafter
undertaken by allowing the
\Uprime\  and MSSM parameters to
run within suitable ranges, taking into account the recent experimental limits.
Throughout this work, particular care will be taken about the decay 
of the $Z'$  into slepton pairs, i.e. charged sleptons or 
sneutrinos, eventually leading to final
states with four charged leptons or two charged leptons and
missing energy, due to neutralinos. In fact, in the complex hadronic environment of the LHC,
leptonic final states are the best channels to perform precise measurements and
searches. Slepton production in $Z'$  decays 
has the advantage that the $Z'$  mass is a further kinematical constrain
on the invariant mass of the slepton pair.
Moreover, the extension of the MSSM by means of the \Uprime\ gauge group provides also an
interesting scenario to study Dark Matter candidates, such as neutralinos
\cite{umssm1,umssm2} or right-handed sneutrinos \cite{belanger}, whose 
annihilation or scattering processes may proceed through
the coupling with a $Z'$ boson.

We shall present results for the $Z'$ production cross sections and
the branching ratios into both Standard Model and supersymmetric final states,
thoroughly scanning the U(1)$'$ and MSSM parameter spaces,
which will enable one to estimate the LHC event rates with sparticle production
in $Z'$ decays. 
We point out that, in order to draw a statement on the feasibility of the
LHC to search for supersymmetry in $Z'$ decays, one should also account
for the Standard Model backgrounds. 
However, in assessing whether the signal can be separated from the
background, one would need to consider exclusive
final states, wherein acceptance cuts on final-state jets, leptons and 
possible missing energy, as well as
detector effects, are expected to play a role. The framework of a Monte Carlo
generator \cite{herwig,pythia}, wherein both signal and background events are provided with
parton showers, hadronization, underlying event and detector simulations,
is therefore the ideal one to carry out such a comparison.
We shall thus defer a detailed  
investigation of the backgrounds to a future study, after the implementation
of our modelling for $Z'$ production and decay in a Monte Carlo code.

The paper is organized as follows. In Section 2 we shall briefly discuss the \Uprime\ gauge
group yielding the $Z'$  boson and the particle content of the MSSM.
Section 3 will be devoted to summarize the new features of the MSSM, once it
is used in conjunction with the \Uprime\  group.
In  Section 4, as a case study, we will choose a specific point of
the MSSM/\Uprime\  parameter space, named 
`Representative Point', and discuss the MSSM spectrum  in this scenario.
In Section 5, we shall present the 
$Z'$  branching ratios  
into SM and BSM particles for several \Uprime\  models and in the
Sequential Standard Model. 
We will first investigate the decay rates in a particular `Reference Point'
of the parameter space and then vary the \Uprime\  mixing angle and
the MSSM parameters. Particular attention will be devoted to the
decays into sleptons and to the dependence of the branching fractions on the
slepton mass.
In Section 6 the leading-order cross section for $Z'$  production
in the \Uprime\ scenarios and in the Sequential Standard Model will be calculated. Besides, 
the number of events with sparticle production in $Z'$ decays will be 
computed for a few energy and luminosity phases of the LHC.
In Section 7 we shall summarize the main results of our study and
make some final remarks on the future developments of the analysis here presented.
In Appendix A the main formulas used to calculate the $Z'$ branching
ratios will be presented.

\section{Modelling {$\mathbf Z'$} production and decay}\label{sec:Ext}

As discussed in the Introduction, we shall consider extensions of the Standard Model 
leading to \Zprime\  bosons, which will be allowed to decay into both SM 
and supersymmetric particles. For the sake of simplicity and
minimizing the dependence of our analysis on unknown parameters, we shall refer to the MSSM.
In this section we wish to briefly review the main aspects of the models used for \Zprime\ 
production and decay.

\subsection{{$\mathbf U(1)'$} models and charges}\label{sec:MSSM}

There are several possible extensions of the SM that can be achieved by adding an extra
\Uprime\ gauge group,
typical of string-inspired GUTs (see, e.g., Refs.~\cite{Langacker:2008yv,rizzo} for a review):
each model is characterized by the coupling constants,
the breaking scale of \Uprime\ 
and the scalar particle responsible for its breaking, 
the quantum numbers of fermions and bosons according to \Uprime.
Throughout our work, we shall focus on the \Uprime\ models
explored by the experimental collaborations. 

Among the \Uprime\ gauge models, special care has been taken about those
coming from a Grand Unification gauge group $\rm{E_6}$, having rank 6,
which breaks according to:
\begin{equation}
{\rm E_6} \ra \rm{SO(10)} \times {\Uprime_\psi},
\label{upsi}
\end{equation}
followed by
\begin{equation}
\rm{SO(10)} \ra  {\rm SU(5)} \times  {\rm \Uprime_\chi}.
\label{uchi}
\end{equation}
The neutral vector bosons associated with the $\Uprime_\psi$ and 
$\Uprime_\chi$ groups are called $Z'_\psi$ and $Z'_\chi$, respectively.
Any other model is characterized by an angle $\theta$ and leads to
a $Z'$ boson which can be expressed as \footnote{In Eq.~(\ref{zpri}) we followed the notation in \cite{Gherghetta:1996yr}
and we shall stick to it throughout this paper. One can easily recover the notation used in
\cite{Langacker:2008yv} by replacing $\theta\to\theta-\pi/2$.}: 
\begin{equation}
\Zprime(\theta)= Z^\prime_\psi \cos \theta-Z^\prime_\chi \sin\theta.
\label{zpri}
\end{equation}
The orthogonal combination to Eq.~(\ref{zpri}) is supposed to 
be relevant only at the Planck scale and can therefore be neglected 
even at LHC energies.
Another model, named U(1)$'_\eta$, is inherited by the direct breaking of 
${\rm E}_6$ to the Standard Model (SM) group, i.e. SU(2)$_{\rm L}
\times$ U(1)$_{\rm Y}$, 
as in superstring-inspired models:
\begin{equation}
{\rm E}_6\to {\rm SM}  \times {\rm U(1)}'_\eta.
\end{equation}
The yielded gauge boson is called $Z'_\eta$ and corresponds to a mixing angle
$\theta=\arccos{\sqrt{5/8}}$ in Eq.~(\ref{zpri}).
The model orthogonal to U(1)$'_\eta$, i.e. $\theta=\arccos{\sqrt{5/8}}-\pi/2$,
leads to a neutral boson which will be referred to as $Z'_{\rm I}$.
Furthermore, in 
the so-called secluded model, a \Uprime$_{\rm S}$ model extends the MSSM with
a singlet field $S$ \cite{erler}. 
The connection with the E$_6$ groups is achieved assuming
a mixing angle $\theta=\arctan(\sqrt{15}/9)-\pi/2$ and a 
gauge boson $Z'_{\rm S}$.

In the Grand Unification group E$_6$ the matter superfields are included in 
the fundamental representation of dimension 27:
\begin{equation}
{\bf 27}=\left(Q,u^c,e^c,L,d^c,\nu^c,H,D^c,H^c,D,S^c\right)_L.
\label{vs}
\end{equation}
In Eq.~(\ref{vs}), $Q$ is
a doublet containing the left-handed quarks, i.e.
\begin{equation}
Q=\begin{pmatrix}
u_L \\ d_L 
\end{pmatrix},
\label{ql}
\end{equation}
whereas $L$ includes the left-handed
leptons: 
\begin{equation}
L=\begin{pmatrix}
\nu_L \\ e_L 
\label{ll}
\end{pmatrix}.
\end{equation}
In Eqs.~(\ref{ql}) and (\ref{ll}), 
$u$, $d$ and $e$ denote generic quark and lepton flavours.
Likewise, $u^c_L$, $d^c_L$, $e^c_L$ and $\nu^c_L$ are singlets, which
are conjugate to the left-handed fields and thus
correspond to right-handed quarks and leptons
\footnote{Following \cite{Martin:1997ns}, the conjugate fields are related to
the right-handed ones via relations like $u^c_L=u^\dagger_R$.}.
In the case of supersymmetric extensions of the Standard Model, such as 
the MSSM, $Q$, $L$, $u^c_L$, $d^c_L$, $e^c_L$ and $\nu^c_L$
will be superfields containing also left-handed sfermions.
Furthermore, in Eq.~(\ref{vs}),
$H$ and $H^c$ are colour-singlet, electroweak doublets which can be
interpreted as Higgs pairs:
\begin{equation}
H=\begin{pmatrix}
\phi^0_1\\ \phi^-_1 
\end{pmatrix}\ \ ,\ \ H^c=\begin{pmatrix}
\phi_2^+\\ \phi^0_2
\end{pmatrix}.
\label{hhc}
\end{equation}
In the MSSM, 
$H$ and $H^c$ are superfields containing also the supersymmetric partners of
the Higgs bosons, i.e. the fermionic higgsinos.
Another possible 
description of the $H$ and $H^c$ fields in the representation {\bf 27}
is that they consist of left-handed exotic leptons (sleptons) $N$ and $L$,
with the same SM 
quantum numbers as the Higgs fields in Eq.~(\ref{hhc}) \cite{Gherghetta:1996yr}
\footnote{In the assumption that $H$ and $H^c$ contain exotic leptons,  
it is: $H=\begin{pmatrix}N_L\\ E_L \end{pmatrix}$
and $H^c=\begin{pmatrix}
E^c_L\\ N^c_L \end{pmatrix}$.}.
Moreover, in Eq.~(\ref{vs}), 
$D$ and $D^c$ are exotic vector-like quarks (squarks) and $S^c$ is a SM singlet
\footnote{A variety of notation is in use in the literature to
denote the exotic fields in the {\bf 27} representation. For example,
in \cite{rizzo,nardi1,nardi2} the exotic quarks $D$ and $D^c$ are
called $h$ and $h^c$.}.
In our phenomenological analysis, as well as in those performed in
Refs.~\cite{Gherghetta:1996yr,baum,chang}, leptons and quarks contained in the
$H$ and $D$ fields are neglected and assumed to be too heavy to 
contribute to $Z'$ phenomenology.
We are nevertheless aware that this is a quite strong assumption and that
in forthcoming BSM investigations one may well assume that such 
exotics leptons and quarks (sleptons and squarks) are lighter than the $Z'$ 
and therefore they can contribute to $Z'$ decays.

When E$_6$ breaks according to Eqs.~(\ref{upsi}) and (\ref{uchi}), the fields in
Eq.~(\ref{vs}) are reorganized according to
SO(10) and SU(5). The SU(5) representations are the following:
\begin{equation}
{\bf 10}=(Q,u^c,e^c)\ ,\  {\bf \bar 5}=(L,d^c)\ ,\  {\bf 1}=(\nu^c)\ ,\ 
{\bf \bar 5}=(H,D^c)\ ,\  {\bf 5}=(H^c, D)\ ,\  {\bf 1}=(S^c).
\label{repr10}
\end{equation}
From the point of view of SO(10), the assignment of the fields in the representations
{\bf 16}, {\bf 10} and {\bf 1} is not uniquely determined. In particular,
there is no actual reason to decide which ${\bf \bar 5}$ representation should
be included in {\bf 16} rather than in {\bf 10}. The usual assignment consists
in having in the  representation {\bf 16} the SM fermions and in the {\bf 10} 
the exotics:
\begin{equation} 
{\bf 16}=(Q,u^c,e^c,L,d^c,\nu^c)\ ,\  {\bf 10}=(H,D^c,H^c, D)\ ,\  
{\bf 1}= (S^c).
\label{rep}
\end{equation}
An alternative description is instead achieved by including $H$ and $D^c$ in
the {\bf 16}, with $L$ and $d^c$ in the {\bf 10}; this 'unconventional' E$_6$
scenario has been intensively studied in Refs.~\cite{nardi1,nardi2,nardi3} and 
leads to a different
$Z'$ phenomenology. In our paper, we shall assume the `conventional'
SO(10) representations, as in Eq.~(\ref{rep}).
Nevertheless, it can be shown \cite{nardi3} that, given a mixing angle $\theta$,
the unconventional E$_6$ scenario can be recovered by applying the
transformation:
\begin{equation}
\theta\to \theta+\arctan{\sqrt{15}}.
\label{unc}
\end{equation}
In fact, in our phenomenological analysis, we shall also
consider the \Uprime$_{\rm N}$ model leading to the so-called $Z'_{\rm N}$ boson,
with a mixing angle $\theta=\arctan{\sqrt{15}}-\pi/2$.
According to Eq.~(\ref{unc}), the $Z'_{\rm N}$ model corresponds to the
$Z'_\chi$ one, but in the unconventional E$_6$ scenario.
Table~\ref{tab:Models} summarizes the \Uprime-based models which will
be investigated throughout this paper, along with the values of the
mixing angle $\theta$.

The \Uprime\ charges of the fields in
Eq.~(\ref{vs}), assuming that they are organized in the SO(10) representations
as in (\ref{rep}),  are listed in Table~\ref{tabq}.
Under a generic \Uprime\ rotation, the charge of a field $\Phi$ is the 
following combination 
of the $\Uprime_\chi$ and $\Uprime_\psi$ charges:
\begin{equation}
Q'(\Phi)= Q'_\psi(\Phi) \cos \theta-Q'_\chi(\Phi) \sin\theta.
\label{qphi}
\end{equation}
\begin{table}[t]
\caption{\Zprime\  models along with the corresponding mixing angle, as given in Eq.~(\ref{zpri}).}
\begin{center}
\begin{tabular}{|c|c|}  
\hline                        
Model & $\theta$\\
\hline
\hline
{\it $\ZPSI$} & 0\\
\hline
{\it $\ZCHI$} & $-\pi/2$\\
\hline
{\it $\ZETA$} & $\arccos\sqrt{5/8}$\\ 
\hline
{\it $\ZS$} & $\arctan(\sqrt{15}/9)-\pi/2$\\
\hline
{\it $\ZI$} & $\arccos\sqrt{5/8}-\pi/2$\\
\hline
{\it $\ZN$} & $\arctan\sqrt{15}-\pi/2$\\
\hline
\end{tabular}
\label{tab:Models}
\end{center}
\end{table}
\begin{table}[htp]
\caption{\Uprime\  charges of the fields in the representation 27 of the Grand Unification
group E$_6$.}
\label{tabq}
\begin{center}
\small
\begin{tabular}{|c|c|c|}
\hline
 & $2\sqrt{10}Q'_\chi$ & $2\sqrt{6}Q'_\psi$\\ 
\hline\hline
$Q$ & -1  & 1   \\
\hline
$u^c$ & -1 & 1  \\
\hline
$d^c$ & 3  & 1 \\
\hline
$L$ & 3 & 1 \\
\hline
$\ell^c$ & -1 & 1  \\
\hline
$\nu_\ell^c$ & -5 & 1 \\
\hline
$H$ & -2& -2 \\
\hline
$H^c$ & 2 & -2 \\
\hline
$S^c$ & 0 & 4 \\
\hline
$D$ & 2 & -2  \\
\hline
$D^c$ & -2 & -2 \\
\hline
\end{tabular}
\end{center}
\end{table}
Besides the \Uprime\ gauge groups, 
another model which is experimentally investigated is the so-called Sequential Standard Model
(SSM), yielding a gauge boson 
$Z'_{\rm{SSM}}$, heavier than the $Z$ boson, 
but with the same couplings to fermions and gauge bosons as in the SM.
As discussed in the Introduction, although the SSM is not based on strong theoretical
arguments, studying the $Z'_{\rm{SSM}}$ phenomenology is very useful, since it depends only
on one parameter, the $Z'$ mass, and therefore it can set a benchmark for the
\Uprime-based analyses. 

In the following, 
the coupling constants of $\rm{U(1)}_Y$, $\rm{SU}(2)_L$ and \Uprime\ will be named $g_1$,
$g_2$ and $g'$, respectively, with $g_1=g_2\tan\theta_W$, $\theta_W$ being the Weinberg angle.  
We shall also assume, 
as occurs in E$_6$-inspired models, a proportionality relation
between the two U(1) couplings, as originally proposed in  \cite{rosner}:
\begin{equation}
g'=\sqrt{\frac{5}{3}}g_1.
\end{equation}

Before closing this subsection, we wish to stress that, in general, 
the electroweak-interaction eigenstates $Z$ and \Zprime\ mix to yield the mass eigenstates, usually
labelled as $Z_1$ and $Z_2$. Ref.~\cite{munir} addressed this issue by using  
precise electroweak data from several experiments and concluded that the mixing angle
$\theta_{ZZ'}$ is very small for any $Z'$ model, namely 
$\sin\theta_{ZZ'}\sim 10^{-3}$-$10^{-4}$.
Likewise, even the $Z\Zprime$ mixing 
associated with the extra kinetic terms due to the two U(1) groups is small and
can be neglected \cite{march}.

\subsection{Particle content of the Minimal Supersymmetric Standard Model}

The  Minimal Supersymmetric Standard Model (MSSM) is the most investigated scenario for supersymmetry,
as it presents a limited set of new parameters and particle content with respect to the Standard Model.
Above all, the MSSM contains the supersymmetric partners of the SM particles: 
scalar sfermions,
such as sleptons $\tilde \ell^\pm$ and $\tilde\nu_\ell$ ($\ell = e,\mu, \tau$) and
squarks $\tilde q$ ($q=u,d,c,s,t,b$), and fermionic
gauginos, i.e. \gluino, ${\tilde W}^\pm$, \susy{Z} and \susy{\gamma}.
It exhibits two Higgs doublets, which, after giving mass to $W$ and $Z$ bosons, lead 
to five scalar degrees of freedom, usually  
parametrized in terms of two CP-even 
neutral scalars, $h$ and $H$, with $h$ lighter than $H$, one  CP-odd 
neutral pseudoscalar $A$, and a pair of charged Higgs
bosons \Hpm. Each Higgs has a supersymmetric fermionic partner, named higgsino.
The light scalar Higgs, i.e. $h$, roughly corresponds to the SM Higgs.

The weak gauginos mix with the higgsinos to form the 
mass eigenstates: two pairs of charginos (\chinoonepm and \chinotwopm)
and four neutralinos (\ninoone, \ninotwo, \ninothree and \ninofour),
where $\ninoone$ is the lightest and $\ninofour$ the heaviest.
Particle masses and couplings in the MSSM are determined after 
diagonalizing the relevant mass matrices.
Hereafter, we assume the conservation of R-parity, 
with the values $\rm{R}_{\rm p}$ = +1 for SM particles and  
$\rm{R}_{\rm p}$ = -1 for their supersymmetric partners. This implies the existence of a 
stable Lightest Supersymmetric Particle (LSP), present in any supersymmetric
decay chain. The lightest neutralino, i.e. \ninoone, is often assumed to be the LSP.

As for the Higgs sector, besides the two Higgs doublets of the MSSM, 
the extra \Zprime\ requires another singlet Higgs to break the \Uprime\ symmetry
and give mass to the \Zprime\ itself. 
Moreover, two extra neutralinos are necessary, since one has a new neutral gaugino, i.e. 
the supersymmetric partner of the \Zprime, and a further higgsino, associated with the above extra Higgs.
As for the sfermions, squark and slepton masses will get an 
an extra contribution to the so-called D-term, 
depending
on the \Uprime\ sfermion charges and Higgs vacuum expectation values.
As will be discussed below, such D-terms, when summed to the soft masses
and to the F-terms, 
will have a crucial impact on sfermion spectra and, whenever
large and negative, they may even lead to discarding some MSSM/\Uprime\ scenarios.

\section{Extending the MSSM with the extra {$\mathbf U(1)'$} group}\label{sec:Model}

In our modelling of \Zprime \ production and decay into SM as well as supersymmetric
particles, the phenomenological analysis in Ref.~\cite{Gherghetta:1996yr} 
will be further expanded and generalized.
In this section  we summarize a few relevant points which are important for our discussion,
referring to the work in \cite{Gherghetta:1996yr} for more details.

\subsection{Higgs bosons in the MSSM and U{$\mathbf (1)'$} models}\label{sec:Higgs}

The two Higgs doublets predicted by the MSSM ($\Phi_1$ and $\Phi_2$) 
can be identified with the
scalar components of the superfields $H$ and $H^c$ in Eq.~(\ref{vs}), 
whereas the extra Higgs ($\Phi_3$), necessary
to break the \Uprime\ symmetry and give mass to the $Z'$, is associated with the
scalar part of the singlet $S^c$. 
The three Higgs bosons are thus two weak-isospin
doublets and one singlet:
\begin{center}
$\Phi_1=\left(\begin{matrix}\phi_1^0\\ \phi_1^- \end{matrix}\right)$\ ,\ 
$\Phi_2=\left(\begin{matrix}\phi_2^+\\ \phi_2^0 \end{matrix}\right)$\ ,\ 
$\Phi_3=\phi_3^0$. \end{center}
The vacuum expectation values 
of the neutral Higgs bosons are given by $\langle \phi_i^0\rangle=v_i/\sqrt{2}$,
with $v_1<v_2<v_3$. From the Higgs vacuum expectation values, one obtains the MSSM
parameter $\tan\beta$, i.e.
\begin{equation}
\tan\beta=v_2/v_1.
\end{equation}
Hereafter, we shall denote the Higgs charges according to the U(1)$'$ symmetry as:
\begin{equation}
Q'_1=Q'(H)\ \ ,\ \ Q'_2=Q'(H^c)\ \ ,\ \  Q'_3=Q'(S^c).
\label{q123}
\end{equation}
Their values can be obtained using the numbers in Table~\ref{tabq} and
Eq.~(\ref{qphi}).

The MSSM superpotential contains a Higgs coupling term giving rise to the well-known
$\mu$ parameter; because of the extra field $\Phi_3$, our model
presents the additional
contribution $W=\lambda\Phi_1\Phi_2\Phi_3$, leading to a trilinear scalar potential
for the neutral Higgs bosons 
\begin{equation}
V_\lambda=\lambda A_\lambda\phi_1^0\phi_2^0\phi_3^0.
\label{vh}
\end{equation}
The parameter $\lambda$ in Eq.~(\ref{vh}) 
is related to the usual $\mu$ term by means of the following relation,
involving the vacuum expectation value of $\phi^0_3$
\cite{Gherghetta:1996yr}:
\begin{equation}
\mu=\frac{\lambda v_3}{\sqrt{2}}.
\end{equation} 
After symmetry breaking and giving mass to $W$, $Z$ and \Zprime\  bosons, one is left with two
charged (\Hpm), and four neutral Higgs bosons, i.e. one pseudoscalar \AA \ and three scalars $h$, $H$  and $H'$
\footnote{We point out that in \cite{Gherghetta:1996yr} the three neutral Higgs bosons
are denoted by $H^0_i$, with $i=1,2,3$ and the pseudoscalar one by $P^0$.}.
Following \cite{barger}, the charged-Higgs mass
is obtained by diagonalizing the mass mixing matrix
\begin{equation}
{\cal M^{\rm 2}_{\rm \Hpm}} = \frac{1}{2}
 \left( \begin{matrix}
(g_2^2/2-\lambda^2)v_1^2+ \lambda A_\lambda v_1v_3/v_2 &(g_2^2/2-\lambda^2)v_1v_2 +\lambda A_\lambda v_3 \\
(g_2^2/2-\lambda^2)v_1v_2+ \lambda A_\lambda v_3 &(g_2^2/2-\lambda^2)v_2^2 +\lambda A_\lambda v_2 v_3/v_1 \\
\end{matrix} \right)
\end{equation}
and is given by
\begin{equation}
m_{\Hpm}^2 = \frac{\lambda A_\lambda v_3}{\sin 2 \beta} + \left( 1- 2 \frac{\lambda^2}{g_2^2} \right) \mW^2.
\label{hpm}
\end{equation}

We refer to \cite{Gherghetta:1996yr} for the mass matrix of the CP-even neutral Higgs bosons: 
the mass eigenvalues are to be evaluated numerically and cannot be
expressed in closed analytical form. One can nonetheless anticipate that the
mass of the heaviest $H'$ is typically about the \Zprime\  mass, and
therefore the \Zprime\  cannot decay into channels containing $H'$.
 
The mass of the pseudoscalar Higgs  \AA \  is obtained after diagonalizing its $2\times 2$
mass matrix and can be computed analytically, as done in
\cite{barger}:
\begin{equation}
\mA ^2 = \frac{\lambda A_\lambda v_3}{\sin 2\beta} \left(1 +\frac{v^2}{4v_3^2}
\sin^2 2 \beta \right), 
\label{amass}
\end{equation}
where $v=\sqrt{v_1^2+v_2^2}$\footnote{In Eqs.~(\ref{hpm}) and (\ref{amass}) we 
have fixed
the typing mistakes contained in Ref.~\cite{Gherghetta:1996yr}, wherein the 
expressions for the masses of charged and pseudoscalar Higgs bosons contain extra factors of 2.}.

\subsection{Neutralinos and charginos}\label{sec:neutralino}

Besides the four neutralinos of the MSSM, $\ninoone, \dots,\ninofour$, 
two extra neutralinos are required, 
namely  $\ninofive$ and $\ninosix$, associated with the \Zprime\  and with the new
neutral Higgs breaking \Uprime.
The $6\times 6$ neutralino mass matrix is typically written in the
basis of the supersymmetric neutral bosons $(-i\tilde B, -i\tilde W_3, -i\tilde B', \tilde \Phi_1,
\tilde \Phi_2, \tilde\Phi_3)$. It depends on the Higgs vacuum expectation values, on
  the soft masses of the gauginos $\tilde B$, $\tilde W_3$ and $\tilde B '$,
named $\Muno$, $\Mdue$ and $\Mprime$ hereafter,
and on the Higgs \Uprime\  charges $Q'_1$, $Q'_2$ and $Q'_3$. 
It reads:
\begin{equation}\label{neumass}
{\cal M}_{\tilde\chi^0}=\left( \begin{matrix}
\Muno& 0    & 0       &  -\frac{1}{2}g_1v_1      & \frac{1}{2}g_1v_2  &   0\\
 0   &\Mdue & 0       &\frac{1}{2}g_2v_1         &\frac{1}{2}g_2v_2    &   0 \\
 0   & 0    & \Mprime & Q^{\prime}_1 g^{\prime}v_1& Q^{\prime}_2 g^{\prime}v_2&Q^{\prime}_3 g^{\prime}v_3 \\
 -\frac{1}{2}g_1v_1& \frac{1}{2}g_2v_1 & Q^{\prime}_1 g^{\prime}v_1& 0& \frac{1}{\sqrt{2}}\lambda v_3&\frac{1}{\sqrt{2}}\lambda v_2 \\
 \frac{1}{2}g_1v_2& -\frac{1}{2}g_2v_2& Q^{\prime}_2 g^{\prime}v_2&\frac{1}{\sqrt{2}}\lambda v_3&0&\frac{1}{\sqrt{2}}\lambda v_1\\
0&0&Q^{\prime}_3 g^{\prime}v_3&\frac{1}{\sqrt{2}}\lambda v_2&\frac{1}{\sqrt{2}}\lambda v_1&0\\
\end{matrix} \right).
\end{equation}
The neutralino mass eigenstates $(\ninoone, \dots, \ninosix)$ and their
masses are obtained numerically after diagonalizing the above matrix.
Approximate analytic expressions for the neutrino masses, 
valid whenever $M_1$, $M_2$, $M^\prime$, $v_1$ and $v_2$ 
are much smaller than $v_3$, can be found in \cite{nandi}.

Since the new \Zprime\  and Higgs bosons are neutral, the chargino sector of the MSSM
remains unchanged even after adding the extra \Uprime\ group. 
The chargino mass matrix is given by \cite{Haber:1984rc}
\begin{equation}
    {\cal M}_{\tilde\chi^\pm}=\left( \begin{matrix}
\Mdue & \sqrt{2} \mW \sin \beta \\
\sqrt{2} \mW \cos \beta & -\mu
\end{matrix} \right)
\end{equation}
and its eigenvalues are
\begin{equation} 
m^2_{\tilde \chi^\pm_1,\tilde \chi^\pm_2}=\frac{1}{2}\left[|M_2|^2+|\mu|^2+2m_W^2\mp \sqrt{\Delta_{\tilde\chi}}\right],
\end{equation}
with
\begin{equation}
\Delta_{\tilde\chi}=(|M_2|^2+|\mu|^2+2 m_W^2)^2-4|\mu M_2-m_W^2\sin 2\beta|^2.
\end{equation}

\subsection{Sfermions}\label{sec:sfermions}

In principle, for the sake of a reliable determination of the sfermion masses, one would
need to carry out a full investigation within models for supersymmetry
breaking, such as gauge-, gravity- or anomaly-mediated 
mechanisms.
Studying supersymmetry-breaking scenarios
goes nevertheless beyond the scopes of the present work.
We just point out that supersymmetry can be spontaneously broken
if the so-called D-term and/or the F-term in the MSSM scalar potential have
non-zero vacuum expectation values, which can be achieved by means of 
the Fayet--Iliopoulos \cite{dterm} or O'Raifeartaigh \cite{fterm}
mechanisms, respectively. \footnote{The scalar potential is given, in terms of
D- and F-terms, by $V(\phi,\phi^*)=F^{*i}F_i+D^aD_a/2$, with
$D^a=-g_a(\phi^*T^a\phi)$ and $F_i=\delta W/\delta\phi_i$,
where $W$ is the superpotential, $\phi_i$ are the scalar (Higgs) fields,
$g_a$ and $T^a$ the coupling constant and the generators
of the gauge group of the theory.}

The sfermion squared masses can thus be
expressed 
as the sum of a soft term $m_0^2$, often set to the same value for both 
squarks and sleptons
at a given scale, and the corrections due to D- and F-terms
\cite{Martin:1997ns}.
The F-terms are proportional to the SM fermion masses and therefore they are
mostly relevant for the stop quarks.
The D-term can be, in principle, important for both light and heavy sfermions and,
for the purpose our study, it consists of two contributions.
A first term is a correction due to the hyperfine splitting 
driven by the
electroweak symmetry breaking, already present in the MSSM. 
For a fermion $a$ of weak isospin $T_{i,a}$, weak hypercharge $Y_a$ and
electric charge $Q_a$, this contribution to the D-term reads:
\begin{equation}
\Delta\tilde m_a^2=(T_{3,a}g_1^2-Y_ag_2^2)(v_1^2-v_2^2)=(T_{3,a}-Q_a\sin^2\theta_W)m_Z^2\cos 2\beta.
\label{d1}
\end{equation}
A second contribution is due to possible 
extensions of the MSSM, such as our \Uprime\  group, and is related to
the Higgs bosons which break the new symmetry:
\begin{equation}
\Delta{\tilde m}_a^{\prime 2}=\frac{g'^2}{2}Q'_a(Q_1'v_1^2+Q_2'v_2^2+Q_3'v_3^2),
\label{dt}
\end{equation}
where $Q'_1$, $Q'_2$ and $Q'_3$
are the Higgs \Uprime\ charges defined in Eq.~(\ref{q123}) and $Q'_a$ 
the charge of sfermion $a$.
When dealing with the Sequential Standard Model \Zprime, only the first
contribution to the D-term, Eq.~(\ref{d1}), must be evaluated.

Left- and right-handed sfermions mix and therefore, in order to obtain the mass
eigenstates, one needs to diagonalize the following squared mass matrix:
\begin{equation}
{\cal M}_{\tilde f}^2=\left( \begin{matrix} (M_{LL}^{\tilde f})^2 \ \ \  
(M_{LR}^{\tilde f})^2\\ (M_{LR}^{\tilde f})^2 \ \ \  
(M_{RR}^{\tilde f})^2  \end{matrix} \right).
\label{smass}
\end{equation}
The value of the soft masses and the scale at which they are evaluated are
in principle arbitrary, as long as the physical sfermion masses, obtained
after diagonalizing the matrix (\ref{smass}),
fall within the current experimental limits for slepton and squark searches.
Following Ref.~\cite{Gherghetta:1996yr}, we  
assume a common soft mass of the order of few TeV for all the sfermions
at the \Zprime\ scale and add 
to it the D- and F-term contributions. 
Another possibility would be, as done e.g. in \cite{Martin:1997ns},
fixing the soft mass at a high ultraviolet scale, such as the Planck mass, and then evolving it
down to the typical energy of the process, 
by means of renormalization group equations.

As an example, we present the expression for the matrix elements in the case of an
up-type squark:
\begin{eqnarray}
( M^{\tilde u}_{LL})^2 &=& (m^0_{\tilde u_{L}})^2 + 
m^2_u + \left(\frac{1}{2}-\frac{2}{3}x_w \right)\mZ ^2\ 
\cos 2\beta + \Delta {\tilde m}^{\prime 2}_{\tilde u_L} \label{mll}\\
(M^{\tilde u}_{RR})^2 &=& 
(m^0_{\tilde u_R})^2 + m^2_u + \left(\frac{1}{2}-
\frac{2}{3}x_w \right)\mZ ^2\ \cos 2\beta + \Delta {\tilde m}^{\prime 2}_{\tilde u_R}
\label{mrr}\\  
( M^{\tilde u}_{LR})^2&=&  m_u\left( A_u - \mu \cot \beta \right).
\label{mlr}
\end{eqnarray}
where  $x_w=\swsq$,  $m^0_{\tilde u_{L,R}}$ is the $\tilde u_{L,R}$ 
soft mass at the \Zprime\  energy scale
and $A_f=m_uA_u$ is the coupling constant entering in the Higgs-sfermion interaction term. 

The dependence on $m_{Z'}$ and on the mixing angle $\theta$
is embedded in the
$\Delta \tilde m'^2_{\tilde u_{L,R}}$ 
term; analogous expressions hold for down squarks and sleptons \cite{Gherghetta:1996yr}.
In the following, the up-squark mass eigenstates will be
named as $\tilde u_1$ and $\tilde u_2$ and their masses as $m_{\tilde u_1}$
and $m_{\tilde u_2}$. Likewise, $\tilde d_{1,2}$, $\tilde \ell_{1,2}$ and
$\tilde\nu_{1,2}$ will be the mass eigenstates for down-type squarks, charged
sleptons and sneutrinos and their masses will be denoted by $m_{\tilde d_{1,2}}$,
$m_{\tilde \ell_{1,2}}$ and $m_{\tilde \nu_{1,2}}$, respectively. 

The terms $m_u^2$ in Eqs.~(\ref{mll}) and 
(\ref{mrr}), as well as the the mixing term (\ref{mlr}) are inherited by the
F-terms in the scalar potential.
As the mass of SM light quarks and leptons is very small, such terms 
are typically 
irrelevant and the mass matrix 
of sleptons and light squarks is
roughly diagonal.
On the contrary, the mixing term  $M_{LR}$ can be relevant for top squarks, 
and therefore the stop mass eigenstates $\tilde t_{1,2}$ can 
in principle be different from the weak eigenstates $\tilde t_{\rm L,R}$.
However, we can anticipate that, as will be seen later on, for a $Z'$ boson
with a mass of the order of a few TeV, much higher than the top-quark mass, even 
the stop mixing term will be negligible.


\section{Representative Point}\label{sec:Representative Point}

The investigation on $Z'$ production and decays into SM and BSM
particles depends on several parameters, such as the $Z'$ or 
supersymmetric particle
masses;
the experimental searches for physics beyond the Standard Model set
exclusion limits on such quantities \cite{pdg}.

In the following, we shall first consider a specific configuration
of the parameter space, which we call 'Representative Point', to study 
the $Z'$\ phenomenology in a scenario yielding
non-zero branching ratios in the more relevant decay channels.
Then, each parameter will be varied individually, in order to
investigate its
relevance on the physical quantities.

The set of parameters chosen is the following:
\begin{equation}
m_{Z^\prime}= 3~{\rm TeV} \ ,\ \theta=\arccos\sqrt{\frac{5}{8}}-\frac{\pi}{2},
\nonumber\end{equation}
\begin{equation}
\mu =200~{\rm GeV}\ ,\ \tan\beta=20\ ,\ A_q=A_\ell=A_\lambda=A_f=500~{\rm GeV}\ ,\nonumber\end{equation}
\begin{equation}
m^0_{\tilde q_L}=m^0_{\tilde q_R}=m^0_{\tilde\ell_L}=m^0_{\tilde\ell_R}=m^0_{\tilde\nu_L}=
m_{\tilde\nu_R}^0=2.5~{\rm TeV},\nonumber\end{equation}
\begin{equation}
M_1=100~{\rm GeV}\ ,\ M_2=200~{\rm GeV}\ ,\ M'=1~{\rm TeV},
\label{reprpoint}
\end{equation}
where the value of $\theta$ corresponds to the $Z'_I$ model and
by $q$ and $\ell$ we have denoted any possible quark and lepton flavour, respectively.
In Eq.~(\ref{reprpoint}) the gaugino masses $M_1$ and $M_2$ 
satisfy, within very good accuracy, the following relation, inspired
by Grand Unification Theories:
\begin{equation}
\frac{M_1}{M_2}=\frac{5}{3}\tan^2\theta_W.
\label{m1m2}
\end{equation}

\subsection{Sfermion masses}\label{sec:Sfermion}

\begin{figure}[ht]
\centerline{\resizebox{0.49\textwidth}{!}{\includegraphics{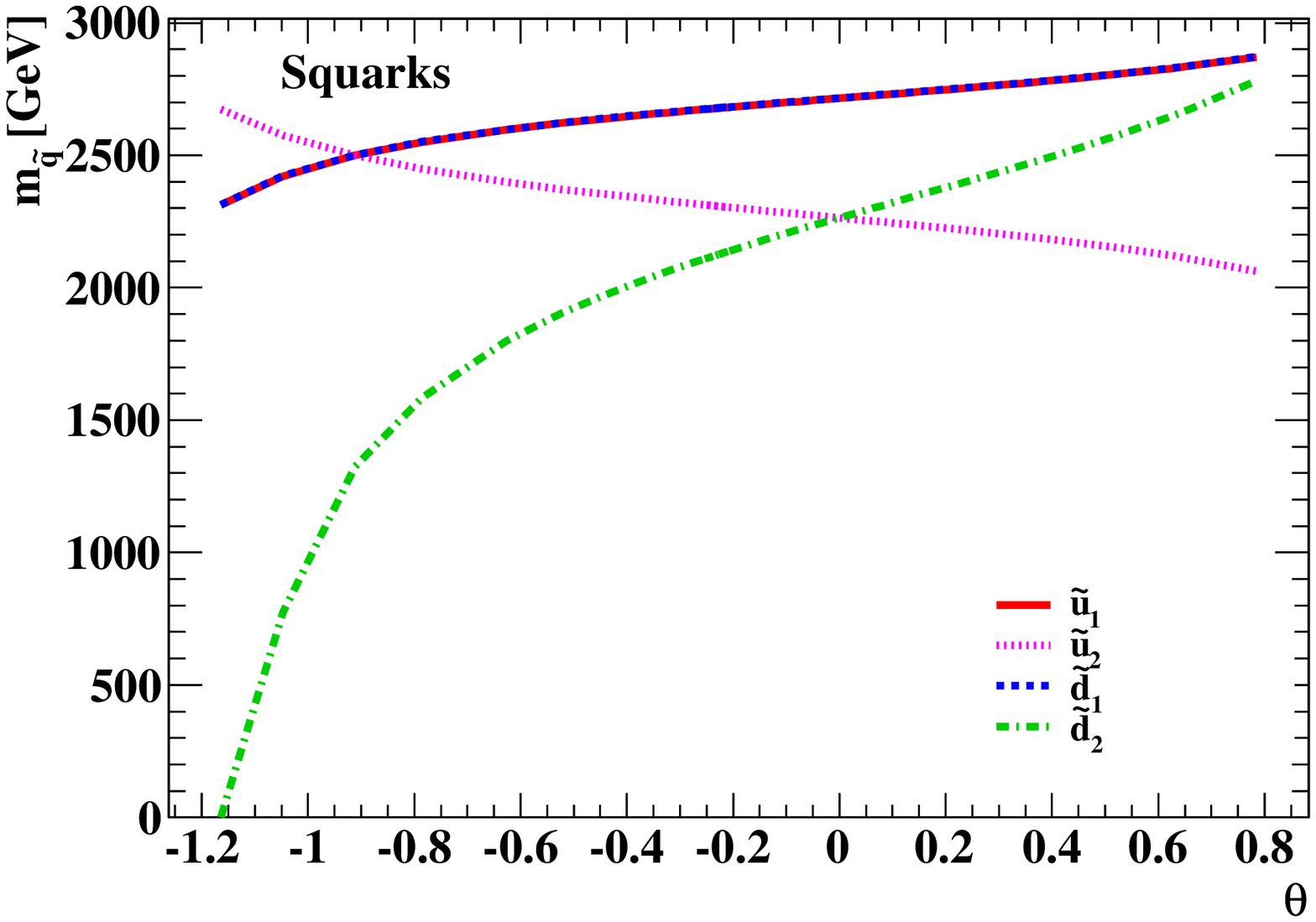}}%
\hfill%
\resizebox{0.49\textwidth}{!}{\includegraphics{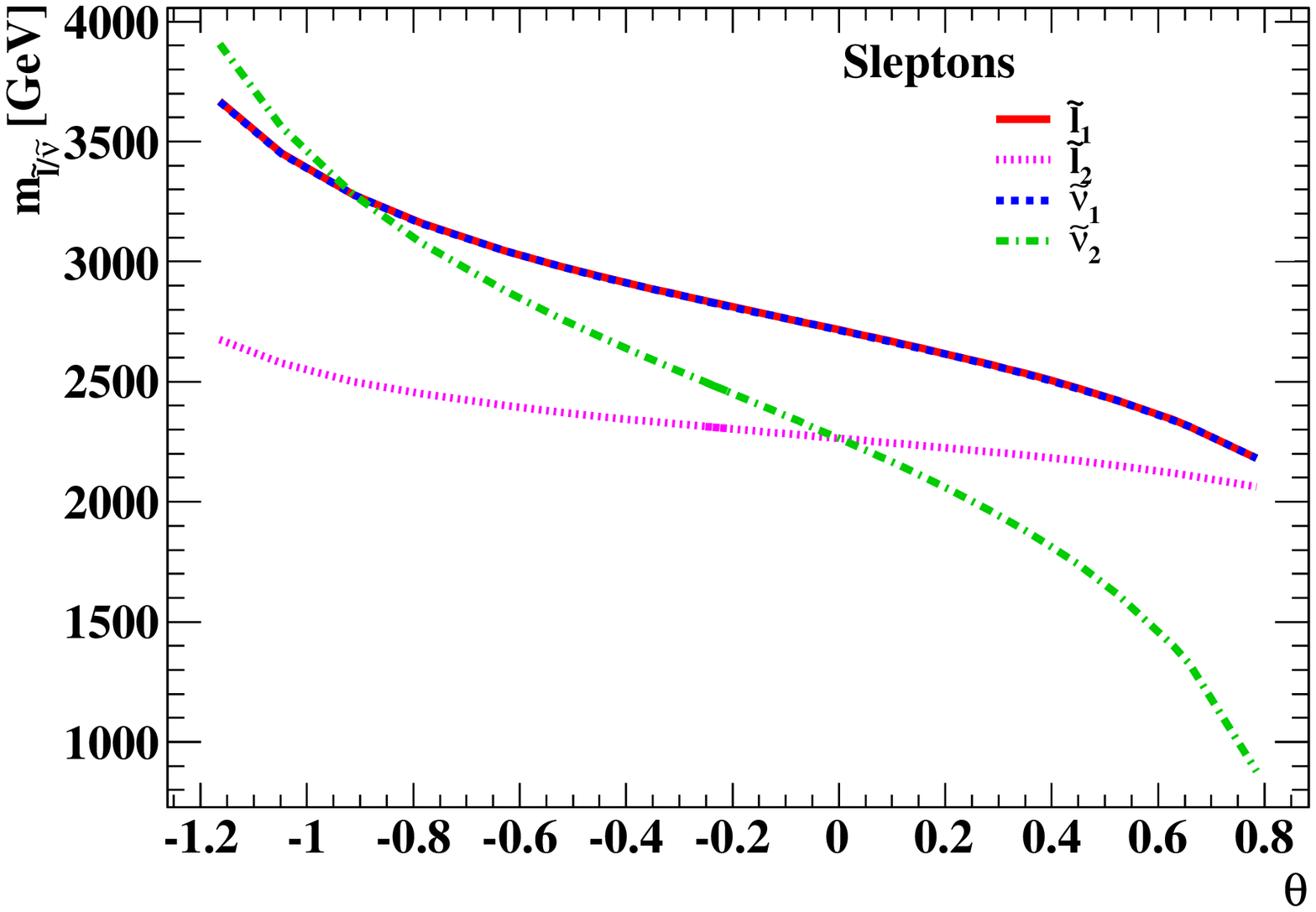}}}
\caption{Dependence on the \Uprime\ mixing angle $\theta$ of squark (left) and 
slepton (right) masses.}
\label{thetamass}
\end{figure}

The sfermion masses are given by 
the sum of a common soft
mass, which we have set to the same values for all squarks 
and sleptons at the $Z'$ scale, as in 
Eq.~(\ref{reprpoint}), and the F- and D-terms, given in Eqs.~(\ref{smass})--(\ref{mlr}).
The D-term, and then the sfermion squared masses, 
is expected to depend strongly on the  
\Uprime\ and MSSM parameters, and can possibly be negative and large, up to the
point of leading to an
unphysical (imaginary) sfermion mass. The F-term, being proportional to
the lepton/quark masses, is significant
only for top squarks.
In Fig.~\ref{thetamass} we study the dependence of  
squark (left) and slepton (right) 
masses on the \Uprime\ mixing angle $\theta$.
The symbols $\tilde u_{1,2}$  $\tilde d_{1,2}$,
$\tilde\ell_{1,2}$ and $\tilde\nu_{1,2}$ stand for 
generic up-, down-type squarks, charged sleptons and 
sneutrinos, respectively.
With the parametrization in Eq.~(\ref{reprpoint}), in particular the fact that the
$Z'$ mass has been fixed to 3 TeV, a value much higher than SM quark and lepton masses, 
the sfermion masses do not depend on the squark or slepton flavour.
In this case, even the stop mixing term is negligible, so that 
the $\tilde t_{1,2}$ masses are roughly equal to those of the other up-type squarks.

In Fig.~\ref{thetamass} 
the mass spectra are presented in the range $-1.2<\theta<0.8$: 
in fact, for $\theta<-1.2$ and $\theta>0.8$ the squared masses of $\tilde d_2$ and 
$\tilde\nu_2$ become negative and thus unphysical, respectively, 
due to a D-tem which is negative and large. This implies that the model $Z'_\chi$,
corresponding to $\theta=-\pi/2$, cannot be investigated within supersymmetry
for the scenario in Eq.~(\ref{reprpoint}),
as it does not yield a meaningful
sfermion spectrum. In the following, we shall still investigate the phenomenology
of the $Z'_\chi$ in a generic Two Higgs Doublet Model, 
but the sfermion decay modes will not contribute to its decay width. 
From Fig.~\ref{thetamass} (left) one can learn that the masses of
$\tilde u_1$ and $\tilde d_1$ are degenerate and
vary from about 2.2 to 3 TeV for increasing values of $\theta$, whereas
the $u_2$ mass decreases from 2.7 to about 2 TeV.
A stronger dependence on $\theta$ is exhibited by $m_{\tilde d_2}$:
it is almost zero for
$\theta\simeq -1.2$ and about 3 TeV for $\theta\simeq 0.8$. 
The slepton masses, as shown in  Fig.~\ref{thetamass} (right), decrease as $\theta$ increases:
the mass of $\tilde\ell_1$ 
is degenerate with $\tilde\nu_1$ and shows a larger variation (from 3.7 to 2.2 TeV) than
$\tilde \ell_2$ (from 2.7 to 2.2 TeV).
Sneutrinos $\tilde\nu_2$ exhibit a remarkable $\theta$ dependence: $m_{\tilde\nu_2}$
can be as high as 4 TeV for $\theta\simeq -1.2$ 
and almost zero for $\theta\simeq 0.8$.
\begin{figure}[ht]
\centerline{\resizebox{0.65\textwidth}{!}{\includegraphics{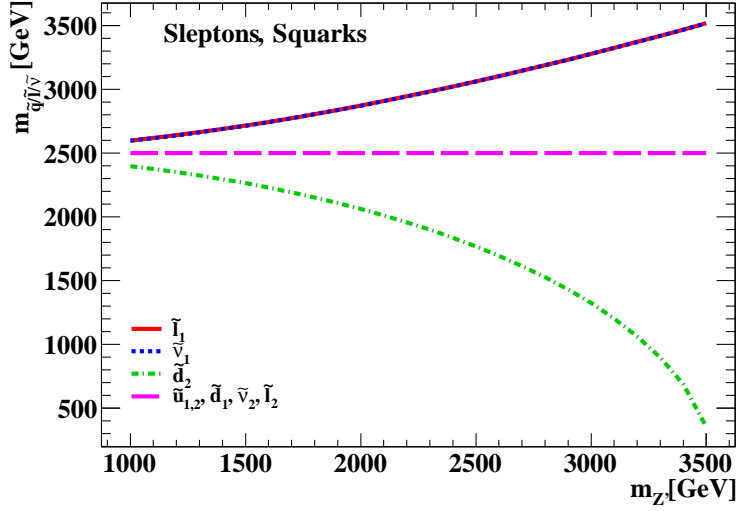}}}
\caption{Sfermion masses as a function of the $Z'$\ mass.}
\label{massmzp}
\end{figure}
\begin{figure}[ht]
\centerline{\resizebox{0.49\textwidth}{!}{\includegraphics{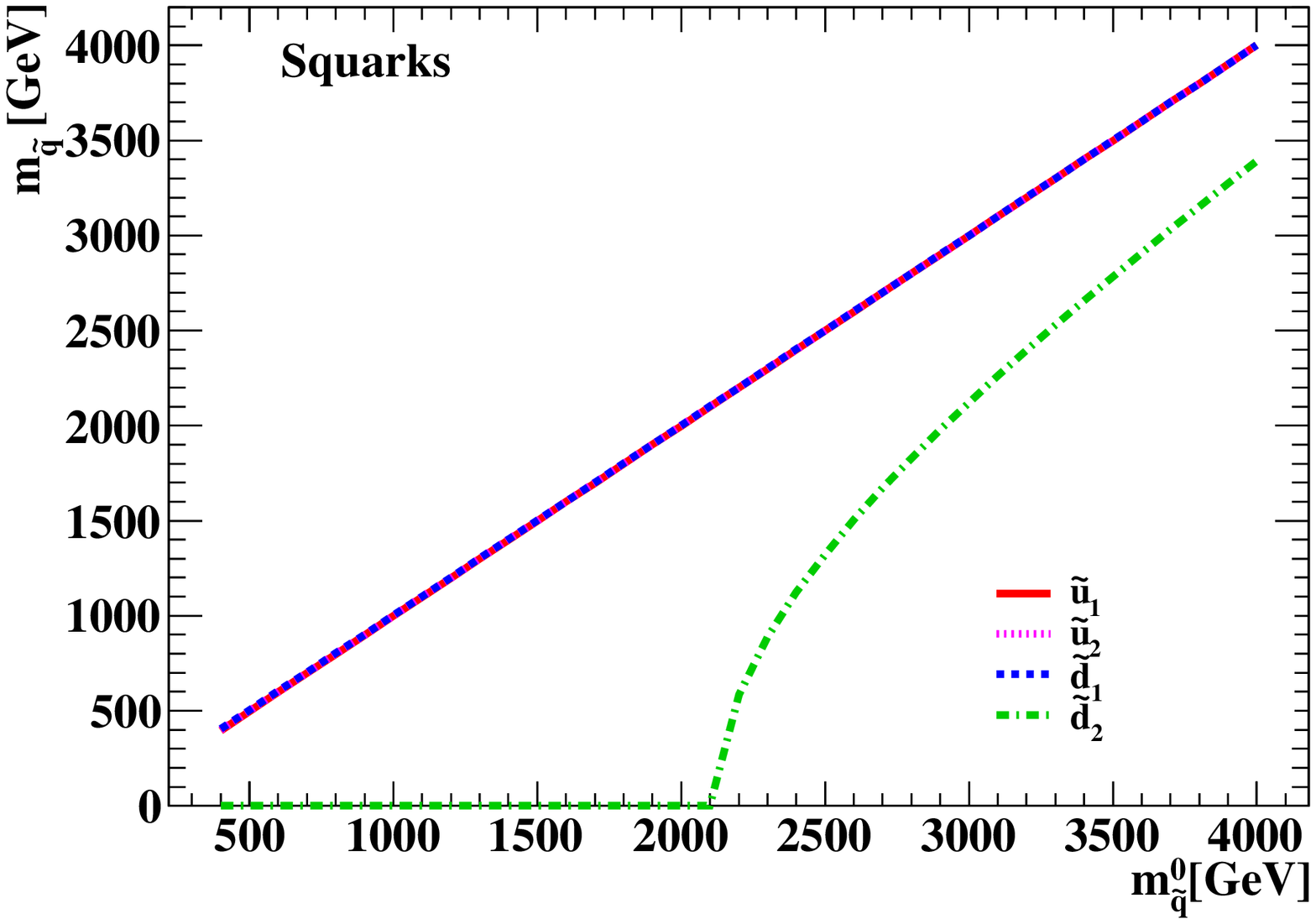}}%
\hfill%
\resizebox{0.49\textwidth}{!}{\includegraphics{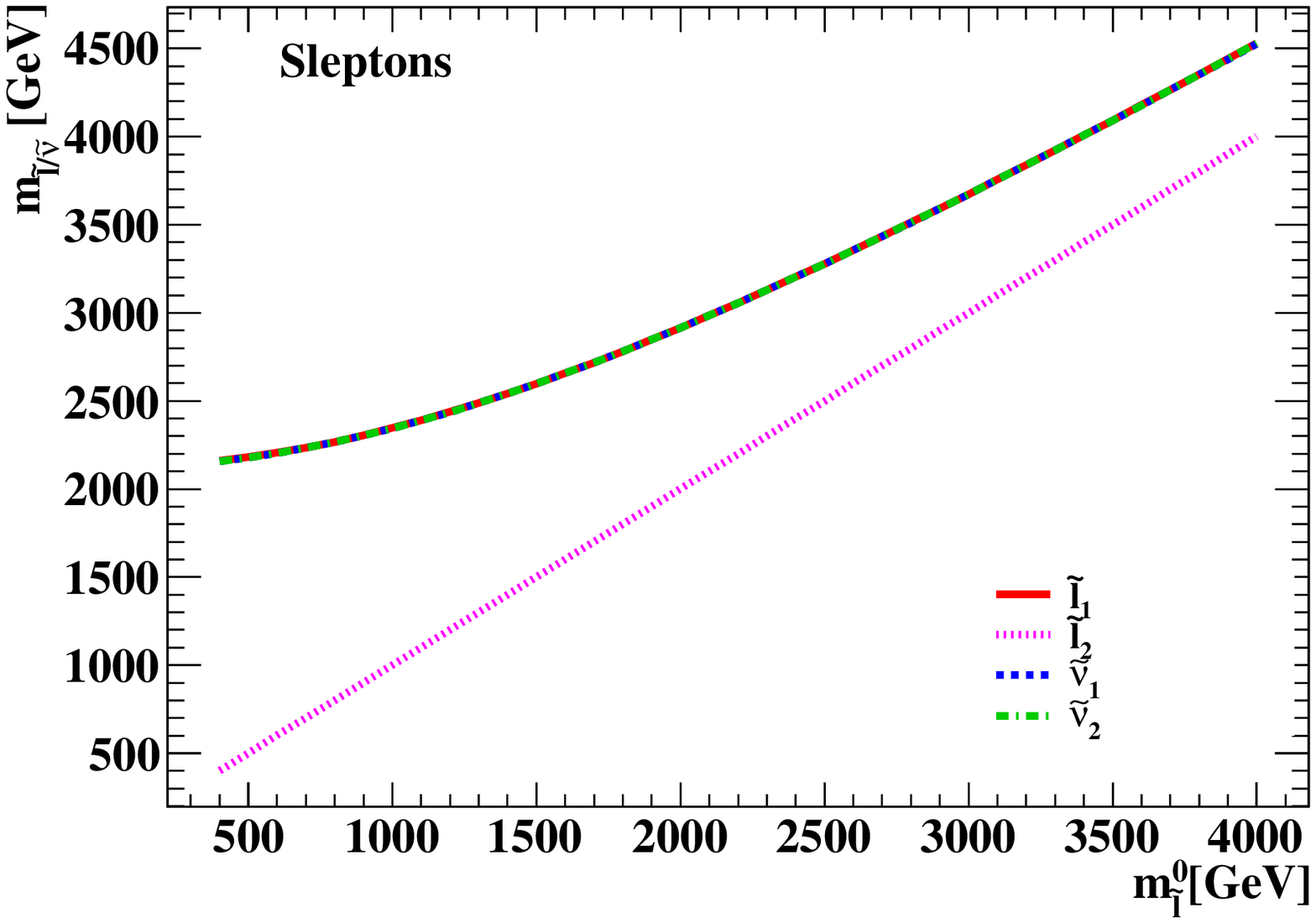}}}
\caption{Sfermion masses as a function of the initial
values $m^0_{\tilde q}$ and $m^0_{\tilde \ell}$, set at the $Z'$ mass scale.
Left: squarks. Right: sleptons.}
\label{m0fig}
\end{figure}

The D-term correction, and therefore the sfermion masses, 
is also function of the $Z'$ mass: this dependence is studied for
the $Z'_{\rm I}$ model and the parameters set as in the Representative Point, 
in the range $1~{\rm TeV}<m_{Z'}<3.5$~TeV. 
In Fig.~\ref{massmzp} 
the squark and slepton masses are plotted with respect to $m_{Z'}$, obtaining quite cumbersome results.
The masses of $\tilde u_{1,2}$, $\tilde d_1$, $\tilde\ell_2$
and $\tilde\nu_2$ are independent of $m_\Zprime$;
on 
the contrary, $m_{\tilde\nu_1}$ and $m_{\tilde\ell_1}$ are degenerate and increase from
2.5 TeV ($m_{Z'}=1$~TeV) to about 3.5 TeV ($m_{Z'}=3.5$~TeV).
The mass of $\tilde d_2$ is $m_{\tilde d_2}\simeq 2.4$~TeV for  $m_{Z'}=1$~TeV
and $m_{\tilde d_2}\simeq 0$ for  $m_{Z'}=3.5$~TeV;
due to the large negative D-term for 
$\tilde d_2$ squarks, no physical solution for
$m_{\tilde d_2}$ is allowed above $m_{Z'}=3.5$~TeV. 

The dependence of the sfermion masses on the initial values 
$m^0_{\tilde q}$ and $m^0_{\tilde\ell}$, set at the $Z'$ mass scale, and varied from
400 GeV to 4 TeV,
is presented in Fig.~\ref{m0fig}.
As expected, given Eqs.~(\ref{mll})-(\ref{mlr}), 
all sfermion masses are monotonically increasing function of $m^0_{\tilde f}$;
in the case of $\tilde u_1$, $\tilde u_2$, $\tilde d_1$
and $\ell_2$, being the D-term negligible, they are degenerate and
approximately equal to $m^0_{\tilde\ell,\tilde q}$ in the whole explored
range.
The mass of the squark $\tilde d_2$ is instead physical only for $m^0_{\tilde q}>$~2.1 TeV
and 
increases up to the value $m_{\tilde d_2}\simeq 3.3$~TeV for $m^0_{\tilde q}=4$~TeV.
The masses
$m_{\tilde\ell_1}$, $m_{\tilde\nu_1}$ and $m_{\tilde\nu_2}$ are also degenerate and 
vary from about 2.1 TeV ($m^0_{\tilde\ell}=400$~GeV) to 4.5 TeV 
($m^0_{\tilde\ell}=4$~TeV).

We also studied 
the variation of the sfermion masses with respect to $\tan\beta$,
in the range $1.5<\tan\beta<5$, and
on the trilinear coupling $A_f$, for $1~{\rm TeV}<A_f<4~{\rm TeV}$, 
but found very little dependence on such parameters. 
Moreover, there is no dependence on  $M_1$, $M_2$ and $M^\prime$,
which do not enter in the expressions of the 
sfermion masses, even after the D-term correction.

\subsection{Neutralino masses}\label{sec:Neutralino}

We wish to study the dependence of the neutralino masses on the parameters playing a role in our
analysis: unlike the sfermion masses, they depend also on the gaugino masses
$M_1$, $M_2$ and $M^\prime$.
Table~\ref{massneu} reports the six
neutralino masses for the parametrization in Eq.~(\ref{reprpoint}).
For $m_{Z'}=3$~TeV, $Z'$ decays into channels containing the
heaviest neutralino $\tilde\chi^0_6$
are not permitted because of phase-space restrictions.
and therefore they can be discarded in the Representative Point scenario.
Being $m_{\tilde\chi^0_5}\simeq 2.54$~TeV, decays into states containing 
$\tilde\chi^0_5$ are kinematically allowed, but one can already foresee
very small branching ratios.
\begin{table}[ht]
\caption{\label{massneu}Neutralino masses for a $Z'$\ mass of 
3 TeV and the parameters of the MSSM and \Uprime\ set as in Eq.~(\ref{reprpoint}).}
\begin{center}
\small
\begin{tabular}{|c|c|c|c|c|c|}
\hline
\Mninoone & \Mninotwo & \Mninothree     & \Mninofour & \Mninofive  &  \Mninosix    \\
\hline
\hline
94.6 GeV & 156.6 GeV & 212.2 GeV & 261.0 GeV & 2541.0 GeV & 3541.0 GeV \\
\hline
\end{tabular}
\end{center}
\end{table}\par
\begin{figure}[ht]
\centerline{\resizebox{0.49\textwidth}{!}{\includegraphics{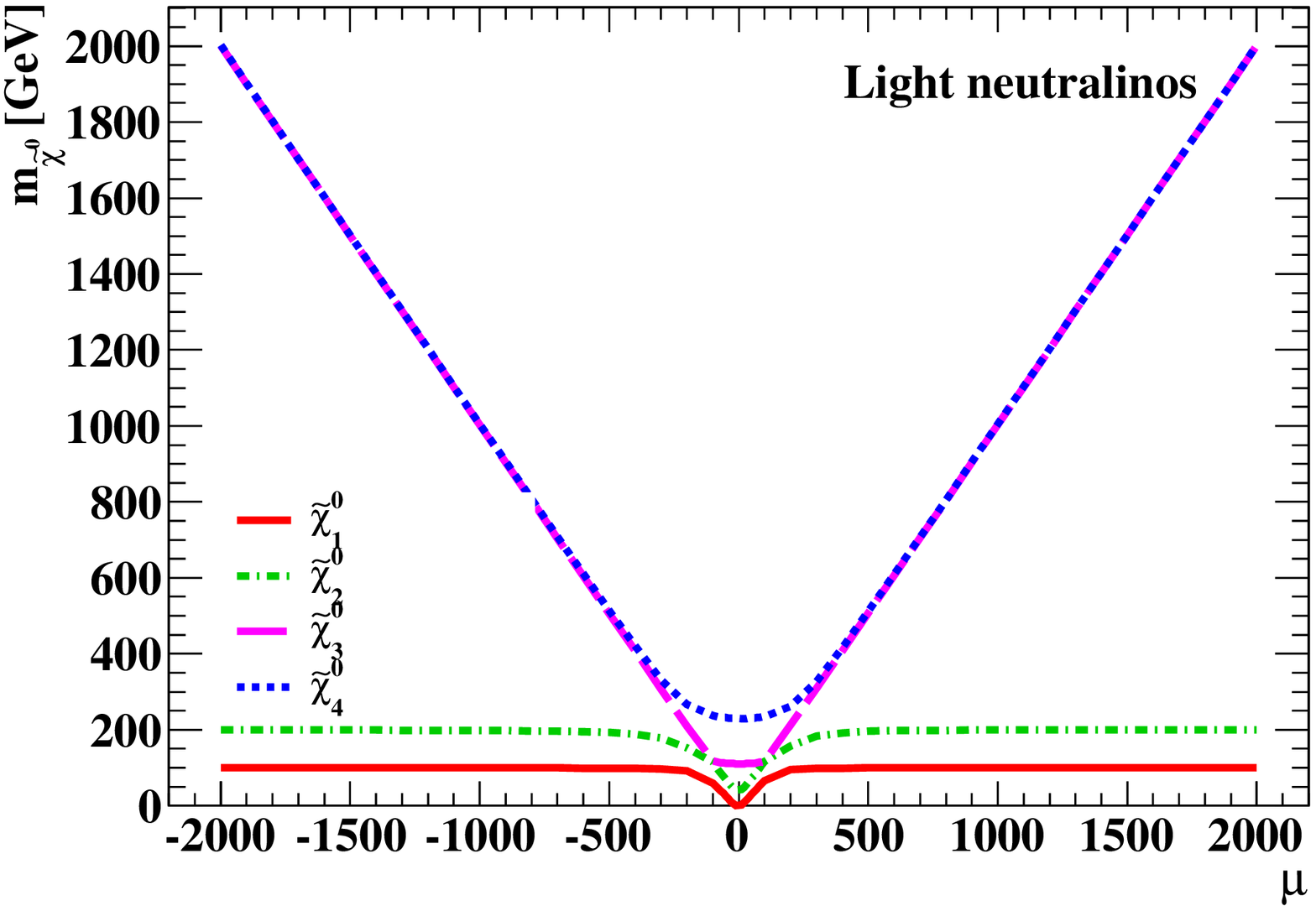}}%
\hfill%
\resizebox{0.49\textwidth}{!}{\includegraphics{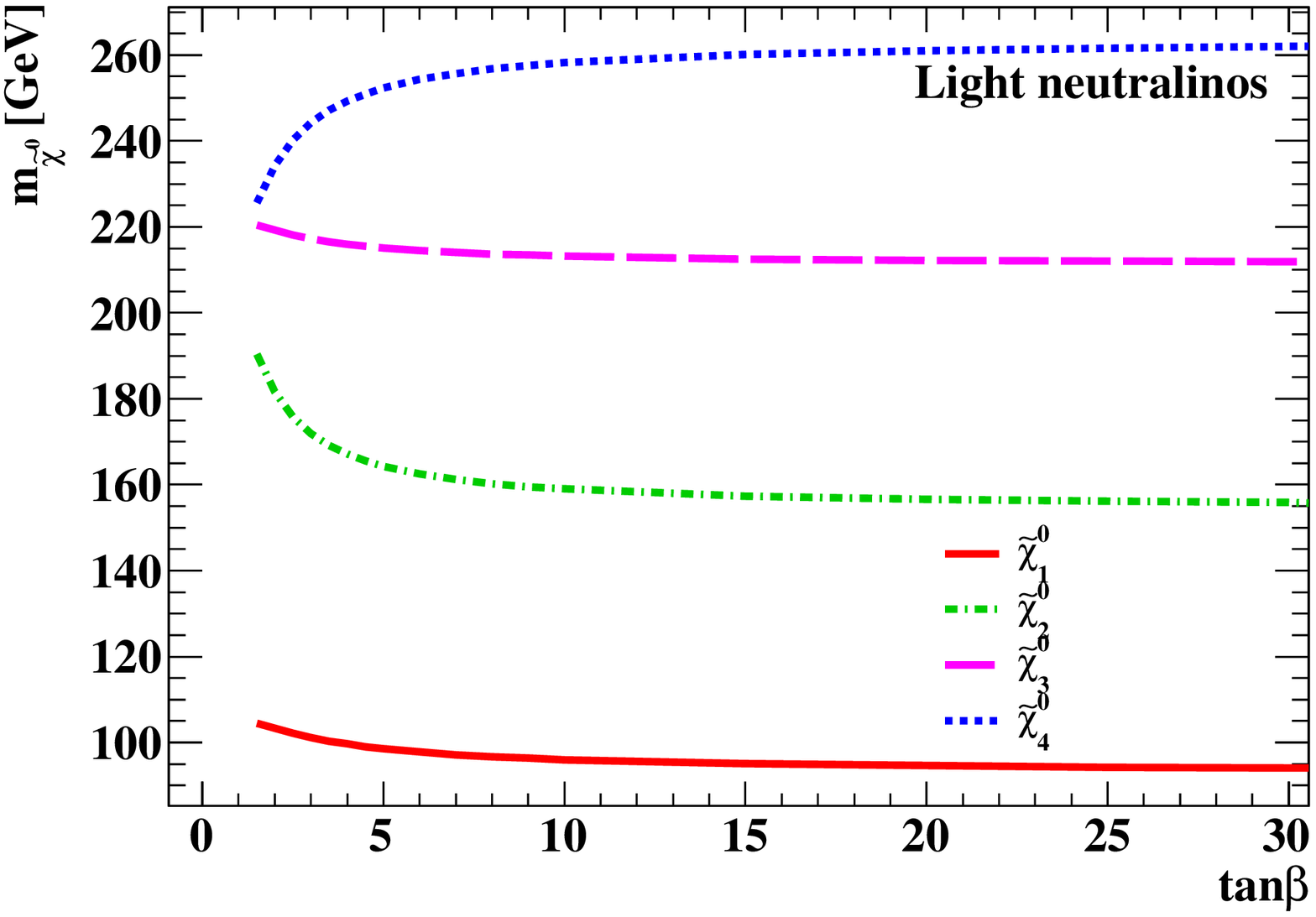}}}
\caption{Dependence of the mass of the four lightest neutralinos on the MSSM parameters $\mu$ (left) and
$\tan\beta$ (right).\label{neumutanb}}
\end{figure}

Figure~\ref{neumutanb} presents the dependence of the mass of the four
lightest neutralinos, i.e.
$\tilde\chi^0_1\dots \tilde\chi^0_4$, on the supersymmetry parameters $\mu$ (left) 
and $\tan\beta$ (right), for  $-2000~{\rm GeV}<\mu<2000~{\rm GeV}$ and $1.5<\tan\beta<30$,
with the others as in Eq.~(\ref{reprpoint}). 
The distribution of the masses of $\tilde\chi^0_1, \dots \tilde\chi^0_4$
is symmetric with respect to
$\mu=0$. Nevertheless, $m_{\tilde\chi^0_1}$ and $m_{\tilde\chi^0_2}$
increase from 0 ($\mu=0$) to about 100 ($m_{\tilde\chi^0_1}$) 
and 200 GeV ($m_{\tilde\chi^0_2}$) 
in the range $|\mu|<300$~GeV, whereas they are almost constant for $300~{\rm GeV}<|\mu|<2000{\rm GeV}$.
On the contrary, the masses of $\tilde\chi^0_3$ and $\tilde\chi^0_4$
exhibit a minimum for $\mu=0$, about 110 and
230 GeV respectively, and increase monotonically in terms of  $|\mu|$,
with a behaviour leading to $m_{\tilde\chi^0_3}\sim m_{\tilde\chi^0_4}\sim |\mu|$
for large $|\mu|$.   
As for $\tan\beta$, a small dependence is visible only in the low $\tan\beta$ range, 
i.e. $1.5<\tan\beta<8$, with the masses of $\tilde\chi^0_1$, 
$\tilde\chi^0_2$ and $\tilde\chi^0_3$ slightly decreasing and the one of 
$\tilde\chi^0_4$ mildly increasing.
Outside this range, the light neutralino masses are
roughly independent of $\tan\beta$.

\begin{figure}[ht]
\centerline{\resizebox{0.49\textwidth}{!}{\includegraphics{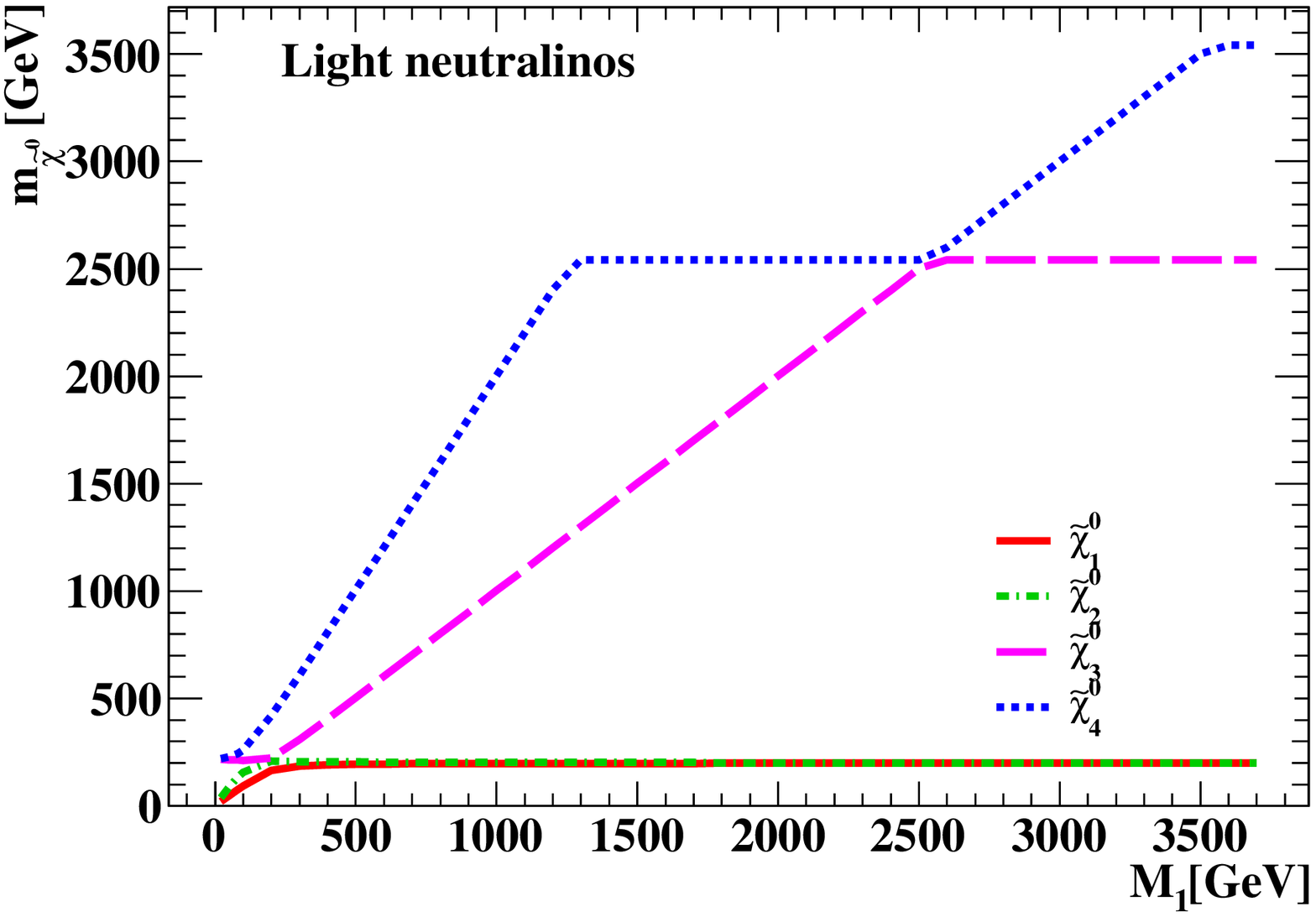}}%
\hfill%
\resizebox{0.49\textwidth}{!}{\includegraphics{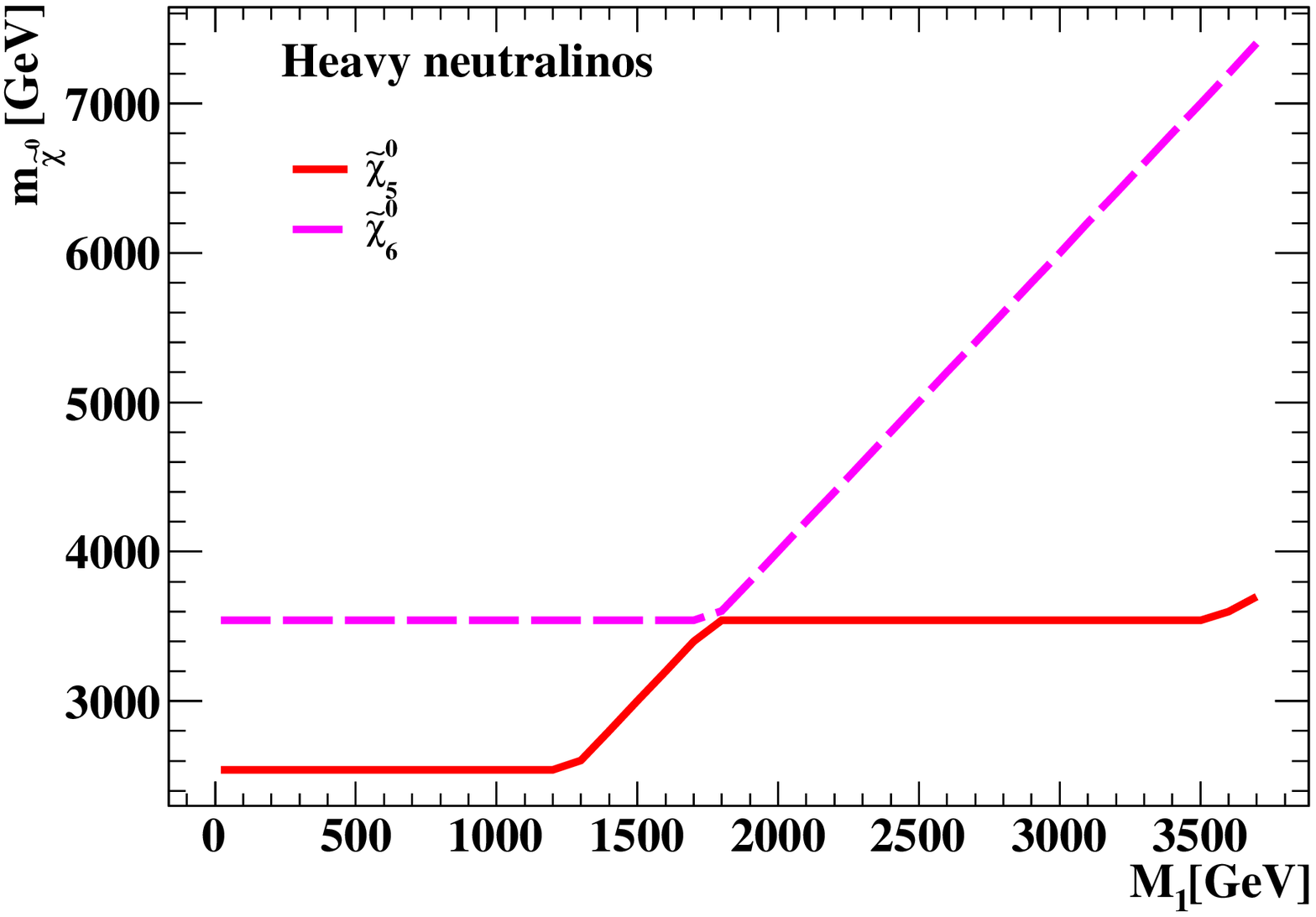}}}
\caption{Dependence of the neutralino masses on the MSSM parameter $M_1$.
Left: light neutralinos. Right: heavy neutralinos.
\label{neum1m2}}
\end{figure}

In Fig.~\ref{neum1m2} we present the dependence of the light (left)
and heavy (right) neutralino masses on the gaugino mass $M_1$ for
$M_1<3.7~{\rm TeV}$.
In the light case, the masses exhibit a step-like behaviour:
$m_{\tilde\chi^0_1}$ and $m_{\tilde\chi^0_2}$ have roughly the same value through
all $M_1$ range, growing for small
$M_1$ and amounting to approximately 200 GeV for 
$M_1>200$~GeV.  The mass
$m_{\tilde\chi^0_3}$ increases in the range $200~{\rm GeV}<M_1<2.5~{\rm TeV}$
and is about $m_{\tilde\chi^0_3}\simeq 2.54$~TeV for $M_1>$~2.5 TeV.
The mass of $\tilde\chi^0_4$ is roughly $m_{\tilde\chi^0_4}\simeq 2M_1$ for
$200~{\rm GeV}<M_1<1.2$~TeV, then 
$m_{\tilde\chi^0_4}\simeq 2.54$~TeV, up to $M_1\simeq 2.5$~TeV,
and ultimately $m_{\tilde\chi^0_4}\simeq M_1$ for larger $M_1$. 
As for the heavy neutralinos, the mass of $\tilde\chi^0_5$ is 
$m_{\tilde\chi^0_5}\simeq 2.54$~TeV for $M_1< 1.3$~TeV, then it increases
linearly in the range $1.3~{\rm TeV}<M_1<1.8$~TeV and it is 
$m_{\tilde\chi^0_5}\simeq 3.54$~TeV for $M_1>1.8$~TeV.
The mass of the heaviest neutralino $\tilde\chi^0_6$ is constant,
namely $m_{\tilde\chi^0_6}\simeq 3.54$~TeV, for $M_1<1.8$~TeV,
then it grows linearly, reaching the value $m_{\tilde\chi^0_6}\simeq 7$~TeV
for $M_1=3.5$~TeV.

\begin{figure}[ht]
\centerline{\resizebox{0.49\textwidth}{!}{\includegraphics{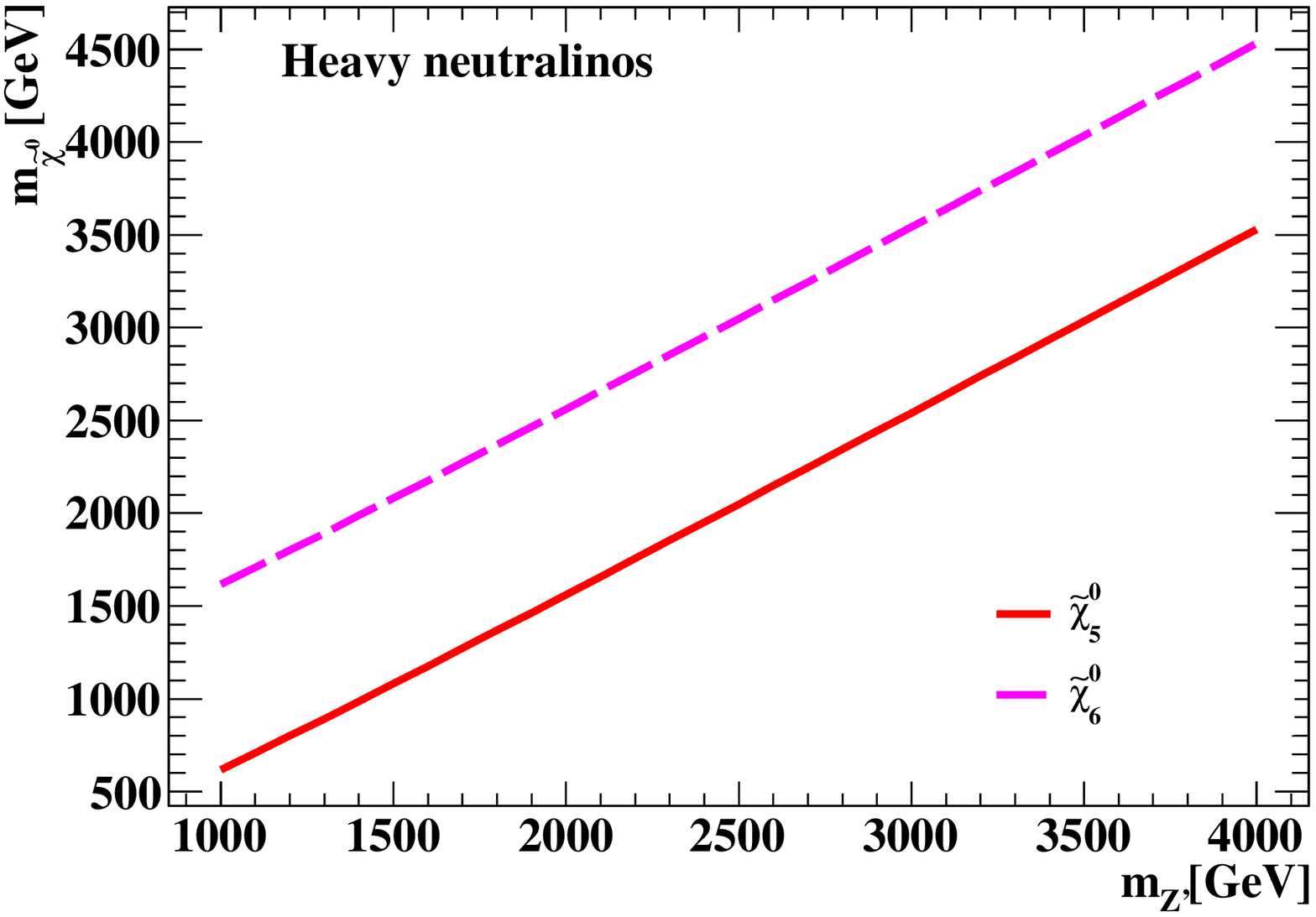}}%
\hfill%
\resizebox{0.49\textwidth}{!}{\includegraphics{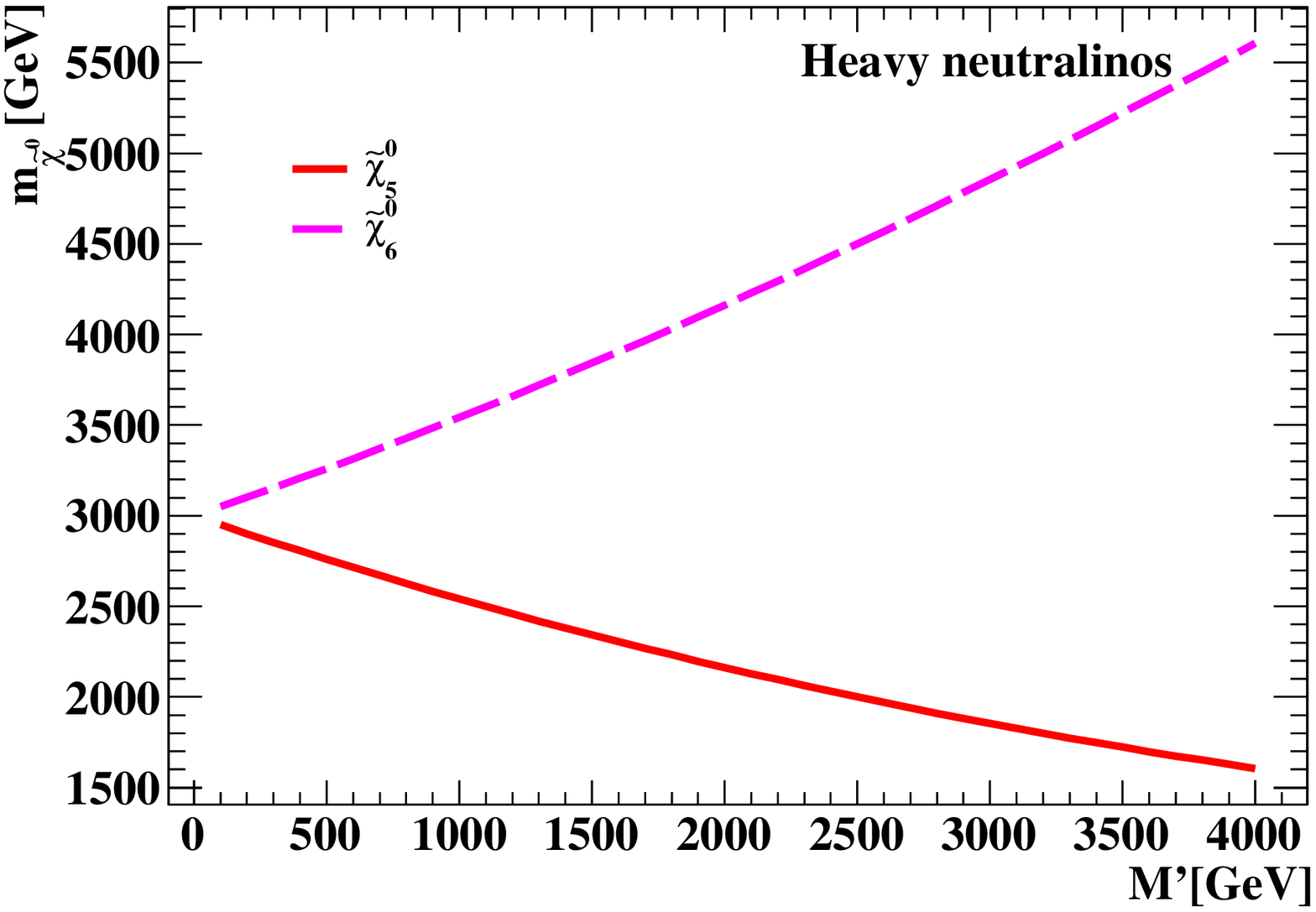}}}
\caption{Mass spectra of the heavy neutralinos as functions of the $Z'$ mass (left) and
the gaugino mass $M'$ (right). \label{neumprime}}
\end{figure}

Figure~\ref{neumprime} presents the masses of $\tilde\chi^0_5$ and
$\tilde\chi^0_6$ with respect to the $Z'$ mass in the range
1 TeV$<m_{Z'}<$~4 TeV (left) and to the $M'$ parameter for
100 GeV$<M'<$ 4 TeV (right). 
The masses of $\tilde\chi^0_5$ and $\tilde\chi^0_6$ grow linearly 
as a function of $m_{Z'}$, whereas they exhibit opposite behaviour with respect to 
$M'$, as $m_{\tilde\chi^0_6}$ increases from 3 to 5.5 TeV and  $m_{\tilde\chi^0_5}$ decreases
from 3 to 1.5 TeV.
The four light-neutralino masses are instead roughly independent of $m_{Z'}$ and $M'$,
as expected.

\begin{figure}[ht]
\centerline{\resizebox{0.49\textwidth}{!}{\includegraphics{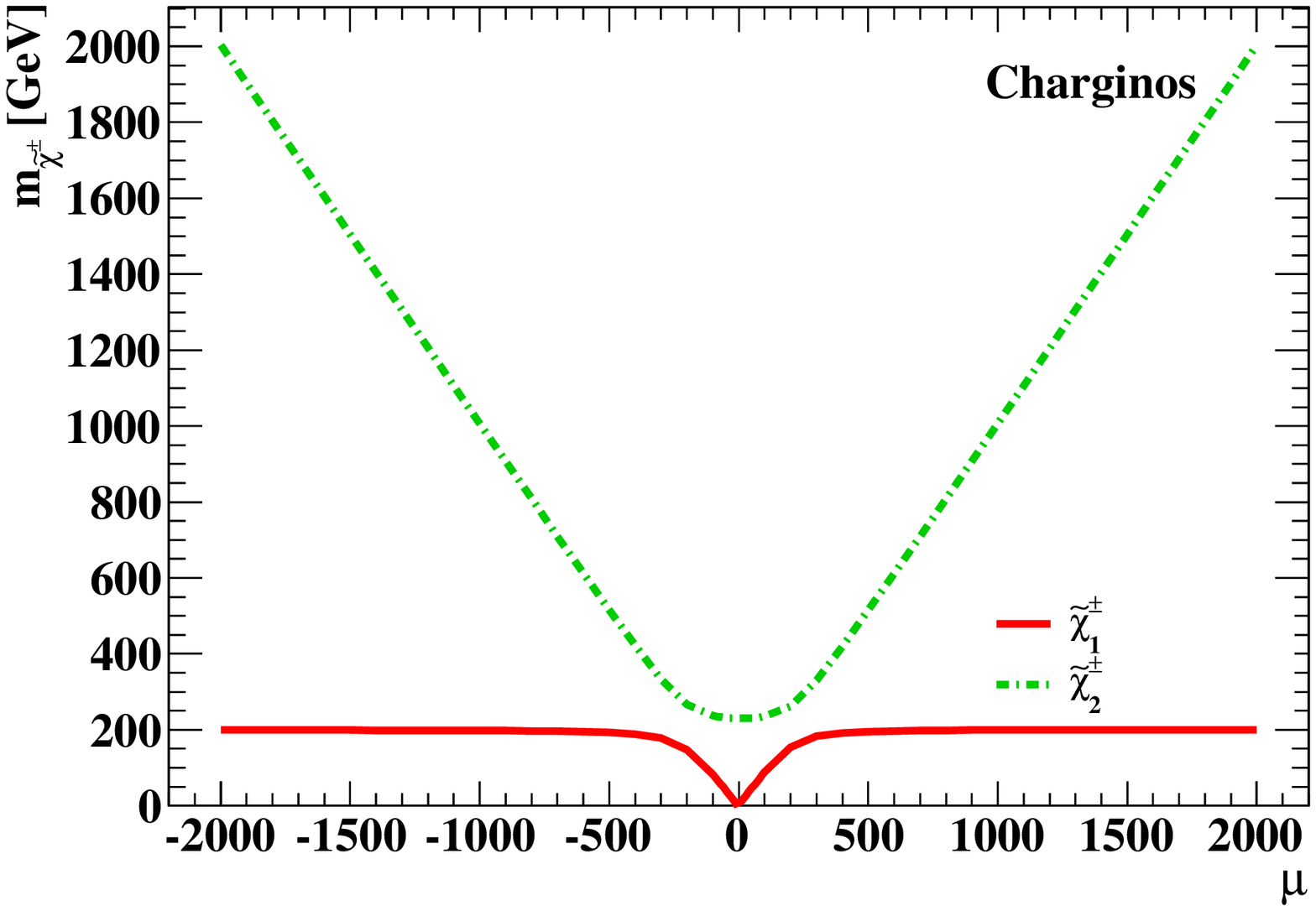}}%
\hfill%
\resizebox{0.49\textwidth}{!}{\includegraphics{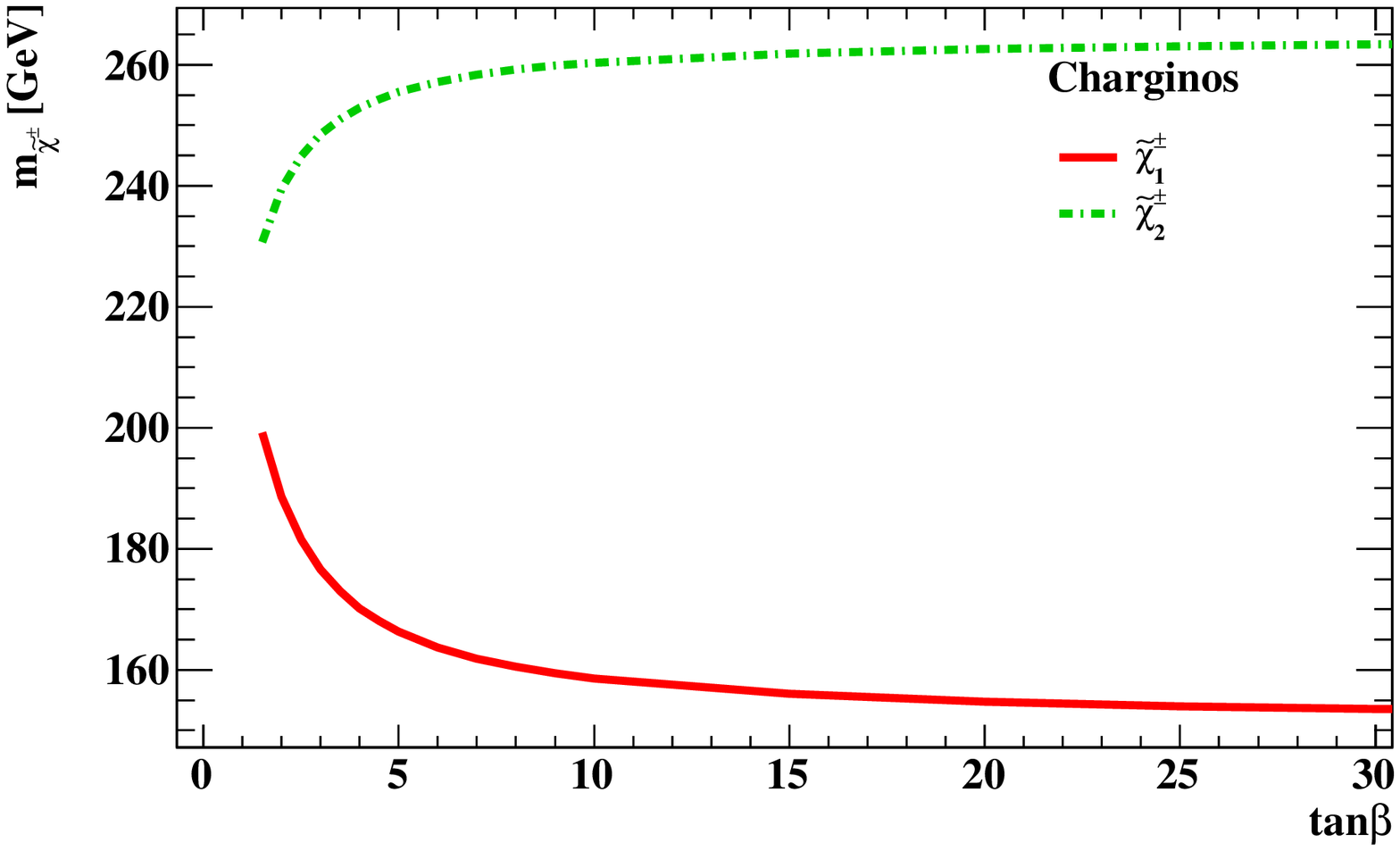}}}
\caption{Chargino masses with respect to the the MSSM parameters $\mu$ (left) and $\tan\beta$ (right).} 
\label{char1}
\end{figure}
\begin{figure}[ht]
\centerline{\resizebox{0.65\textwidth}{!}{\includegraphics{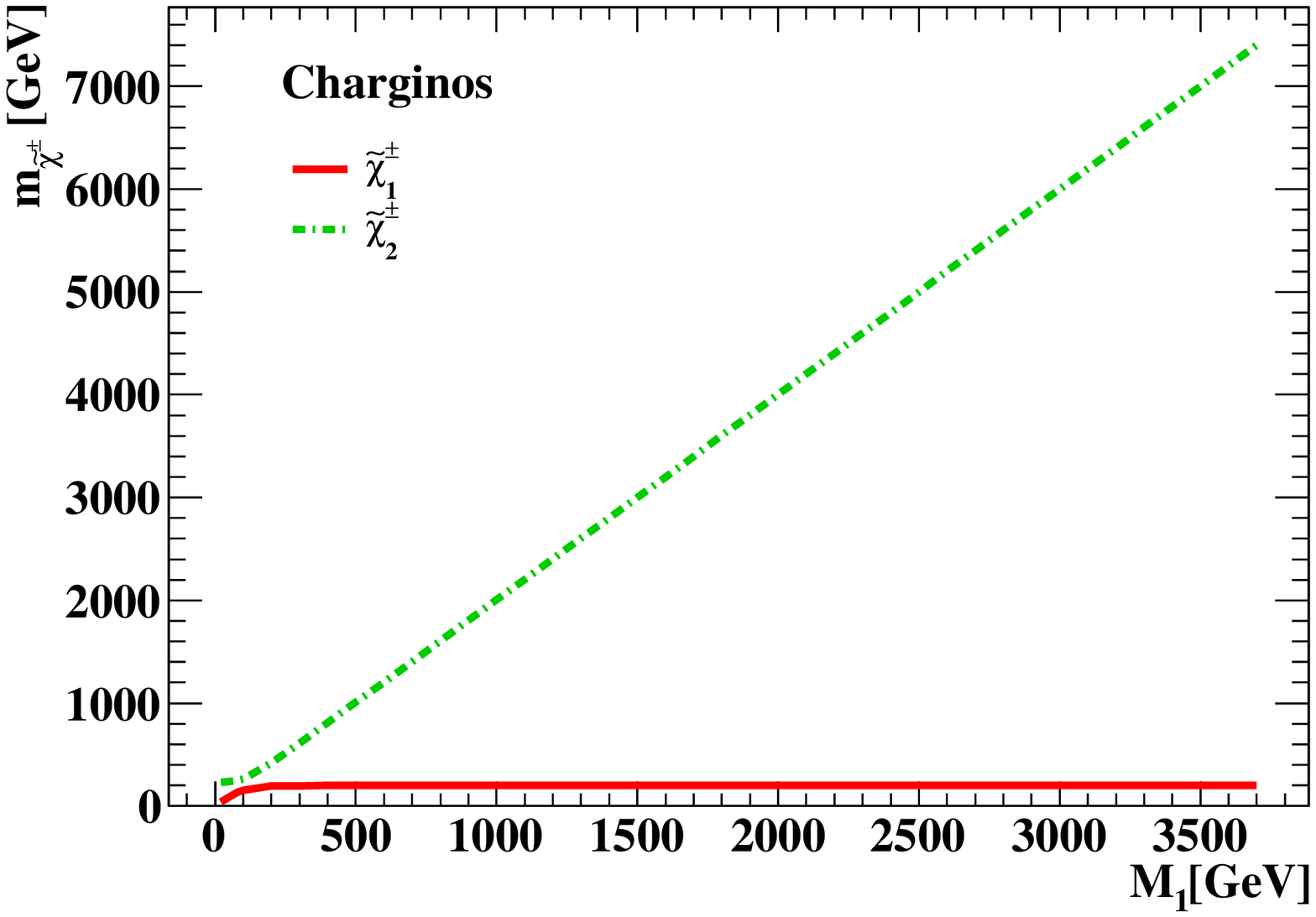}}}
\caption{Dependence of the chargino masses on the MSSM parameter $M_1$.}
\label{char2}
\end{figure}

\subsection{Chargino masses}
As discussed before, the chargino sector remains 
unchanged after the introduction of the extra group \Uprime.
Therefore, the chargino masses do not depend on the U(1)$'$ new parameter $M'$ and
on $m_{Z'}$, but just on the MSSM parameters $\mu$, $\tan\beta$ and $M_{1}$.
Figures~\ref{char1} and \ref{char2} show the dependence on such quantities,
which are varied individually, with the other parameters
fixed as in Eq.~(\ref{reprpoint}).

The dependence on $\mu$, displayed in Fig.~\ref{char1} (left), is symmetric with respect to 
$\mu=0$. In particular, $m_{\tilde\chi_1^\pm}$ varies 
significantly, from about 3 to 200 GeV,  only for $|\mu|<300$~GeV,
whereas the heavier chargino mass 
exhibits a behaviour $m_{\tilde\chi_1^\pm}\sim |\mu|$ and is as large as 
2 TeV for $|\mu|\simeq 2000$~GeV. As for $\tan\beta$, Fig.~\ref{char1} (right), 
the mass of the heavy 
chargino $\tilde\chi_2^\pm$ increases quite mildly from 230 to about 263 GeV,
whereas $m_{\tilde\chi^\pm_1}$ decreases from almost 200 GeV ($\tan\beta=1.5$) 
to about 154 GeV ($\tan\beta=30$).

The variation with respect to $M_1$, presented in
Fig.~\ref{char2}, is instead quite different for the 
two charginos. The mass of the lighter one 
changes very little only for $M_1<200$~GeV, whereas for larger $M_1$ it is about
$m_{\tilde\chi_1^\pm}\simeq 200$~GeV.
The mass of $\tilde\chi^\pm_2$ increases almost linearly with $M_1$ and 
and is $m_{\chi^\pm_2}\simeq 2 M_1\simeq M_2$ for large $M_1$.

\subsection{Higgs masses}\label{sec:Higgsmasses}

As pointed out before in the paper, after 
adding the \Uprime\ symmetry, one has an extra
neutral scalar Higgs, named $H'$, besides the Higgs sector of
the MSSM, i.e. the bosons $h$, $H$, $H^\pm$ and $A$.
The $Z'$ phenomenology will thus depend on the three Higgs masses and 
vacuum expectation values $v_1$, $v_2$ and $v_3$.	 
In the Representative Point parametrization, the lightest $h$ has a mass 
$m_h\simeq 90$~GeV, $H$, $A$ and $H^\pm$ are degenerate and have a mass of about 1190 GeV, 
whereas the \Uprime-inherited $H'$ is about 3 TeV, like the $Z^\prime$.
Therefore, in this scenario 
the $Z'$ is not capable of decaying into final states containing $H'$.

\begin{figure}
\centerline{\resizebox{0.49\textwidth}{!}{\includegraphics{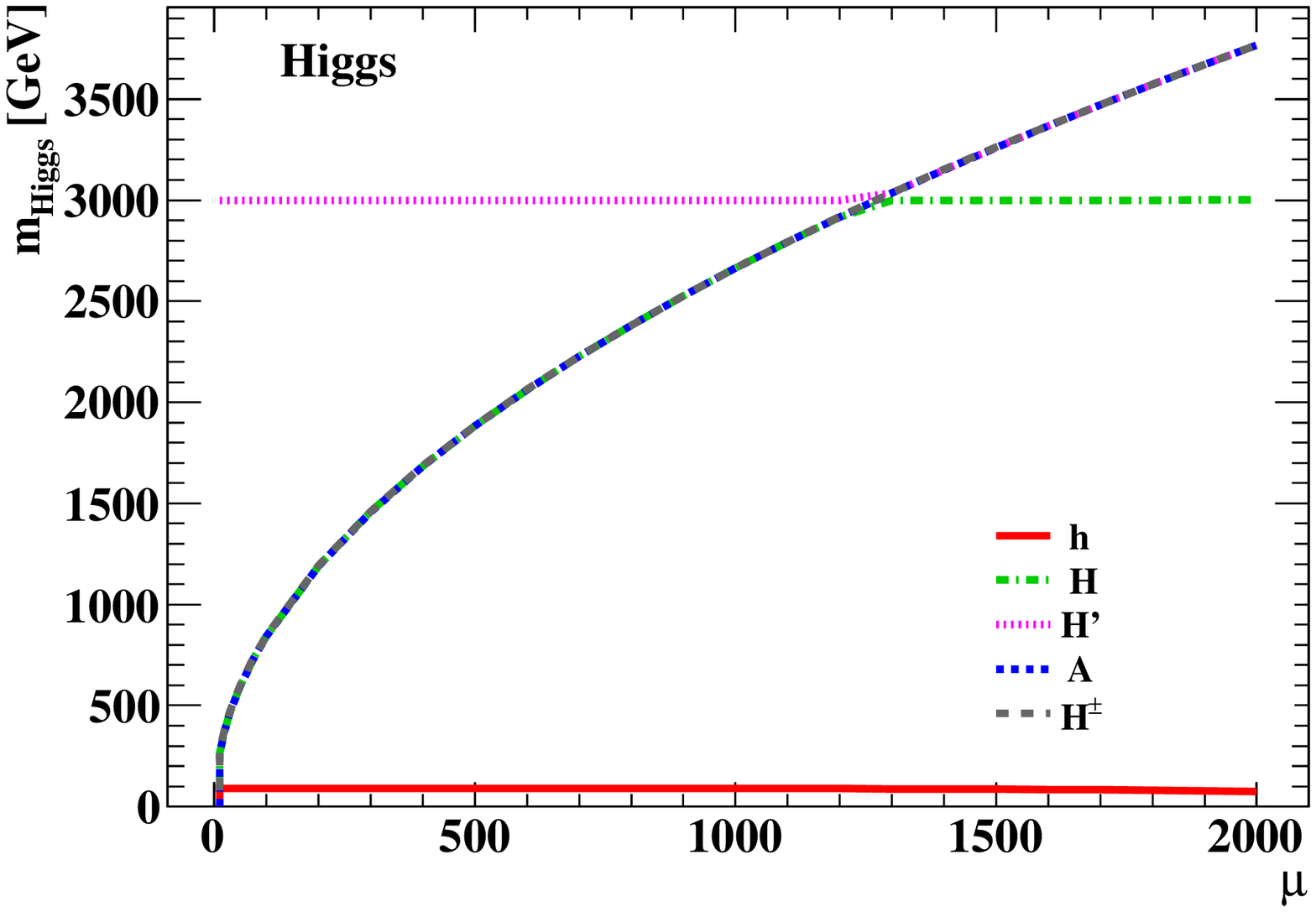}}%
\hfill%
\resizebox{0.57\textwidth}{!}{\includegraphics{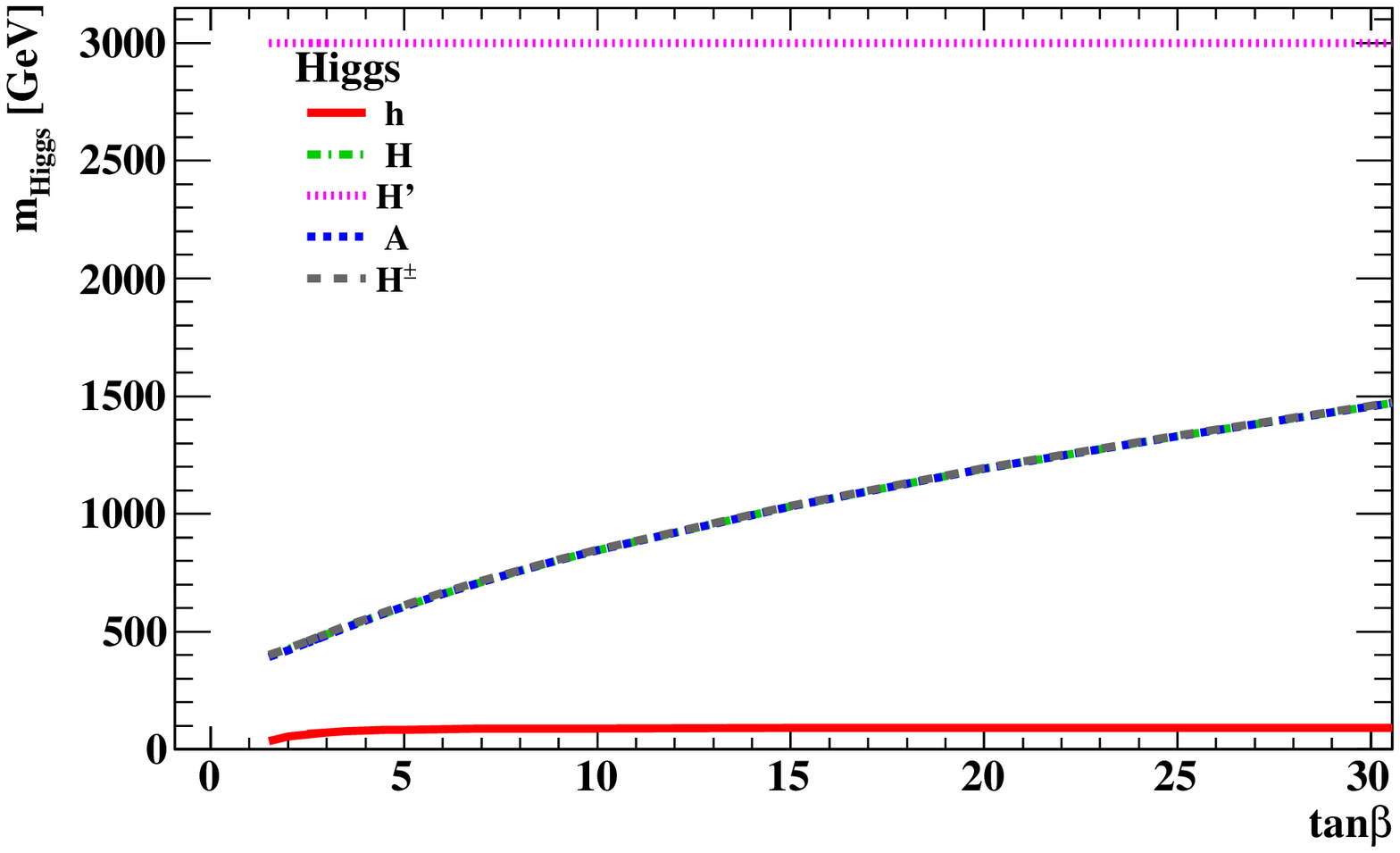}}}
\caption{Dependence of the mass of the Higgs bosons $h$, $H$, $A$, $H^\prime$ and
$H^\pm$ on the MSSM quantities $\mu$ (left) and $\tan\beta$ (right).
\label{higmutanb}}
\end{figure}

\begin{figure}
\centerline{\resizebox{0.49\textwidth}{!}{\includegraphics{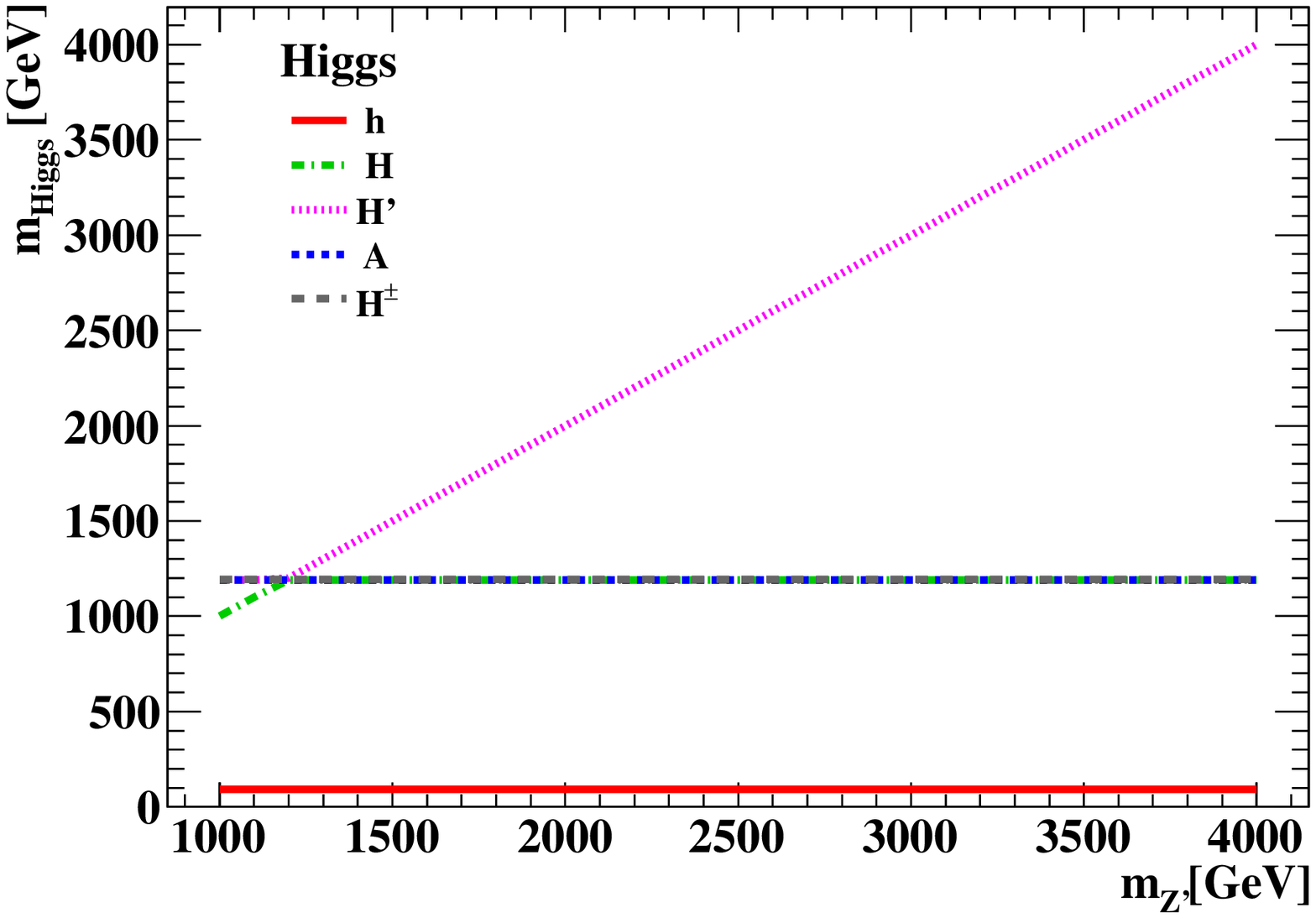}}%
\hfill%
\resizebox{0.49\textwidth}{!}{\includegraphics{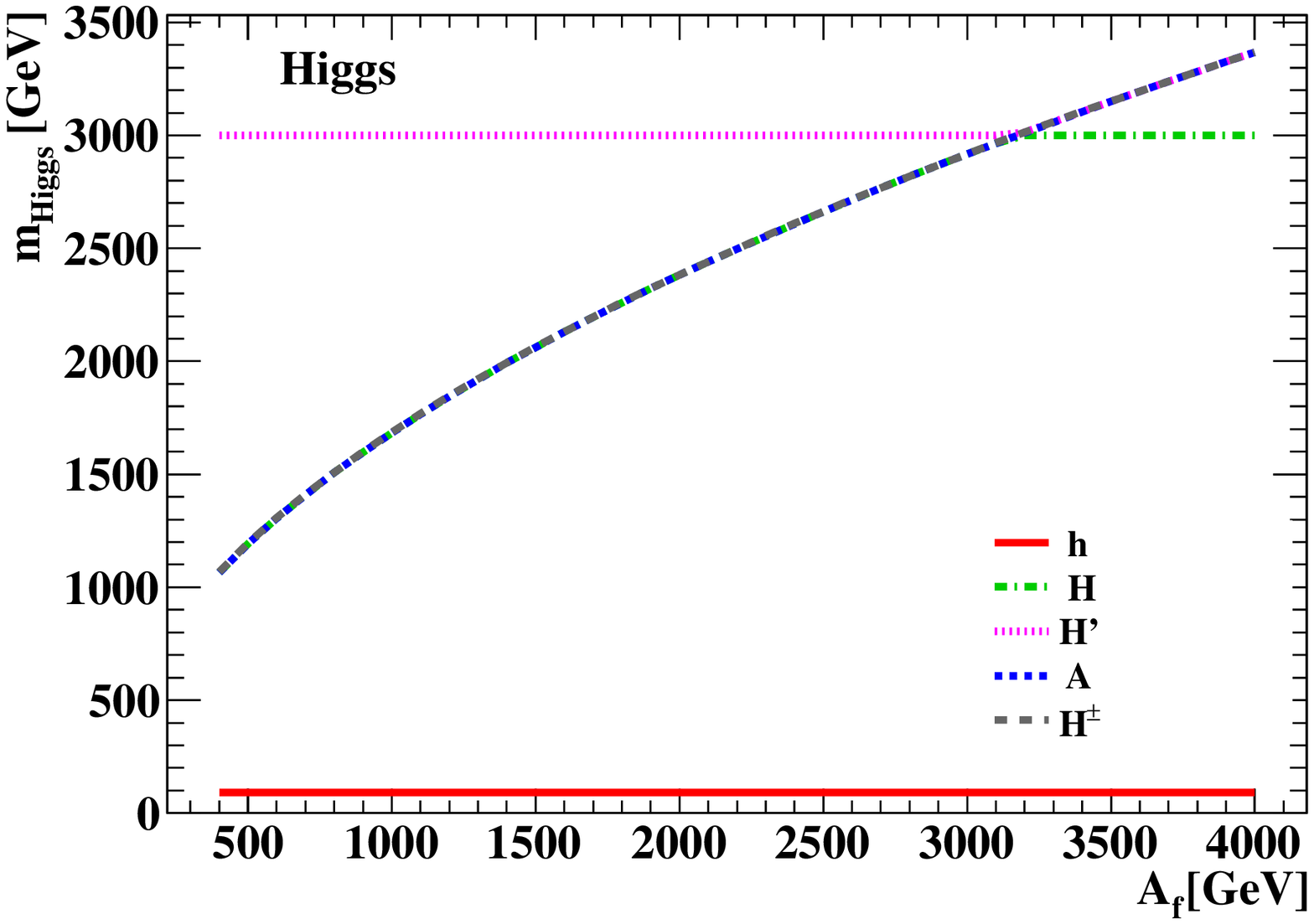}}}
\caption{Higgs mass spectra with respect to  
the $Z'$ mass (left) and the trilinear coupling $A_f$ (right).
\label{higmzpa}}
\end{figure}

Figure~\ref{higmutanb} presents the variation of the Higgs masses in terms of
$\mu$ (left) and $\tan\beta$ (right); Fig.~\ref{higmzpa}
shows the dependence on $m_{Z'}$ (left) and $A_f$ (right).
One can immediately notice that the mass of the lightest $h$ is 
roughly independent of these quantities and it is $m_h\simeq m_Z\simeq
90$~GeV through the whole $\mu$, $\tan\beta$, $m_{Z'}$ and $A_f$ ranges.
Since the supersymmetric light Higgs $h$ should roughly play the role of  
the SM Higgs boson, a value of about 90 GeV for its mass is
too low, given the current limits from LEP \cite{leph}
and Tevatron \cite{tevh} experiments and the recent LHC
results \cite{atlash,cmsh} on the observation of a new Higgs-like particle
with a mass about 125 GeV. 
This is due to the fact that the $h$ mass obtained after diagonalizing the
neutral Higgs mass matrix is just a tree-level result; the possible inclusion
of radiative corrections should increase the light Higgs mass
value in such a way to be consistent with the experimental limits. 
In fact, the Representative Point 
will be used only to illustrate the features of the particle spectra in
the MSSM, after one adds the extra \Uprime\ symmetry group.
Any realistic analysis of $Z'$ decays in supersymmetry should 
of course use values of the Higgs masses accounting for 
higher-order corrections and in agreement with the 
experimental data.

The heavy MSSM scalar Higgs $H$ is physical, i.e. its squared mass 
positive definlite, only for positive values of $\mu$, therefore in
Fig.~\ref{higmutanb} the Higgs masses are plotted for $\mu>0$.
The mass of $H$ increases monotonically from 0 ($\mu=0$) to 
3~TeV ($\mu\simeq 1260$~GeV), and then it is $m_H\simeq m_{Z'}$
also for larger $\mu$-values. As for the \Uprime-inherited $H'$,
its mass is about $m_{H'}\simeq m_{Z'}$ for $0<\mu<1260$~GeV;
for larger $\mu$ it increases monotonically, up to
$m_{H'}\simeq 3.75$~TeV, value reached for $\mu=2000$~GeV.
In other words, for $\mu>1260$~GeV, $H$ and $H'$ behave as if they
exchanged their roles, with increasing $m_{H'}$ and
constant $m_H=m_{Z'}$. 
The masses of $A$ and $H^\pm$ exhibit instead the same behaviour
and increase monotonically with respect to $\mu$ in the whole
range. It is also interesting to notice that, for $0<\mu<1260$~GeV,
one has $m_H\simeq m_{H^\pm}\simeq m_A$.
As for the dependence on $\tan\beta$, presented in Fig.~\ref{higmutanb} (right), 
the masses of $H$, $A$ and $H^\pm$ are almost degenerate and
increase from about  400 GeV ($\tan\beta=1.5$) 
to approximately 1.5 TeV ($\tan\beta=30$). The mass
of $H'$ is instead $m_{H'}\simeq m_{Z'}=3$~TeV
for any value of $\tan\beta$. 

The dependence of the Higgs masses on the $Z'$ mass in the range 1 TeV$<m_{Z'}<4$~TeV
is presented in Fig.~\ref{higmzpa} (left). 
$A$ and $H^\pm$ are degenerate and their mass is constantly equal to 1.19 TeV
in the whole explored region. The $H$ mass is $m_H\simeq 1$~TeV for $m_{Z'}=1$~TeV, 
then it slightly increases and amounts to $m_H\simeq 1.19$~TeV in the range 
1.2~TeV$<m_{Z'}<4$~TeV. 
Figure~\ref{higmzpa} (right) shows the Higgs masses as functions of 
the trilinear coupling
$A_f$ for $500~{\rm GeV}<A_f<4~{\rm TeV}$.
The masses of the charged and pseudoscalar Higgs bosons are degenerate and
increase from 1.1 TeV ($A_f=500$~GeV) to about 3.4 TeV ($A_f=4$~TeV).
The mass of the scalar neutral $H$ is degenerate with the ones of $A$ and $H^\pm$ for
$500~{\rm GeV}<A_f<3.2~{\rm TeV}$, then it is $m_H=m_{Z'}=3$~TeV for
$A_f$ between 3.2 and 4 TeV.  The $H'$ mass is constant, i.e.
$m_{H'}=m_{Z'}=3$~TeV for $500~{\rm GeV}<A_f<3.2~{\rm TeV}$, then it increases
in the same manner as the masses of $H^\pm$ and $A$.
As already observed for the $\mu$ dependence, $H$ and $H'$ exchange their roles
for $A_f>3.2$~TeV.

\subsection{Consistency of the MSSM masses with ISAJET}\label{sec:MSSM_masses}

An experimental search for supersymmetric $Z'$ decays demands the 
implementation of our MSSM/\Uprime\ scenario in a Monte Carlo event generator.
Therefore, it is essential to
verify whether our mass spectra are consistent with those provided by 
the codes typically 
used to compute masses and decay rates in supersymmetry.
For this purpose, a widely used program is the ISAJET package \cite{Paige:2003mg}, containing
all the MSSM data; the supersymmetric particle masses and branching ratios obtained by
running ISAJET are then used by programs, such as HERWIG \cite{herwig} and PYTHIA \cite{pythia},
which simulate hard scattering,
parton showers, hadronization and underlying event,
for an assigned MSSM configuration.
It is thus crucial assessing whether such an approach can still be employed
even after the inclusion of the extra $Z'$ boson. 
Squark and slepton masses, corrected by the D-term contribution, 
can be directly given as an input to ISAJET.
Moreover, the chargino spectrum is unchanged, being the $Z'$ neutral,
whereas the extra $H'$, being too heavy, is not relevant for the $Z'$ phenomenology.
Besides,
the masses of the MSSM Higgs bosons $h$, $H$, $A$ and $H^\pm$  depend
very mildly on the \Uprime\ parameters.
In the neutralino sector, the two additional $\tilde\chi^0_5$ and $\tilde\chi^0_6$ 
are also too heavy to be phenomenologically relevant. 
However, the neutralino 
mass matrix, Eq.~(\ref{neumass}), depends also on extra new parameters, such as 
$M^\prime$, $g^\prime$ and the \Uprime\ charges $Q'_{1,2,3}$. Therefore, 
even the mass of the four light neutralinos can potentially feel
the effect of the presence of the $Z'$.

We quote in Table~\ref{reprneu}
the eigenvalues of the neutralino
mass matrix, Eq.~(\ref{neumass}), along with the masses yielded
by ISAJET, for the parameter configuration corresponding to the Representative Point, Eq.~(\ref{reprpoint}).
For the sake of completeness, we also present the Higgs and chargino mass values 
obtained in our framework (\Uprime\ and MSSM), to investigate whether they
agree with the ISAJET results (only MSSM).

\begin{table}
\caption{\label{reprneu}Mass values in GeV for neutralinos, charginos and Higgs bosons
in our model, based on \Uprime\ and the MSSM, and according
to the ISAJET code, which implements only the MSSM.}
\begin{center}
\small
\begin{tabular}{|l|cccccccccc|}
\hline
Model & \Mninoone & \Mninotwo  & \Mninothree     & \Mninofour & $m_h$ & $m_H$ & $m_A$ & $m_{H^\pm}$ 
& $m_{\tilde\chi_1^\pm}$ & $m_{\tilde\chi_2^\pm}$ \\   
\hline
\hline
\Uprime/MSSM & 94.6 & 156.6 & 212.2 & 261.0 & 90.7 & 1190.0 & 1190.0 &
1190.0 & 155.0 & 263.0 \\
MSSM  & 91.3 & 152.2  & 210.2 & 266.7 & 114.1 & 1190.0 & 1197.9 & 1200.7 & 147.5 & 266.8\\
\hline
\end{tabular}
\end{center}
\end{table}
From Table~\ref{reprneu} one learns that the masses 
of the neutralinos agree within 5\%;
a larger discrepancy is instead found, about 20\%, for the mass of the lightest Higgs, i.e.
$h$; as pointed out before, this difference is due to the fact that, unlike ISAJET,
our calculation is just a tree-level one and does not include radiative
corrections.
Both Higgs masses are nevertheless much smaller than the 
$m_{Z'}$, fixed to 3 TeV in the Representative
Point; therefore, $Z'$ decays into Higgs bosons will not be significantly affected by this 
discrepancy. 

As for the chargino masses, the difference between our analytical calculation and the
prediction of ISAJET is approximately 5\% for $\tilde\chi_1^\pm$ and 1\% for  $\tilde\chi_2^\pm$.
Overall, one can say that some differences in the spectra yielded by our computations
and ISAJET are visible, but they should not have much impact on $Z'$ phenomenology.
The implementation of the \Uprime\ model in HERWIG or PYTHIA, 
along with the employment of a standalone program like ISAJET for masses and branching ratios
in supersymmetry, may thus provide a useful tool to explore $Z'$ phenomenology
in an extended MSSM.

\subsection{{$\mathbf Z'$} decays in the Representative Point}\label{sec:Br}

Before concluding this section, 
we wish to present the branching ratios of the $Z'$ boson into both SM 
and new-physics particles.
If BSM decays are competitive with the SM ones, then
the current limits on the $Z^\prime$ mass will have to be reconsidered.
We shall first present the branching ratios in the Representative Point parametrization,
Eq.~(\ref{reprpoint}), i.e. a $Z'_{\rm I}$ boson with mass 3 TeV,
and then we will vary the quantities entering in our analysis.

\subsubsection{Branching ratios in the Representative Point}

In Table~\ref{massrep} we summarize, for the reader's convenience, 
the masses of the BSM particles for
the parameters in Eq.~(\ref{reprpoint}), in such a way to figure out the 
decay channels which are kinematically permitted. 
\begin{table}[htp]
\caption{\label{massrep}
Masses in GeV of BSM particles in the MSSM/\Uprime\ scenario, with 
the parameters set as in Eq.~(\ref{reprpoint}).}
\begin{center}
\small
\begin{tabular}{|cccccccc|}
\hline
$m_{\tilde u_1}$ &  $m_{\tilde u_2}$ & $m_{\tilde d_1}$ & $m_{\tilde d_2}$ &  
$m_{\tilde\ell_1}$ & $m_{\tilde\ell_2}$ &   $m_{\tilde\nu_1}$ & $m_{\tilde\nu_2}$   \\
\hline\hline
 2499.4  & 2499.7 & 2500.7 & 1323.1 & 3279.0 & 2500.4 & 3278.1 &   3279.1 \\
\hline
 \Mninoone & \Mninotwo & \Mninothree     & \Mninofour & \Mninofive  &  \Mninosix & \Mchinoonepm & \Mchinotwopm\\
\hline
94.6  &  156.5 & 212.2 &        260.9 &        2541.4        &        3541.4  &      154.8 &        262.1       \\
\hline\hline
 $m_h$ & $m_A$ &$m_H$ & $m_{H'}$ & $m_{H^\pm}$   &&&    \\
\hline\hline
90.7 &                1190.7     &        1190.7        &    3000.0        &  1193.4        &&&   \\
\hline
\end{tabular}
\end{center}
\end{table}
At this point it is possible to calculate the $Z^\prime$ widths 
into the kinematically allowed decay channels. 
The $Z^\prime$ SM decay channels are the same as the $Z$ boson, i.e. quark or lepton pairs,
with the addition of the $W^+W^-$ mode, which is accessible due to the higher $Z'$ mass.
However, since the $Z'$ has no direct coupling to $W$ bosons, the
$Z'\to W^+W^-$ occurs only via $ZZ'$ mixing and therefore one can already foresee
small branching ratios.
Furthermore, the extended MSSM
allows $Z^\prime$ decays into squarks, i.e. 
$\tilde q_i\tilde q_i^*$
($q=u,d$ and $i=1,2$), charged sleptons $\tilde\ell_i\tilde\ell_i$,
sneutrinos $\tilde\nu_{i,\ell}\tilde\nu_{i,\ell}^*$ ($\ell=e$, $\mu$, $\tau$,
$i=1,2$),  
neutralino, chargino, or Higgs ($hh$, $HH$, $hH$, $hA$, $HA$, 
$H'A$, $H^+H^-$) pairs,
as well as into states with Higgs bosons associated with $W/Z$, such as 
$Zh$, $ZH$ and $W^\pm H^\mp$.

We refer to \cite{Gherghetta:1996yr} for the analytical form of such widths,
at leading order in the
\Uprime\ coupling constant, i.e. ${\cal O}(g'^2)$; in
Appendix A the main formulas will be summarized.
Summing up all partial rates, one can thus obtain
the $Z^\prime$ total width and the branching ratios into the allowed 
decay channels. 

In Table~\ref{br1} we quote the $Z'$ branching ratios in the 
Representative Point parametrization.
Since, at the scale of 3 TeV, one does not distinguish the quark or lepton flavour, the
quoted branching ratios are summed over all possible flavours and
 $u\bar u$, $d\bar d$, $\ell^+\ell^-$ and $\nu\bar\nu$ denote any possible up-, down-type quark,
charged-lepton or neutrino pair. Likewise, $\tilde u\tilde u^*$, $\tilde d\tilde d^*$, 
$\tilde\ell^+\tilde\ell^-$ and
$\tilde\nu\tilde\nu^*$ are their supersymmetric
 counterparts. We present separately the branching ratios into 
all possible different species of charginos and neutralinos, as they yield different 
decay chains and final-state configurations.
\begin{table}[ht]
\caption{Branching ratios of the $Z^\prime$ with the parameters fixed as in Eq.~(\ref{reprpoint}). The branching ratios into fermions and sfermions 
have been summed over all the possible flavours, e.g. $u\bar u$ ($\ell^+\ell^-$) denotes the sum of the rates into 
up, charm and top (electron, muon and tau) pairs. } 
\begin{center}
\small
\begin{tabular}{|c|c||c|c|}
\hline
Final state & BR $(\%)$  & Final state & BR $(\%)$ \\
\hline\hline
$u\bar u$ & 0.00& $\tilde\chi_1^0\tilde\chi_1^0$ & 0.07 \\
$d \bar d$ & 40.67&$\tilde\chi_1^0\tilde\chi_2^0$ & 0.43  \\
$\ell^+ \ell^-$ & 13.56  & $\tilde\chi_1^0\tilde\chi_3^0$ & 0.71 \\
$\nu\bar \nu$ & 27.11 &$\tilde\chi_1^0\tilde\chi_4^0$ & 0.27  \\
$\tilde u\tilde u^*$ & 0.00 &$\tilde\chi_1^0\tilde\chi_5^0$ & ${\cal O} (10^{-6})$ \\
$\tilde d\tilde d^*$ & 9.58 &$\tilde\chi_2^0\tilde\chi_2^0$ & 0.65 \\
$\tilde \ell^+\tilde \ell^-$ & 0.00 &$\tilde\chi_2^0\tilde\chi_3^0$ & 2.13 \\
$\tilde \nu\tilde \nu^*$ & 0.00 &$\tilde\chi_2^0\tilde\chi_4^0$ & 0.80 \\
$W^+W^-$ & ${\cal O}(10^{-5})$ & $\tilde\chi_2^0\tilde\chi_5^0$ & ${\cal O} (10^{-6})$\\
$H^+H^-$ & 0.50 &$\tilde\chi_3^0\tilde\chi_3^0$ & 1.75 \\
$hA$ & ${\cal O}(10^{-3})$ &$\tilde\chi_3^0\tilde\chi_4^0$ & 1.31 \\
$HA$ & 0.51 &$\tilde\chi_3^0\tilde\chi_5^0$ & ${\cal O} (10^{-6})$ \\
$ZH$ & ${\cal O} (10^{-3})$ & $\tilde\chi_4^0\tilde\chi_4^0$ & 0.25\\
$Zh$ & ${\cal O} (10^{-5})$ &$\tilde\chi_4^0\tilde\chi_5^0$ & ${\cal O} (10^{-7})$ \\
$ZH'$ & 0.00 & $\tilde\chi_5^0\tilde\chi_5^0$ & 0.00\\
$H'A$ & 0.00 &  $\sum_i\tilde\chi_i^0\tilde\chi_6^0$ & 0.00  \\
$W^\pm H^\mp$ & ${\cal O} (10^{-3})$ & $\tilde\chi_1^+\tilde\chi_1^-$ &  1.76 \\
&                  &      $\tilde\chi_1^\pm\tilde\chi_2^\mp $ & 1.95 \\
&                  &  $\tilde\chi_2^+\tilde\chi_2^- $ & 0.54\\
\hline
\end{tabular}
\label{br1}
\end{center}
\end{table}
In Table~\ref{br1}, several branching ratios are zero or very small: the decays
into up-type squarks and sleptons, heavy neutralinos $\tilde\chi^0_6$ and the \Uprime-inherited 
$H'$ are kinematically forbidden for a $Z'$ of 3 TeV.
The only allowed decay into sfermion pairs is the one into 
down-type squarks $\tilde d_2\tilde d_2^*$.
Despite being kinematically permitted, the width into up-type quarks 
vanishes, since, as will be clarified in Appendix A, in the $Z'_{\rm I}$ model
the vector ($v_u$) and vector-axial ($a_u$) couplings, contained in the 
in the interaction Lagrangian 
of the $Z'$ with up quarks, are zero. 
From Table~\ref{br1} we learn that, at the Representative Point,
the SM decays account for roughly the 77\% of the total $Z^\prime$ width and
the BSM ones for the remaining 23\%. 
As for the BSM modes, the rate into down squarks is 
about 9\% of the total rate, the ones into charginos and neutralinos
4.2\% and 8.4\%, respectively.
In the gaugino sector, the channels $\tilde\chi^0_2\tilde\chi_3^0$ and
$\tilde\chi_1^\pm\tilde\chi^\mp_2$ have the highest branching ratios.
The decay into $\tilde\chi_1^0\tilde\chi_1^0$ has a very small branching fraction
and is experimentally undetectable if $\tilde\chi_1^0$ is the 
lightest supersymmetric particle (LSP).
The final states with Higgs bosons are characterized by
very small rates: 
the branching fractions into $H^+ H^-$ and $HA$ are about 0.5\%,
the one into $H^\pm W^\mp$ roughly 0.1\% and an even lower rate, ${\cal O}(10^{-7})$,
is yielded by the modes $hZ$, $ZH$ and $hA$.

These considerations, 
obtained in the particular configuration of the Reference Point, Eq.~(\ref{reprpoint}), 
can be extended to a more general context.
We can then conclude that 
the \Zprime\ BSM  branching fractions are not negligible and should
be taken into account in the evaluation of the mass limits.

\subsubsection{Parameter dependence of the branching ratios}

In this subsection we wish to investigate how 
the $Z'$ branching fractions into SM and supersymmetric
particles fare with respect to the \Uprime\ and MSSM parameters. 
As in Section 3, the study will be carried out at the Representative Point,
varying each parameter individually.

\begin{figure}[ht]
\centerline{\resizebox{0.49\textwidth}{!}
{\includegraphics{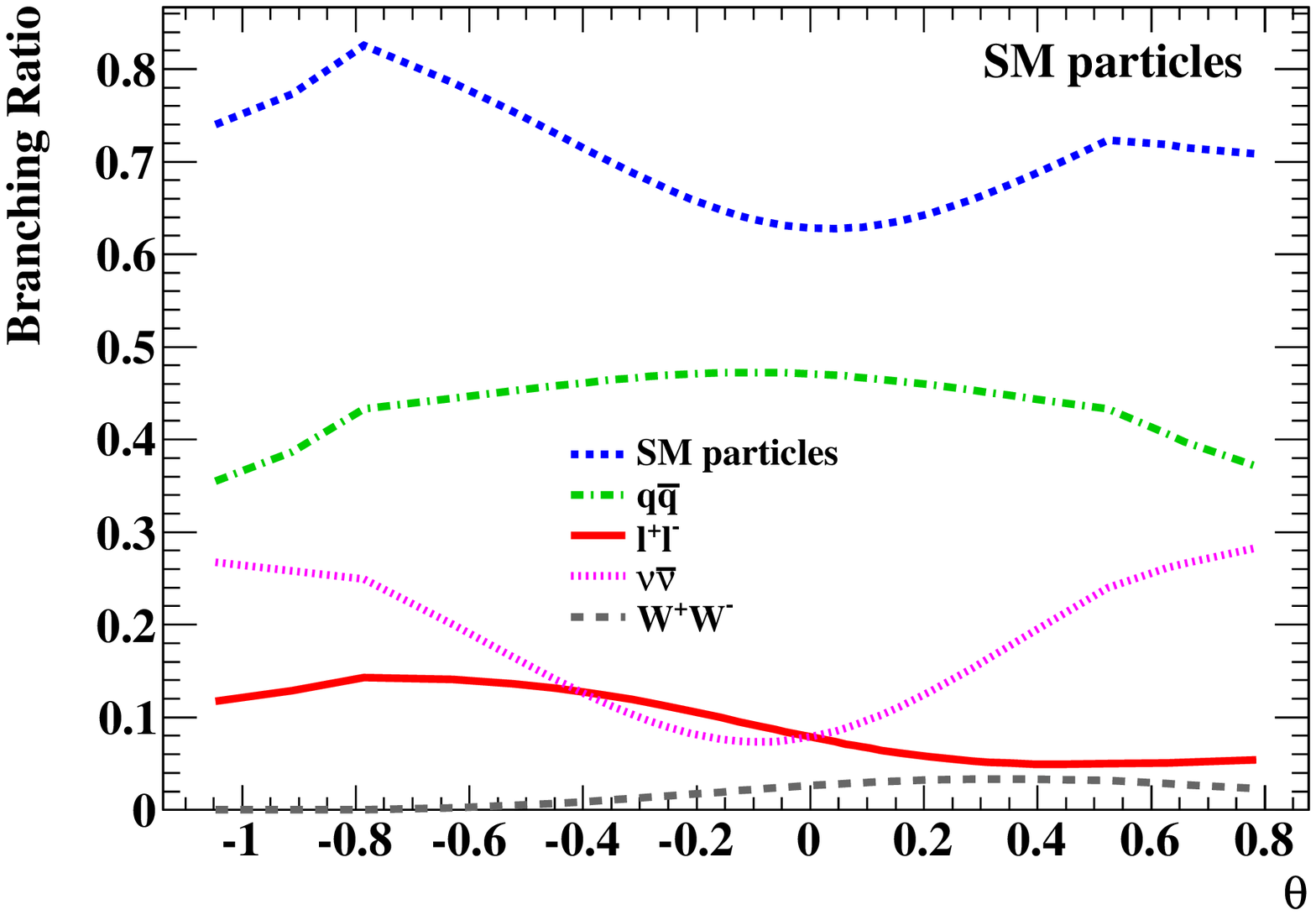}}%
\hfill%
\resizebox{0.49\textwidth}{!}
{\includegraphics{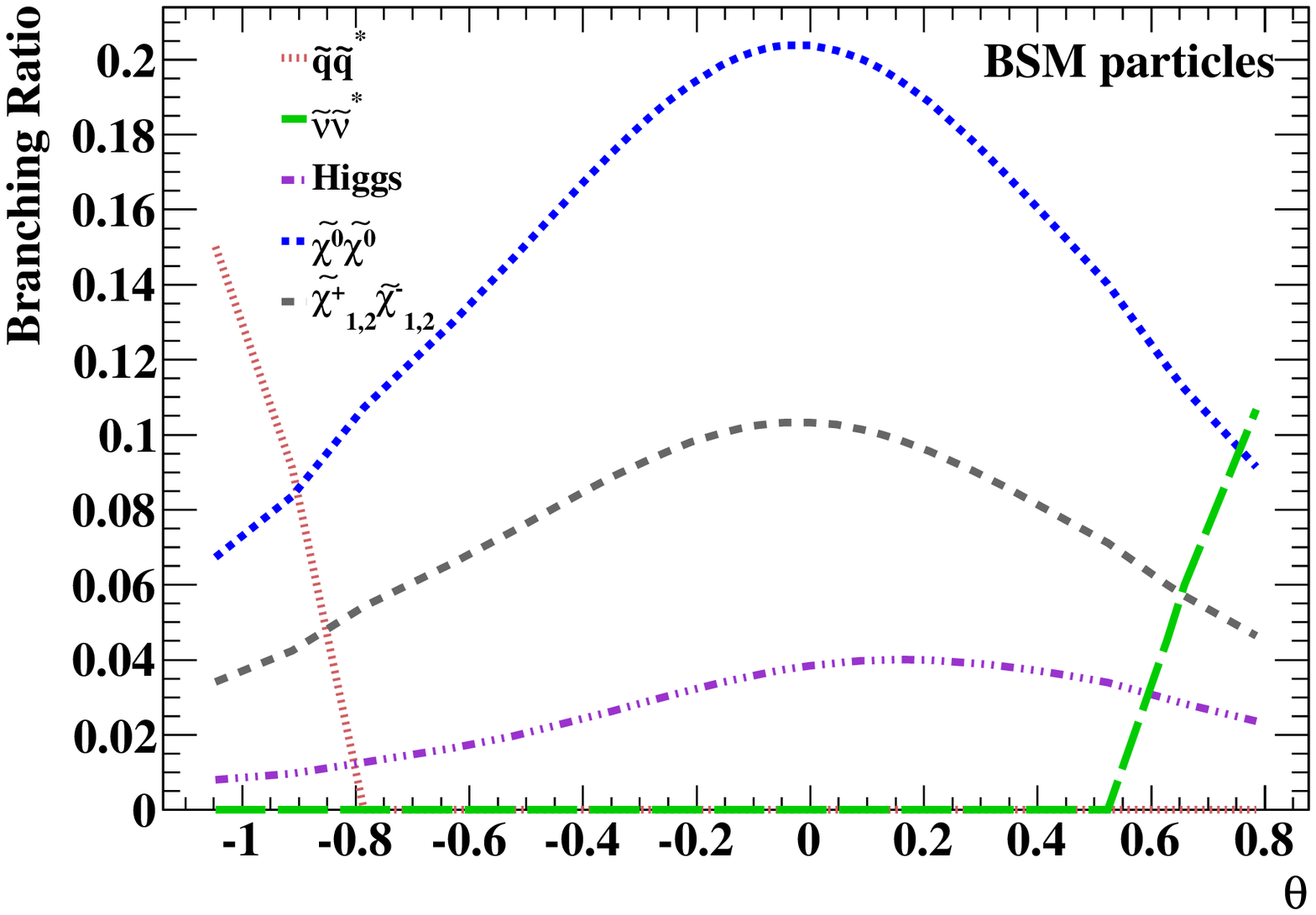}}}
\caption{Dependence of the $Z'$ decay rates on the
\Uprime\ mixing angle $\theta$. Left: SM modes;
right: BSM channels. \label{thbr}}
\end{figure}

In Fig.~\ref{thbr}, the dependence of the branching ratios on the
mixing angle $\theta$ is presented for SM (left) and BSM (right) 
decay modes, in the range $-1<\theta<0.8$;
for the SM channels, we have also plotted the total 
branching ratio.
The $Z'$ decay rate into quarks exhibits a quite flat distribution,
amounting to about 40\% for central values of $\theta$ 
and slightly decreasing for large 
$|\theta|$.
The branching ratio into neutrino pairs is enhanced for 
$\theta$ at the edges of the explored range, being about 25\%, and presents a minimum 
for $\theta\simeq -0.1$. 
The rate into charged leptons varies between 5 and
15\%, with a small enhancement around $|\theta|\simeq 0.8$;
the branching fraction into $W^+W^-$ is below 2\% in the whole $\theta$
range. 

As for the BSM channels, described in Fig.~\ref{thbr} (right), 
the neutralino, chargino, and Higgs modes have a similar behaviour,
with a central broad maximum around $\theta=0$ and branching ratios
about 20\%, 10\% and 3\%, respectively
The sneutrino modes give a
non-negligible contribution only for $\theta>0.5$, reaching 
about 10\%, at the boundary of the investigated $\theta$ region, i.e.
$\theta\simeq 0.8$. The squark-pair channel has a significant rate, 
about 15\%,  for negative mixing angles, i.e. $\theta\simeq=-1$. 
The rates in the Higgs channels lie between the neutralino
and chargino ones and exhibit a maximum value, about 10\%, for $\theta=0$.

\begin{figure}[ht]
\centerline{\resizebox{0.49\textwidth}{!}
{\includegraphics{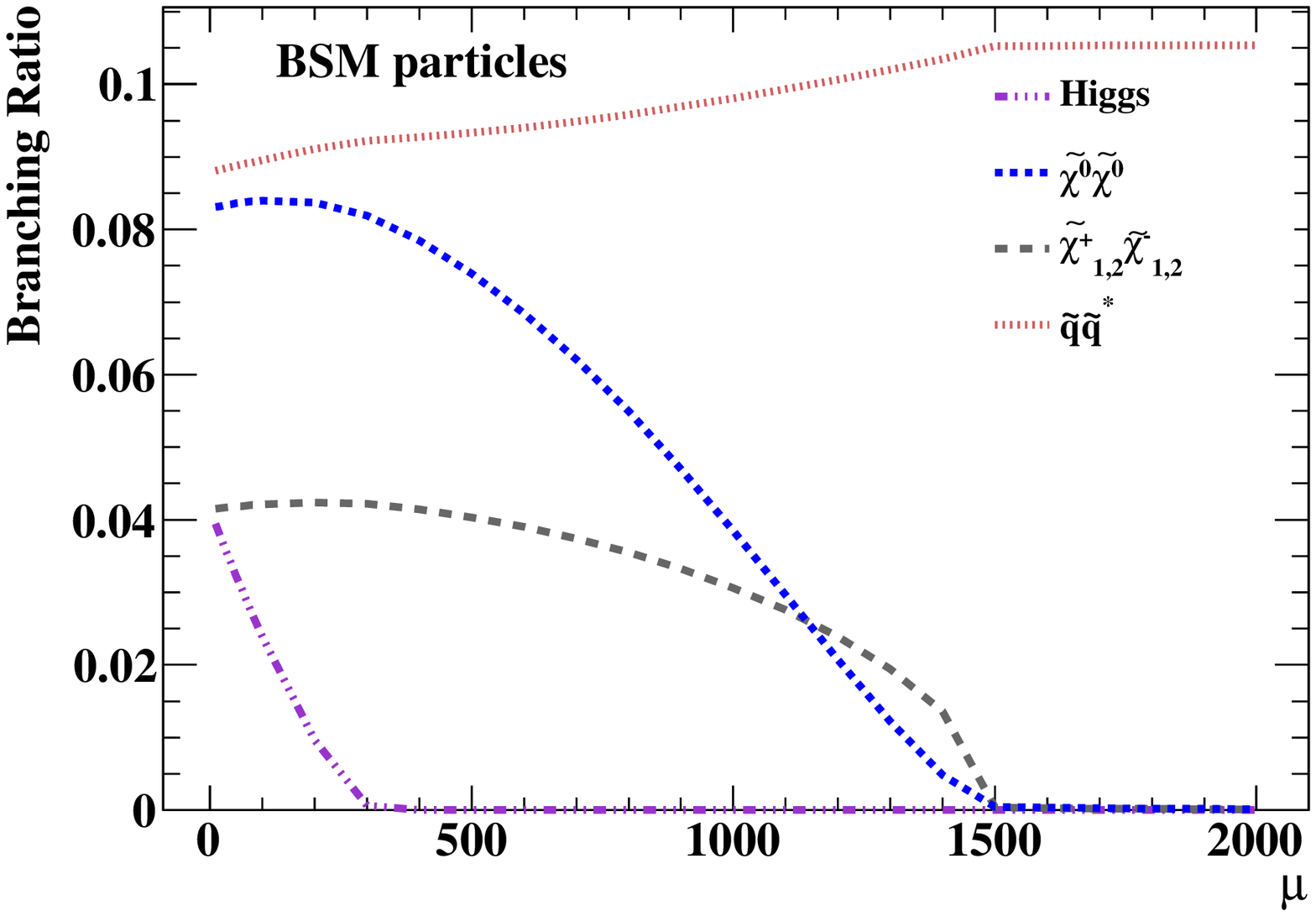}}%
\hfill%
\resizebox{0.49\textwidth}{!}{\includegraphics{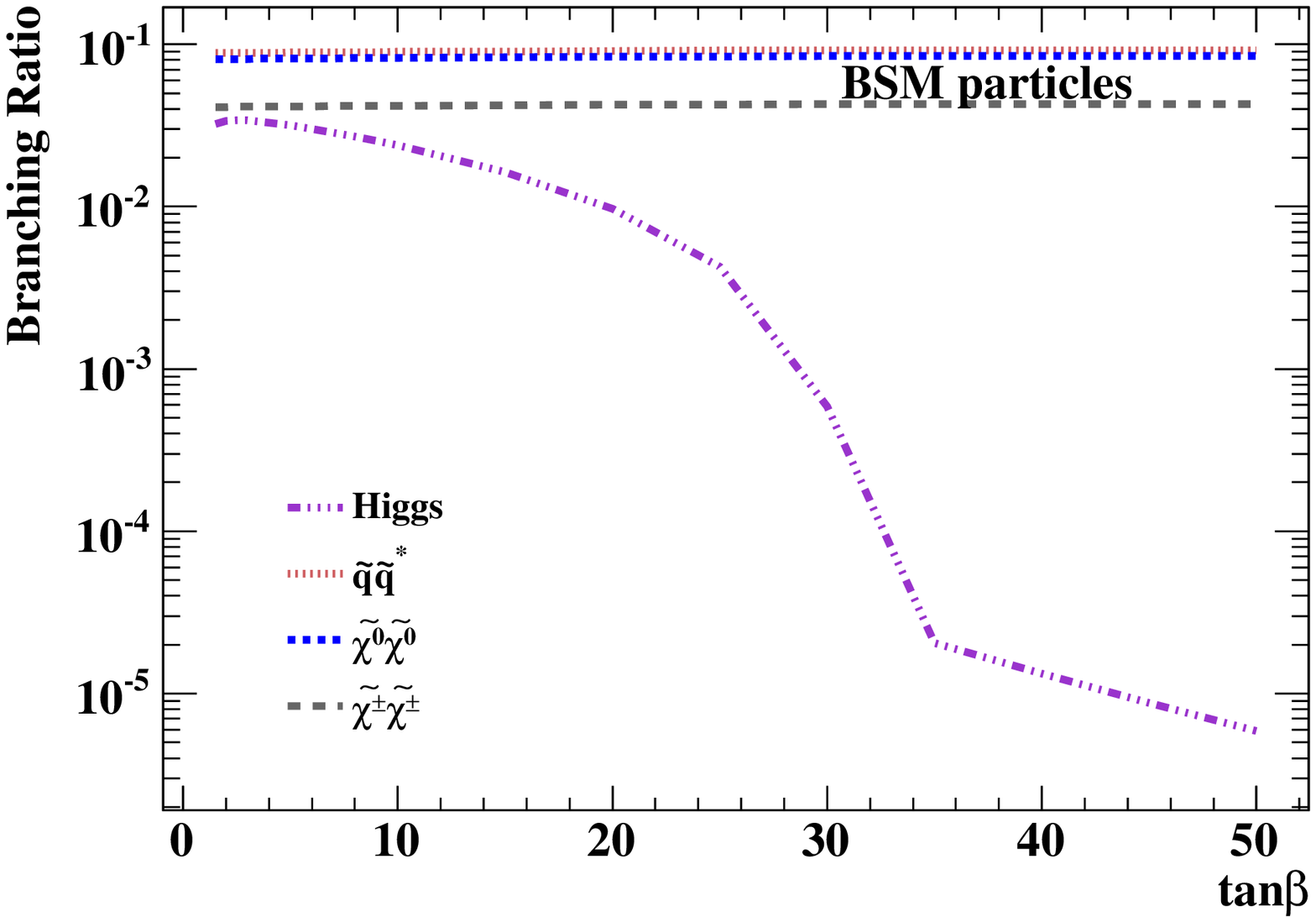}}}
\caption{BSM branching ratios with respect to the MSSM parameters $\mu$ (left) 
and $\tan\beta$ (right).
\label{tanmubr}}
\end{figure}

Figure~\ref{tanmubr} presents the dependence of the BSM $Z'$ branching ratios
on the MSSM parameters $\mu$ (left) and $\tan\beta$ (right). 
The SM rates are not shown, since their dependence on these parameters is
negligible.
The decay rate into squarks slightly increases from
9 to 10\% in the explored $\mu$ range; the neutralino branching
ratio decreases quite rapidly from about 8\% ($\mu=0$) to
zero ($\mu\simeq 1500$~GeV).
The rate into charginos is about 4\% for small values of $\mu$, then it
smoothly decreases, being negligible for $\mu>1500$~GeV. The branching fraction into 
Higgs modes is almost 4\% at $\mu=0$ and rapidly becomes nearly zero
for $\mu>300$~GeV. As for $\tan\beta$, the $\tilde q\tilde q^*$,
$\tilde\chi^+\tilde\chi^-$ and $\tilde\chi^0\tilde\chi^0$ modes
are roughly independent of it, with rates about 9\% (squarks),
8\% (neutralinos) and 4\% (charginos).
The decays into states with Higgs bosons account for
4\% of the $Z'$ width at small $\tan\beta$ and are below $1\%$ for $\tan\beta>20$. 

\section{{$\mathbf Z'$} decays into final states with leptons}
\label{sec:Reference Points}

\begin{figure}[htp]
\begin{center}
\includegraphics[angle=0,bb=80 748 316 694,width=0.45\textwidth]{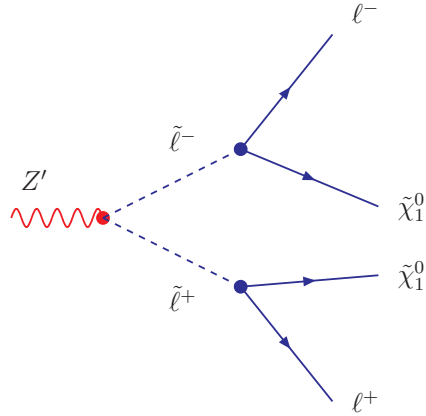}
\vspace{6.cm}
\caption[]{\label{zsl}Diagram for the decay of the $Z'$ into a charged-slepton pair,
yielding a final state with two charged leptons and missing energy.}
\end{center}
\end{figure}

Leptonic final states are considered as golden channels 
from the viewpoint of the LHC experimental searches.
To exploit these features, this study will be focused 
on the decays of the \Zprime\ boson \ into  supersymmetric particles, leading to
final states with leptons and missing energy, due to the
presence of neutralinos or neutrinos.
Final states with two charged leptons and missing energy
come from primary decays 
$Z'\to\tilde\ell^+\tilde\ell^-$, 
presented in Fig.~\ref{zsl}, with the charged sleptons decaying into
a lepton and a neutralino.

\begin{figure}[htp]
\begin{center}
\includegraphics[angle=0,bb=80 748 316 694,width=0.45\textwidth]
{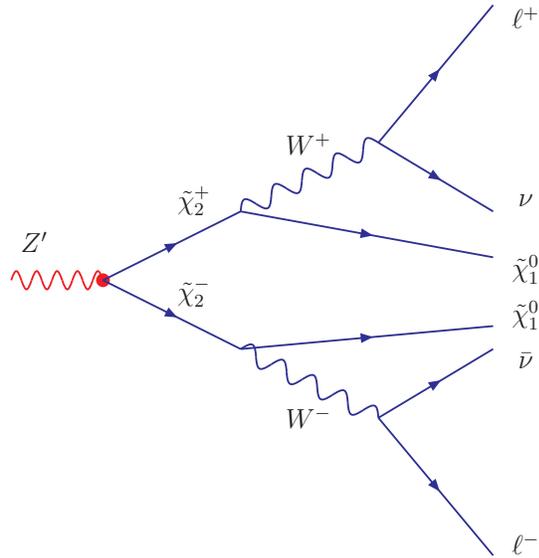}
\vspace{7.cm}
\caption[]{\label{zch}Final state with two charged leptons and missing energy,
through a primary decay of the $Z'$ into a chargino pair.} 
\end{center}
\end{figure}

Furthermore, primary decays into charginos
$Z'\to\tilde\chi_2^+\tilde\chi_2^-$, followed by $\tilde\chi_2^\pm\to W^\pm\tilde\chi_1^0$
and $W^+\to \ell^+\bar\nu$ ($W^-\to \ell^-\nu$),
as in Fig.~\ref{zch}, yield final states with two charged leptons and
missing energy as well.
With respect to  the
direct production in $pp$ collisions, where the partonic centre-of-mass energy is
not uniquely determined, the production of charginos in $Z'$ decays has the advantage that the
$Z'$ mass sets a kinematic constrain on the chargino invariant mass.

\begin{figure}[htp]
\begin{center}
\centerline{\resizebox{0.4\textwidth}{!}{\includegraphics{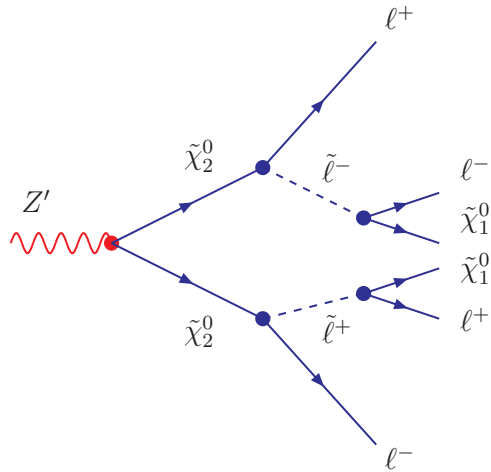}}}
    \caption{\label{zneu}Decays of $Z'$ bosons into neutralinos,
leading to final states with four charged leptons and missing energy.}
\end{center}
\end{figure}

\begin{figure}[htp]
\begin{center}
\vspace{1.cm}\includegraphics[angle=0,bb=80 748 316 694,width=0.45\textwidth]{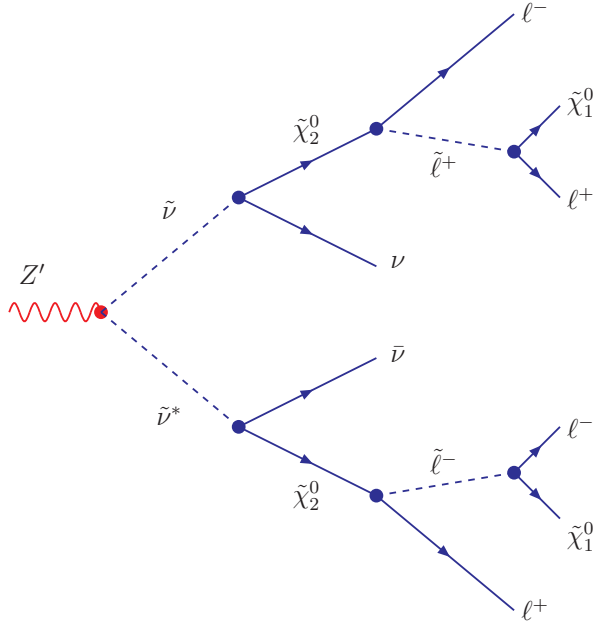}
\vspace{7.cm}
\caption[]{\label{zsneu}Final state with four charged leptons and missing energy,
due to the presence of neutrinos and neutralinos, yielded by a primary $Z'\to \tilde\nu\tilde\nu^*$ decay.}
\end{center}
\end{figure}

A decay chain, leading to four charged leptons and missing energy, 
is yielded by $Z'$ decays into neutralinos $Z'\to\tilde\chi_2^0\chi_2^0$, with
subsequent $\tilde\chi_2^0\to \ell^\pm\tilde\ell^\mp$ and 
$\tilde\ell^\pm\to\ell^\pm\tilde\chi_1^0$, as
in Fig.~\ref{zneu}.
Finally, we shall also investigate the decay into sneutrino pairs, such as
$Z'\to\tilde\nu_2\tilde\nu_2^*$, followed by  $\tilde\nu_2\to  \tilde\chi^0_2\nu$
and $\tilde\chi_2^0\to \ell^+\ell^-\tilde\chi^0_1$, with an intermediate charged slepton
(see Fig.~\ref{zsneu}).
The final state of the latest decay chain is made of four
charged leptons plus missing energy, due to neutrinos and neutralinos.

In the following, we wish to present a study of $Z'$ decays into leptonic final states 
for a given set of the MSSM and \Uprime\  parameters. In particular,
we shall be interested in understanding the behaviour of such rates as a function of the
slepton mass, which will be treated as a free parameter.
In order to increase the rate into sleptons, with respect to the
scenario yielded by the Representative Point, the squark mass at the $Z'$ scale will be  
increased to 5 TeV, in such a way to suppress $Z'$ decays into hadronic jets. 

In our study we consider 
the models in Table~\ref{tab:Models} and vary the initial slepton mass
 $\Mslep^0$ for several fixed values of $m_{Z'}$, with the goal of determining
an optimal combination of 
$m^0_{\tilde\ell}$ and $m_{Z'}$, 
enhancing the rates into leptonic final states,
i.e. the decay modes containing primary sleptons, charginos or neutralinos.
The other parameters are set to the following Reference Point:
\begin{eqnarray}
&\ &
\mu =200~{\rm GeV}\ ,\ \tan\beta=20\ ,\ A_q=A_\ell=A_f=500~{\rm GeV}\ ,\nonumber \\
&\ &m^0_{\tilde q}=5~{\rm TeV}\ ,\ 
 M_1=150~{\rm GeV}\ ,\ M_2=300~{\rm GeV}\ ,\ M^\prime=1~{\rm TeV}.
\label{refpoint}
\end{eqnarray}
Any given parametrization will be taken into account only if the sfermion masses
are physical after the addition of the D-term. 
Hereafter, we denote by \Brqqbar, \Brll, \Brnu\ and \BrWW\ the branching ratios
into quark, charged-lepton, neutrino and $W$ pairs, with BR$_{\rm SM}$ being
the total SM decay rate. Likewise, 
BR$_{\tilde q\tilde q^*}$,  BR$_{\tilde\ell^+\tilde\ell^-}$ and 
BR$_{\tilde\nu\tilde\nu^*}$ are the rates into
squarks, charged sleptons and sneutrinos,   
\Brchinochino,\ \Brninonino,\ \BrHpHm,\ \BrhA,\
\BrHA\ are the ones into chargino, neutralino, charged- and neutral-Higgs pairs, 
\BrWmpHpm\ the branching fraction into $W^\mp H^\pm$.
Moreover, for convenience, \BrZh\ is the sum of the branching ratios into $Zh$ and $ZH$ and
BR$_{\rm BSM}$ the total BSM branching ratio.



\newpage
\subsection{Reference Point: Model {$\mathbf\ZETA$}}\label{sec:ZETA} 

An extra \Uprime\  group with a mixing angle $\theta=\arccos\sqrt{5/8}$
leads to a new neutral boson labelled as $Z'_\eta$. 
In Table~\ref{lepeta} we list the masses of charged ($\msluno$ and $\msldue$)
and neutral ($\msnuno$ and $\msndue$) sleptons, for various $m_{Z'}$ 
and for the values of $m^0_{\tilde\ell}$ which, as will be clarified later,
yield a physical sfermion spectrum and 
a maximum and minimum rate into sneutrinos. 
From Table~\ref{lepeta} we learn that the decays into pairs of charged sleptons are
always kinematically forbidden, whereas the decay into  $\tilde\nu_2$ pairs is accessible.
The effect of the
D-term on the mass of $\tilde\nu_2$ is remarkable: 
variations of $m^0_{\tilde\ell}$ of few hundreds GeV
induce in $m_{\tilde\nu_2}$ a change of 1 TeV or more, especially for large values of the 
$Z'$ mass.
\begin{figure}[htp]
\centerline{\resizebox{0.65\textwidth}{!}
{\includegraphics{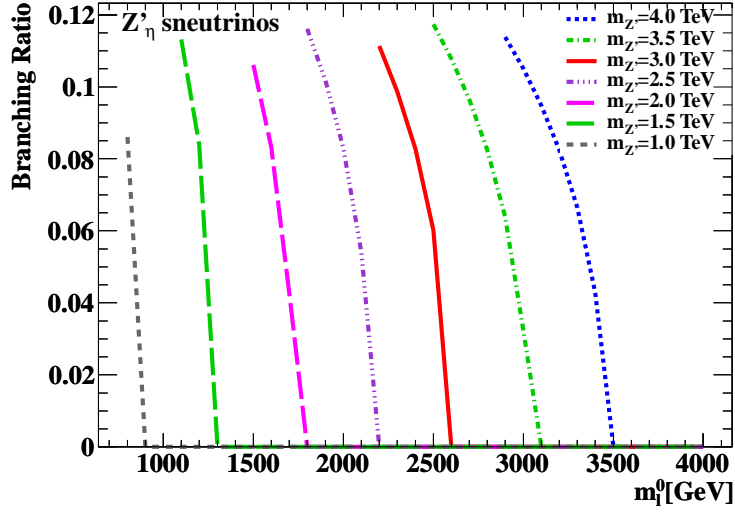}}}
\caption{Branching ratio of the $Z'_\eta$ boson into sneutrino pairs $\tilde\nu_2\tilde\nu_2^*$, as a function
of the slepton mass $m^0_{\tilde\ell}$, for several values of $m_{Z'}$.}
\label{etafig}
\end{figure}
\begin{table}[htp]
\caption{Slepton masses at the Reference Point for the $Z'_\eta$ model, varying $m_{Z'}$ and $m^0_{\tilde\ell}$. $m_{\tilde\ell_{1,2}}$ and $m_{\tilde\nu_{1,2}}$ are the
charged-slepton and sneutrino mass eigenvalues, as discussed in the text.
All masses are given in GeV.}
\label{lepeta}
\begin{center}
\small
\begin{tabular}{|c|c|cccc|}
\hline
\MZprime& $\Mslep^0$    &       \msluno & \msldue    &   \msnuno   &  \msndue  \\
\hline\hline
1000 & 800 & 	        736.9	&	 665.9	&	 732.6	&	 379.3	\\
1000 & 900 &             844.4 & 783.2 & 840.6 & 560.2 \\
1500 & 1100 &		994.0  	& 873.8	& 990.8	& 298.0	\\
1500 & 1300 &             1211.6 & 1115.1 & 1209.0 &754.2 \\
2000 & 1500 &		 1361.2 & 1205.6 & 1358.9 & 503.8 \\
2000 & 1800 &              1686.1 & 1563.1 & 1684.2 & 1115.3\\  
2500 & 1800 &		 1618.0	&	 1411.9	&	 1616.1	&	 344.7	\\
2500 & 2200 &            2053.8 & 1895.6 & 2052.2 & 1311.0 \\       
3000 & 2200 &		 1985.7	&	 1744.6	&	 1984.1	&	 586.4	\\
3000 & 2600 &          2421.4 & 2227.9 & 2420.0 & 1504.6     \\
3500 & 2500 &		 2242.3	&	 1950.2	&	 2240.9	&	 358.9	\\
3500 & 3100 &            2896.2 & 2676.5 & 2895.1 &  1867.8\\
4000 & 2900 &		 2610.2	&	 2283.3	&	 2608.9	&	 643.3	\\
4000 & 3500 &            3263.9 & 3008.9 & 3262.9 & 2062.5\\
\hline
\end{tabular}
\end{center}
\end{table}
\begin{table}[htp]
\caption{Branching ratios of the $Z'_\eta$ boson into
SM and BSM channels, varying $m_{Z'}$ and 
$m_{\tilde\ell}^0$, given in TeV, along the lines described in the text. 
$\rm{BR}_{\rm{SM}}$ and $\rm{BR}_{\rm{BSM}}$ denote the 
total branching fractions, respectively.}
\label{tabeta}
\begin{center}
\small
\begin{tabular}{|c|c|cccccccc|c|c|}
\hline
\MZprime & $\Mslep^0 $   & \Brqqbar   &\Brlep &\Brnu&        
BR$_{W^+W^-}$    & \BrZh  & \Brchinochino &\Brninonino & BR$_{\tilde\nu\tilde\nu^*}$ & $\rm{BR}_ {\rm{SM}}$ & $\rm{BR}_{\rm{BSM}}$	\\
\hline\hline
1.0 & 0.8& 39.45 & 5.24 & 27.26 & 3.01	 & 2.91 & 4.92 & 8.64 &  8.54 & 74.97 & 25.03  \\
1.0 & 0.9 & 43.14 & 5.73 & 29.81 & 3.30 & 3.18 & 5.38 & 9.45 & 0.00 & 81.98 & 18.02 \\
1.5 & 1.1  & 37.82 & 4.93 & 25.63 & 2.71 & 2.67 &  5.16 & 9.76 & 11.31 & 71.10 & 28.90 \\
1.5 & 1.3 & 42.65 & 5.56 & 28.90 & 3.06 & 3.01 & 5.82 & 11.00 & 0.00 & 80.16 & 19.84\\
2.0 & 1.5  &  37.97 & 4.91 & 25.54 &  2.66 & 2.64 & 5.33 & 10.33 & 10.61 & 71.48 & 28.52 \\
2.0 & 1.8  & 42.47 & 5.49 & 28.57 & 2.98 & 2.95 & 5.96 & 11.56 & 0.00 & 79.52 & 20.48\\ 
2.5 & 1.8  & 37.46 & 4.83 & 25.12 & 2.60 & 2.59 & 5.33 & 10.44 & 11.61 & 70.02 & 29.98 \\
2.5 & 2.2  & 42.39 & 5.47 & 28.42 & 2.94 & 2.93 & 6.02 & 11.81 & 0.00 & 79.21 & 20.79\\
3.0 & 2.2  & 37.60 & 4.84 & 25.17 & 2.59 & 2.59 & 5.38 & 10.61 &  11.14 & 70.19 & 29.81 \\
3.0 & 2.6  & 42.31 & 5.45 & 28.32 & 2.92 & 2.91 & 6.06 & 11.94 & 0.00 &    78.64 & 21.36 \\
3.5 & 2.5  & 37.30 & 4.80 & 24.94 & 2.56 & 2.56 & 5.36 & 10.61 & 11.73 & 69.59 & 30.41 \\ 
3.5 & 3.1  & 42.26 & 5.43 & 28.25 & 2.90 & 2.90 & 6.07 & 12.02 & 0.00 & 78.84 & 21.16 \\ 
4.0 & 2.9  & 37.41 & 4.81 & 25.00 & 2.56 & 2.56 & 5.39 & 10.70 &  11.38 & 69.78 & 30.22 \\
4.0 & 3.5  & 42.22 & 5.43 & 28.21 & 2.89 & 2.89 & 6.08 & 12.07 & 0.00 & 78.74 & 21.26\\
\hline								
\end{tabular}							   
\end{center}							
\end{table}							
Table~\ref{tabeta} summarizes the
branching ratios into all allowed SM and BSM channels, for the same $m_{Z'}$ and
$m^0_{\tilde\ell}$ values as in Table~\ref{lepeta}, whereas 
Fig.~\ref{etafig} presents the branching ratio $Z'_\eta\to\tilde\nu_2\tilde\nu_2^*$ as a function 
of $m^0_{\tilde\ell}$ and for 1 TeV~$<m_{Z'}<$~4 TeV. 
The branching fraction into sneutrinos can be as large as about 11\%  
for any value of $m_{Z'}$;
for larger $m^0_{\tilde\ell}$ the sneutrino rate decreases, as
displayed in Fig.~\ref{etafig}. 
Furthermore, Table~\ref{tabeta} shows that, within the scenario
identified by the Reference Point, even the decays into charginos and neutralinos are accessible,
with branching ratios about 5-6\% (charginos) and 10-12\% (neutralinos).
Decays into $W^+W^-$ pairs or Higgs bosons associated with $Z$'s are also permitted,
with rates about 3\%.
The decrease of the sneutrino rate for 
large $m^0_{\tilde\ell}$ results 
in an enhancement of the SM branching ratios into $q\bar q$ and neutrino pairs.
As a whole, summing up the contributions from sneutrinos, 
charginos and neutralinos, the branching ratio into BSM particles 
runs from 24 to 33\%, thus displaying
the relevance of those decays in any analysis on $Z'$ production in a
supersymmetric scenario.

\subsection{Reference Point: {$\mathbf\ZPSI$}}\label{sec:ZPSI}
An extra group \Uprime\  with a mixing angle $\theta=0$ leads to a neutral
vector boson labelled as $Z'_\psi$ (Table~\ref{tab:Models}). 
In Table~\ref{tabpsi}, we quote the slepton masses for a few values of $m_{Z'}$ and
$m^0_{\tilde\ell}$: as before, the results are presented for the 
two values of $m^0_{\tilde\ell}$ 
which are found to enhance and  minimize the slepton rate. 
For any mass value, the D-term enhances by few hundreds GeV
the masses of $\tilde\ell_1$ and $\tilde\nu_1$ and 
strongly decreases $m_{\tilde\ell_2}$ and  $m_{\tilde\ell_2}$,
especially for small $m^0_{\tilde\ell}$ and large $m_{Z'}$.
In Table~\ref{brpsi} we
present the branching ratios into all channels, for the same values
of $m_{Z'}$ and $m^0_{\tilde\ell}$ as in Table~\ref{tabpsi}.
Unlike the $Z'_\eta$ case,  supersymmetric decays into charged-slepton pairs are
allowed for $\theta=0$, with a branching ratio, about 2\%, roughly equal to
the sneutrino rate. 
Furthermore, even the decays into gauginos are relevant,
with rates into $\tilde\chi^+\tilde\chi^-$ and
$\tilde\chi^0\tilde\chi^0$ about 10 and 20\%, respectively. 
The decays into boson pairs, i.e. $Zh$ and $W^+W^-$, are also non-negligible and account
for about 3\% of the total $Z'_\psi$ width. 

\begin{figure}[htp]
\centerline{\resizebox{0.49\textwidth}{!}{\includegraphics{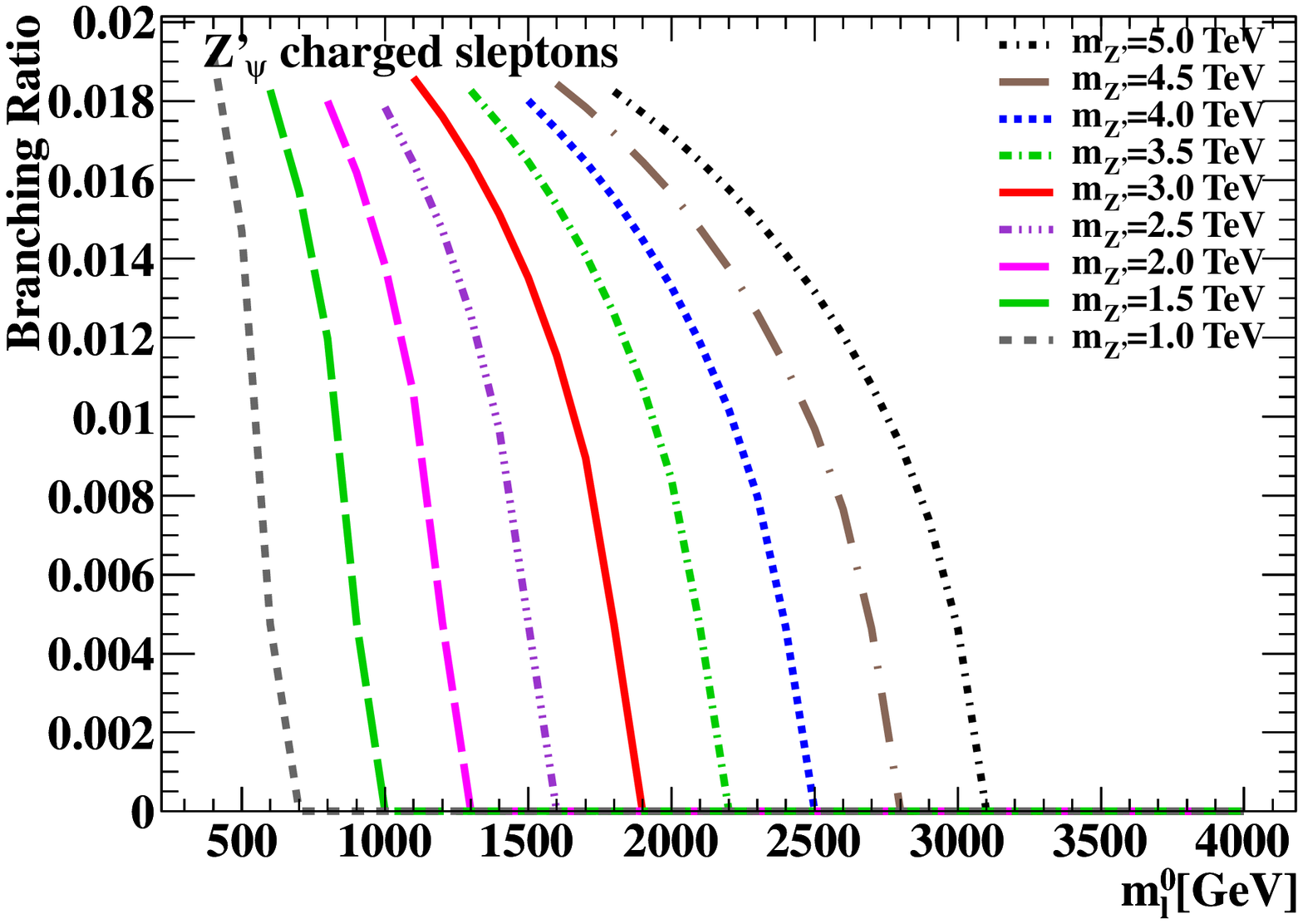}}%
\hfill%
\resizebox{0.49\textwidth}{!}{\includegraphics{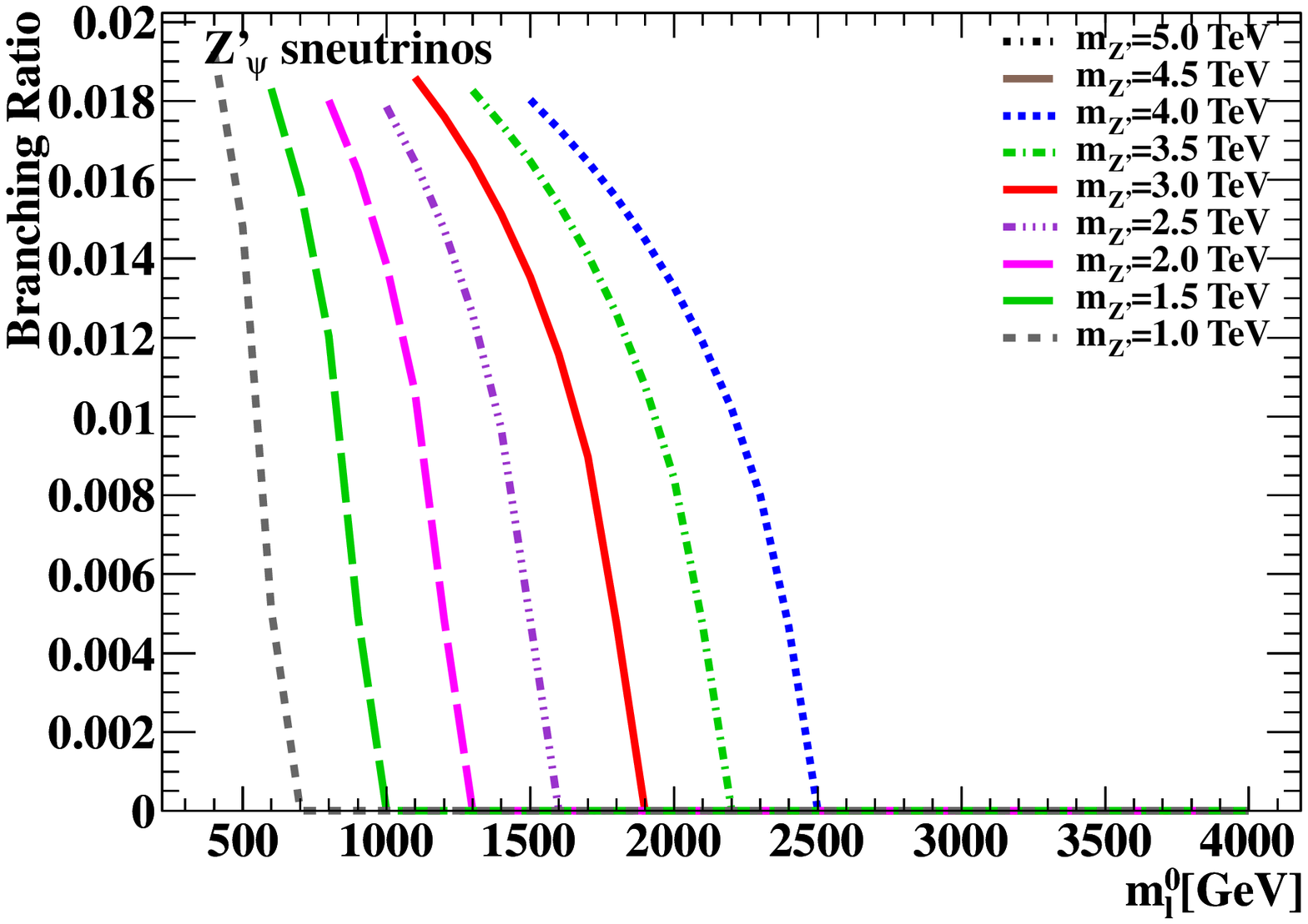}}}
\caption{Dependence of the $Z'_\psi$ branching ratio into charged sleptons (left)
and sneutrinos (right) as a function of $m^0_{\tilde\ell}$, for several values of
$m_{Z'}$.}
\label{figpsi}
\end{figure}

As a whole, the $Z'_\psi$ modelling above depicted yields branching ratios of the order of
35-40\% into BSM particle, and therefore it looks like being a
promising scenario to investigate $Z'$ production within the MSSM.
Figure~\ref{figpsi} finally displays the branching ratios into sneutrinos and charged sleptons
as a function of $m^0_{\tilde\ell}$ and for several values of $m_{Z'}$.
\begin{table}[htp]
\caption{Slepton masses in GeV at the Reference Point for the model $Z'_\psi$ and
a few values of $m_{Z'}$ and $m^0_{\tilde\ell}$. \label{tabpsi}}
\begin{center}
\small
\begin{tabular}{|c|c|cccc|}
\hline
\MZprime& $\Mslep^0$    &       \msluno & \msldue    &   \msnuno   &  \msndue  \\
\hline\hline
1000 & 400  &	 535.2	&	 194.2	&	 529.2	&	 189.2	\\                                   
1000 & 700 & 785.1 & 606.4 & 781.0 & 604.8 \\
1500 & 600  &	 801.7	&	 285.4	&	 797.7	&	 282.0	\\
1500 & 1000 &    1132.6 & 849.4 & 112.7 & 848.3 \\
2000 & 800  &	 1068.4	&	 377.8	&	 1065.4	&	 375.2	\\
2000 & 1300 & 1480.3 & 1092.1 & 1478.2 & 1091.2 \\
2500 & 1000 &	 1335.2	&	 470.6	&	 1333.8	&	 468.6	\\
2500 & 1600 &    1828.3 & 1334.7 & 1826.6 & 1334.0 \\
3000 & 1100 &	 1528.5	&	 296.2	&	 1526.4	&	 292.9	\\
3000 & 1900 &        2176.3 & 1577.2 & 2174.9 & 1576.6\\ 
3500 & 1300 &	 1795.2	&	 401.8	&	 1793.4	&	 399.4	\\
3500 & 2200 &     2524.4 & 1819.7 & 2523.2 & 1819.2\\
4000 & 1500 &	 2061.9	&	 502.7	&	 2060.4	&	 500.8	\\
4000 & 2500 &  2872.5 & 2062.2 & 2871.4 & 2061.7\\
4500 & 1600 &	 2256.7	&	 177.4	&	 2255.3	&	 171.9	\\
4500 & 2800 &  3220.7 & 2304.7 & 3219.7 & 2304.2 \\        
5000 & 1800 & 2523.2 &	 343.1	&	 2521.9	&	 340.3	\\
5000 & 3100 & 3568.8 & 2547.1 & 3567.9 & 2546.7 \\
\hline
\end{tabular}
\end{center}
\end{table}
\begin{table}[htp]
\caption{Branching ratios of the $Z'_{\psi}$ boson into
SM and BSM channels, varying $m_{Z'}$ and 
$m_{\tilde\ell}^0$. The masses are expressed in TeV.}
\label{brpsi}
\begin{center}
\small
\begin{tabular}{|c|c|ccccccccc|c|c|}
\hline
\MZprime & $\Mslep^0 $   & \Brqqbar   &\Brlep & \Brnu &        \BrWW   & \BrZh  
&\Brchinochino & \Brninonino & BR$_{\tilde\nu\tilde\nu^*}$ &
BR$_{\ell\tilde\ell^*}$ &  $\rm{BR}_{\rm{SM}}$ & $\rm{BR}_{\rm{BSM}}$\\
\hline\hline
1.0    & 0.4 & 48.16 & 8.26 & 8.26 & 3.00 & 2.89 & 9.13 & 16.53 & 
1.91 & 1.90 &67.69 & 32.31 \\
1.0    & 0.7	& 50.07 & 8.59 & 8.59 & 3.08 & 2.99 & 9.49 & 17.18 & 0.00 & 
0.00 & 70.33 & 29.67 \\ 
1.5    & 0.6	& 46.78 & 7.90 & 7.90 & 2.71 & 2.69 & 9.73 & 18.64 & 
1.83 & 1.83 & 65.28 & 34.72\\
1.5    & 1.0 & 48.55 & 8.20 & 8.20 & 2.81 & 2.79 & 10.10 & 19.35 & 0.00 & 
0.00 & 67.76 & 32.24 \\
2.0   & 0.8 & 46.30 & 7.77 & 7.77 & 2.62 & 2.62 & 9.92 & 19.37 & 
1.80 & 1.80 & 64.47 & 35.53\\
2.0   &1.3 & 48.03 & 8.06 & 8.06 & 2.72 & 2.72 & 10.29 & 20.10 & 0.00 & 
0.00 & 66.88 & 33.12 \\
2.5   &1.0 & 46.01 & 7.70 & 7.70 & 2.58 & 2.59 & 9.99 & 19.68 &
1.79 & 1.78 & 64.00 & 36.00 \\
2.5   &1.6	& 47.72 & 7.99 & 7.99 & 2.67 & 2.68 & 10.36 & 20.41 & 
0.00 &    0.00 & 66.37 & 33.63 \\
3.0 & 1.1 & 45.35 & 7.58 & 7.58 & 2.53 & 2.54 & 9.92 & 19.63 & 1.86 
& 1.86 & 63.04 & 36.96 \\
3.0 & 1.9 & 47.10 & 7.88 & 7.88 & 2.62 & 2.64 & 10.30 & 20.39 & 0.00 & 
0.00 & 65.47 & 34.53 \\
3.5 & 1.3 & 44.91 & 7.50 & 7.50 & 2.49 & 2.51 & 9.86 & 19.58 & 
1.83 & 1.83 & 62.41 & 37.59	\\	
3.5 & 2.2 & 46.61 & 7.79 & 7.79 & 2.59 & 2.61 & 10.24 & 20.32 & 0.00 & 
0.00 & 64.78 & 35.22 \\
4.0 & 1.5 & 44.60 & 7.45 & 7.45 & 2.47 & 2.49 & 9.82 & 19.53 & 
1.80 & 1.80 & 61.96 & 38.04\\
4.0 & 2.5 & 46.26 & 7.72 & 7.72 & 2.56 & 2.58 & 10.19 & 20.26 & 0.00 & 
0.00 &    64.27 & 35.73 \\
4.5 & 1.6 & 44.32 & 7.40 & 7.40 & 2.45 & 2.47 & 9.78 & 19.47 & 
1.84 & 1.84 & 61.56 & 38.44\\
4.5 & 2.8 & 46.01 & 7.68 & 7.68 & 2.54 & 2.57 & 10.15 & 20.21 & 0.00 & 0.00 & 
63.91 & 36.09\\
5.0 & 1.8 & 44.16 & 7.37 & 7.37 & 2.44 & 2.46 & 9.76 & 19.44 & 
1.82 & 1.82 & 61.33 & 38.67\\
5.0 & 3.1 & 45.83 & 7.65 & 7.65 & 2.53 & 2.55 & 10.13 & 20.18 & 0.00 & 
0.00 & 63.65 & 36.35\\
\hline\end{tabular}			\end{center}
\end{table}								       		

\subsection{Reference Point: {$\mathbf\ZN$}}\label{sec:ZN}
In this subsection we investigate the phenomenology of the $Z'_{\rm N}$ boson, i.e.
a \Uprime\  gauge group with a  mixing angle $\theta=\arctan\sqrt{15}-\pi/2$ (Table~\ref{tab:Models}),
along the lines of the previous sections. As discussed above, the $Z'_{\rm N}$ model
is interesting since it corresponds to the $Z'_\chi$ model, but with the 
unconventional assignment of the SO(10) 
representations. Referring to the notation in Eq.~(\ref{vs}), in the unconventional
E$_6$ model the fields $H$ and $D^c$ are in
the representation {\bf 16} and $L$ and $d^c$ in the {\bf 10} of SO(10).

Table~\ref{tabn} presents
the slepton masses varying $m_{Z'}$ and for the values of $m_{\tilde{\ell}}^0$
which minimize and maximize the slepton rate. The D-term addition to $m^0_{\tilde\ell}$ increases the mass of  $\tilde\ell_1$ and $\tilde\nu_1$
and decreases the mass of $\tilde\ell_2$; its impact on $\tilde\nu_2$ is negligible and
one can assume  $m_{\tilde\nu_2}\simeq m^0_{\tilde\ell}$.
Both decays into  $\slepton_2^+\slepton_2^-$ and $\snu_2\snu_2^*$
are kinematically allowed, whereas  $\tilde\ell_1$ and $\tilde\nu_1$ are too heavy to
contribute to the $Z'$ width.

\begin{figure}
\centerline{\resizebox{0.65\textwidth}{!}{\includegraphics{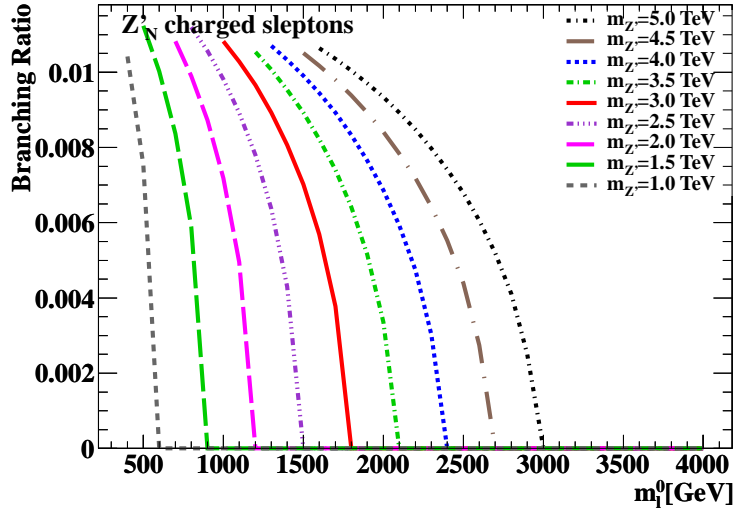}}}
\caption{Slepton branching ratios of the $Z'_{\rm N}$ boson as a function of $m^0_{\tilde\ell}$.}
\label{Nfig}
\end{figure}

Table~\ref{brn} quotes the branching ratios for the $Z'_{\rm N}$,
computed for the same values of $m_{Z'}$ and $m^0_{\tilde\ell}$ as in Table~\ref{tabn}. 
Although $Z'_{\rm N}\to \tilde\nu_2\tilde\nu_2^*$ is kinematically allowed,
the coupling of the $Z'_{\rm N}$ to 
sneutrinos is zero for $\theta=\arctan\sqrt{15}-\pi/2$,
since, as will be discussed in Appendix A, the rate into right-handed sfermions 
vanishes for equal vector and vector-axial coupling, i.e. $v_{\tilde\nu}=a_{\tilde\nu}$:
therefore, this decay mode can be  discarded. 
As for the other supersymmetric channels, the rates into 
charginos and neutralinos are quite significant and amount to about 9\% and
28\%, respectively. The decays into $W^+W^-$ and $Zh$ states account for
approximately 1-2\%,
whereas the branching ratio into charged-slepton pairs is about 1\%,
even in the most favourable case.
As a whole, the rates into BSM final states run from
18 to about 35\% and therefore are a relevant contribution
to the total $Z'$ cross section. 
Figure~\ref{Nfig} finally presents the variation of the charged-slepton branching ratio as
a function of $m^0_{\tilde\ell}$, for a few values of $m_{Z'}$.
\begin{table}[htp]
\caption{Slepton masses in the $\ZN$ model, varying $m_{Z'}$ and $m^0_{\tilde\ell}$, as discussed in
the text. All masses are given in GeV.}\label{tabn}
\begin{center}
\small
\begin{tabular}{|c|c|cccc|}
\hline
\MZprime& $\Mslep^0$    &       \msluno & \msldue    &   \msnuno   &  \msndue  \\
\hline\hline
1000& 400  &	 601.1 & 249.7 & 595.8 & 400.0\\                             
1000 & 600 &   749.2 & 512.2 & 745.0 & 600.0\\
1500& 500  &   837.4 & 165.4 & 833.6 & 500.0  \\
1500 & 900 & 1123.1 & 766.4 & 1120.2 & 900.0 \\
2000& 700  &  1136.4 &	303.9 &	 1133.6	&	 700.0	\\
2000 & 1200 & 1497.1 & 1021.0 & 1495.0 &  1200.0 \\ 
2500& 800  &   1375.8	& 131.8	& 1372.9 & 800.0 \\
2500 & 1500 & 1871.2 & 1275.7 & 1869.5 & 1500.0 \\
3000 & 1000 &   1673.7	&	 319.9	& 1671.8 & 1000.0	\\
3000 & 1800 & 2245.3 & 1530.4 & 2243.9 & 1800.0 \\ 
3500& 1200 &  1972.6	&	 466.2	& 1971.0	&	 1200.0	\\
3500 & 2100 & 2619.4 & 1785.3 & 2618.2 & 2100.0\\
4000& 1300 &  2211.6	&	 303.9	& 2210.2 & 1300.0	\\
4000 & 2400 & 2993.6 & 2040.2 & 2992.5 & 2400.0 \\
4500& 1500 & 2510.2 & 476.8 & 2509.0 & 1500.0\\ 
4500 & 2700 & 3367.7 & 2295.1 & 3366.7 & 2700.0 \\
5000& 1600 & 2749.8 & 249.7	& 2748.6 & 1600.0 \\
5000 & 3100 & 3822.5 & 2666.9 & 3821.6 & 3100.0 \\
\hline
\end{tabular}
\end{center}
\end{table}
\begin{table}[htp]
\caption{Branching ratios of the $Z'_{\rm N}$ boson in
SM and BSM channels, varying $m_{Z'}$ and 
$m_{\tilde\ell}^0$. Slepton and $Z'$ masses are quoted in TeV.}
\label{brn}
\begin{center}
\small
\begin{tabular}{|c|c|cccccccc|c|c|}
\hline
\MZprime & $\Mslep^0 $   & \Brqqbar   &\Brlep                &\Brnu&        \BrWW    & \BrZh  & \Brchinochino &\Brninonino &
BR$_{\tilde\ell^+\tilde\ell^-}$ & $\rm{BR}_ {\rm{SM}}$ & $\rm{BR}_{\rm{BSM}}$	\\
\hline
\hline
1.0  & 0.4 & 49.51 & 11.98 & 9.59 & 1.71 & 1.68 & 8.71 & 15.78 &  
1.04 & 72.79 & 27.21 \\	 
1.0   & 0.6	& 50.03 & 12.11 & 9.69 & 1.73 & 1.69 & 8.80 & 15.94 & 0.00 & 
73.56 & 26.44\\	 
1.5   & 0.5	& 47.99 & 11.51 & 9.21 & 1.57 & 1.57 & 9.26 & 17.76 &
1.12 & 70.28 & 29.72\\	
1.5   & 0.9	& 48.53 & 11.64 & 9.31 & 1.59 & 1.59 & 9.36 & 17.96 & 0.00 & 
71.08 & 28.92\\
2.0   & 0.7	& 47.50 & 11.36 & 9.08 & 1.53 & 1.54 & 9.44 & 18.46 &
1.08 & 69.47 & 30.53\\
2.0 & 1.2 & 48.02 & 11.48 & 9.18 & 1.54 & 1.55 & 9.55 & 18.66 & 0.00 &  
70.22 & 29.78\\	
2.5 & 0.8 & 47.16 & 11.26 & 9.01 & 1.50 & 1.52 & 9.50 & 18.73 &
1.12 & 68.92 & 31.08\\
2.5 & 1.5 & 47.69 & 11.38 & 9.11 & 1.52 & 1.53 & 9.61 & 18.94 & 0.00 &
69.70 & 30.30\\	
3.0 & 1.0 & 46.43 & 11.30 & 8.86 & 1.47 & 1.49 & 9.43 & 18.66 &  
1.08 & 67.83 & 32.17 \\
3.0 & 1.8 & 46.94 & 11.20 & 8.96 & 1.49 & 1.50 & 9.53 & 18.86 & 0.00 &
68.58 &  31.42\\
3.5 & 1.2 & 45.85 & 10.93 & 8.74 & 1.45 & 1.47 & 9.35 & 18.56 &
1.05 & 66.98 & 33.02\\
3.5 & 2.1 & 46.34 & 11.05 & 8.84 & 1.46 & 1.48 & 9.45 & 18.76 & 0.00 &  
67.68 & 32.32 \\
4.0 & 1.3 & 45.42 & 10.83 & 8.66 & 1.43 & 1.45 & 9.29 & 18.47 &
1.07 & 66.34 & 33.66\\
4.0 & 2.4 & 45.91 & 10.94 & 8.75 & 1.45 & 1.47 & 9.39 & 18.67 & 0.00 &
67.06 & 32.94\\
4.5 & 1.5 & 45.13 & 10.75 & 8.60 & 1.42 & 1.44 & 9.24 & 18.41 & 
1.05 & 65.90 & 34.10\\	
4.5 & 2.7 & 45.60 & 10.87 & 8.70 & 1.44 & 1.46 & 9.34 & 18.60 & 0.00 &  
66.61 & 33.39\\
5.0 & 1.6 & 44.90 & 10.70 & 8.56 & 1.41 & 1.43 & 9.21 & 18.35 &
1.06 & 65.56 & 34.44 \\
5.0 & 3.1 & 45.38 & 10.81 & 8.65 & 1.43 & 1.45 & 9.31 & 18.55 & 0.00 & 
66.27 & 33.73\\
\hline								
\end{tabular}
\end{center}       				
\end{table}

\subsection{Reference Point: {$\mathbf\ZI$}}\label{sec:ZI}

The \Uprime-based model leading to a $Z'_{\rm I}$, i.e. 
a mixing angle $\theta=\arccos\sqrt{5/8}-\pi/2$, has been extensively
discussed, as it corresponds to the Representative Point.
It exhibits the property that 
the initial slepton mass $m^0_{\tilde\ell}$ can be as low as a few GeV,
still preserving a physical scenario for the sfermion masses.
In the following, we shall assume a lower limit of $m^0_{\tilde\ell}=200$~GeV
and present results also for 1~TeV, in order to give an estimate of the dependence on $m^0_{\tilde\ell}$.

In Table~\ref{tabi} the charged- and
neutral-slepton masses are listed for a few values of $m_{Z'}$ and $m^0_{\tilde\ell}$.
We already noticed, when discussing the Representative Point and Fig.~\ref{m0fig},
that the D-term correction to the slepton mass is quite important
for $\tilde\ell_1$, $\tilde\nu_1$ and $\tilde\nu_2$, especially for small
values of $m^0_{\tilde\ell}$: this behaviour is confirmed by  Table~\ref{tabi}.
The D-term turns out to
be positive and quite large and the only kinematically
permitted decay into sfermions is $Z'_{\rm I}\to \tilde\ell_2\tilde\ell_2^*$.
However, as in the $Z'_{\rm N}$ case,
the vector and vector-axial coupling are equal, i.e. $v_{\tilde\ell}=a_{\tilde\ell}$,
thus preventing this decay mode for the reasons which will be clarified in
Appendix A.
The conclusion is that in the Reference Point scenario, the $Z'_{\rm I}$ boson can decay into neither
charged nor neutral sleptons.
Therefore, the dependence of the branching ratios on 
$m^0_{\tilde\ell}$ is not interesting and Table~\ref{bri} just reports
the decay rates for fixed
$m^0_{\tilde\ell}=1$~TeV. The total BSM branching 
ratio lies between 12 and 17\% and is mostly due to decays into chargino 
($\sim 4\%$) and neutralino ($\sim 8$-9\%) pairs. Decays involving 
supersymmetric Higgs bosons, such as $H^+H^-$, 
$W^\pm H^\mp$ and $HA$ final states, are possible, but with a total branching ratio
which is negligible for small $Z'$ masses and at most $3\%$
for $m_{Z'}>4$~TeV. As for the decay into SM quarks, it was already pointed out
in Table~\ref{br1} that the rate into $u\bar u$ pairs is zero since the
couplings $v_u$ and $a_u$ (see also Appendix A) vanish for  
$\theta=\arccos\sqrt{5/8}-\pi/2$. Therefore, in Table~\ref{bri}, 
BR$_{q\bar q}$ only accounts for  decays into down quarks.
\begin{table}[htp]
\caption{Slepton masses in the \Uprime\  scenario corresponding to a $Z'_{\rm I}$ boson,
for a few values of $m_{Z'}$ and $m^0_{\tilde\ell}$. All masses are expressed in GeV.}
\label{tabi}
\begin{center}
\small
\begin{tabular}{|c|c|cccc|}
\hline
\MZprime& $\Mslep^0$    &       \msluno & \msldue    &   \msnuno   &  \msndue  \\
\hline\hline
1000& 200 & 736.3	&	 204.7	&	 732.0	&	 734.8	\\
1000& 1000 & 1226.6	&	 1001.0	&	 1223.0	&	 1224.7	\\
\hline
1500&  200     & 1080.4	&	 204.7	&	 1077.4	&	 1079.3	\\
1500&  1000    & 1458.5	&	 1001.0	&	 1456.3	&	 1457.7	\\
\hline
2000&  200     & 1429.1	&	 204.7	&	 1426.8	&	 1428.3	\\
2000&  1000    &  1732.7	&	 1001.0	&	 1730.8	&	 1732.0	\\
\hline
2500& 200 & 1779.7	&	 204.7	&	 1777.9	&	 1779.0	\\
2500& 3000 & 3482.4	&	 3000.3	&	 3481.5	&	 3482.1	\\
\hline
3000 & 200 & 2131.5	&	 204.7	&	 2129.7	&	 2130.7	\\
3000& 3000 & 3674.5	&	 3000.3	&	 3673.7	&	 3674.2	\\
\hline
3500 & 200 & 2483.4	&	 204.7	&	 2482.1	&	 2482.9	\\
3500 & 3000 & 3889.4	&	 3000.3	&	 3888.5	&	 3889.1	\\
\hline
4000& 200 & 2836.9	&	 204.7	&	 2834.8	&	 2835.5	\\
4000& 3000 & 4123.4	&	 3000.3	&	 4122.6	&	 4123.1	\\
\hline
4500& 200 & 3188.6	&	 204.7	&	 3187.6	&	 3188.3	\\
4500& 3000& 4373.5	&	 3000.3	&	 4372.7	&	 4373.2	\\
\hline
5000& 200& 3541.5	&	 204.7	&	 3540.6	&	 3541.2	\\
5000& 3000& 4637.0	&	 3000.3	&	 4636.4	&	 4636.8	\\
\hline
\end{tabular}
\end{center}
\end{table}
\begin{table}[htp]
\caption{Branching ratios of the $Z'_{\rm I}$ into SM and BSM particles for $m^0_{\tilde\ell}=1$~TeV
and varying $m_{Z'}$. The $Z'$ mass is expressed in TeV.
\label{bri}}
\begin{center}
\small
\begin{tabular}{|c|c|cccccccc|c|c|}
\hline
\MZprime & $\Mslep^0$   &\Brqqbar  &\Brlep  & \Brnu   &\BrHpHm &\BrWmpHpm &\BrHA      
&\Brchinochino & \Brninonino &  $\rm{BR}_{\rm{SM}}$ & $\rm{BR}_{\rm{BSM}}$	\\
\hline
\hline
1.0 & 1.0 & 44.06 & 14.69 & 29.37 & 0.00 & ${\cal O}(10^{-3})$ & ${\cal O}(10^{-4})$ & 4.31 & 
7.58 & 88.11 & 11.89 \\
1.5 & 1.0 & 43.39 & 14.46 & 28.93 & 0.00 & ${\cal O} (10^{-4}$ & ${\cal O} (10^{-4})$
&  4.56 & 8.65 & 86.78 & 13.22 \\	 
2.0 & 1.0 & 43.16 &  14.38 & 28.77 & 0.00 & ${\cal O} (10^{-4})$ & ${\cal O} (10^{-3})$ &
4.65 & 9.03 & 86.31 & 13.69 \\
2.5 & 1.0 & 42.99 & 14.33 & 28.66 & 0.06 & ${\cal O}(10^{-3})$ & 0.07 & 
4.68 & 9.19 & 85.98 & 14.02 \\	 
3.0 & 1.0 & 42.53 & 14.18 & 28.36 & 0.53 &${\cal O}(10^{-3})$  & 0.53 & 4.66 & 
9.20 & 85.07 & 14.93 \\
3.5 & 1.0 &  42.16 & 14.05 & 28.11 & 0.91 & ${\cal O}(10^{-3})$ & 0.92 & 4.64 &
9.19 & 84.33 & 15.67 \\
4.0 & 1.0 &  41.90 & 13.96 & 27.93 & 1.20 & ${\cal O}(10^{-3})$  & 1.21 & 4.62 &
9.17 & 83.79 & 16.21 \\
4.5 & 1.0 & 41.70 & 13.90 & 27.80  & 1.40 & ${\cal O}(10^{-3})$  & 1.41 & 4.61 &
9.16 & 83.40 & 16.60 \\
5.0 & 1.0 &  41.56 & 13.85 & 27.71 & 1.56 & 0.01 & 1.57 & 4.60 &
9.15 & 83.12 & 16.88 \\
\hline
\end{tabular}								
\end{center}								 
\end{table}

\subsection{Reference Point: {$\mathbf\ZS$}}\label{sec:ZS}

The $Z'_{\rm S}$ boson corresponds to a 
a mixing angle $\theta=\arctan(\sqrt{15}/9)-\pi/2$.
As in the $Z'_{\rm I}$ model, one can set a small value
of the initial slepton mass,
such as $m^0_{\tilde\ell}=200$~GeV, and still have a meaningful supersymmetric spectrum.
The results on slepton masses and branching ratios are summarized in
Tables~\ref{tabs} and \ref{brs}. Since the $Z'_{\rm S}$ decay
rates are roughly independent of the slepton mass, in Table~\ref{brs}
the branching ratios are quoted only for $m^0_{\tilde\ell}=200$~GeV.
From Table~\ref{tabs} we learn that the D-term contribution
to slepton masses is positive and that
$Z'_{\rm S}\to\tilde\ell_2\tilde\ell_2^*$ is the only decay
kinematically allowed, at least for relatively small values of
$m^0_{\tilde\ell}$. 
However, as displayed in Table~\ref{tabs}, the branching ratio into such
charged sleptons is very small, about 0.1\%, even for low $\Mslep^0$ values.
As for the other BSM decay modes, the most relevant ones 
are into chargino (about 3\%) and
neutralino (about 6-7\%) pairs, the others being quite negligible.
It is interesting, however, noticing that for $m_{Z'}=5$~TeV 
the branching ratio into squark pairs starts to play a role, 
amounting to roughly 8\%. In fact, although we set a high value like
$m^0_{\tilde q}=5$~TeV, for relatively large $Z'$ masses, i.e. $m_{Z'}>3.8$~TeV,
the D-term for $\tilde d_2$-type squarks starts to be negative, in such a way that  
$\tilde d_2\tilde d_2^*$ final states are kinematically permitted.
As a whole, one can say that, at the Reference Point, for $m_{Z'}<5$~TeV the 
BSM decay rate is
about 10-12\%, but it becomes much higher for larger $Z'$ masses, even above
20\%, due to the opening 
of the decay into squark pairs. However, since the experimental signature of 
squark production is given by jets in the final state, it is quite difficult separating them
from the QCD backgrounds. This scenario seems therefore not very promising for a
possible discovery of supersymmetry via $Z'$ decays.
\begin{table}[htp]
\caption{Slepton masses at the Reference Point with a $Z'_{\rm S}$
gauge boson and for few values of $m_{Z'}$ and $m^0_{\tilde\ell}$, given in GeV, as debated in the text.  
\label{tabs}}
\begin{center}
\small
\begin{tabular}{|c|c|cccc|}
\hline
\MZprime& $\Mslep^0$    &       \msluno & \msldue    &   \msnuno   &  \msndue  \\
\hline\hline                                                           
1000&  200   & 917.9	&	 376.8	&	 914.4	&	 1020.0	\\
1000&  1000  &	1342.6	&	 1049.7	&	 1340.2	&	 1414.3	\\	
\hline
1500&    200  &1357.4	&	 516.7	&	 1355.0	&	 1513.4	\\
1500&   1000  &1674.1	&	 1107.7	&	 1672.2	&	 1802.9	\\
\hline
2000&    200 & 1800.7	&	 664.8	&	 1798.9	&	 2010.0	\\
2000&   1000 & 2050.0	&	 1184.0	&	 2048.4	&	 2236.1	\\
\hline
2500&    200 &2245.5	&	 816.7	&	 2244.1	&	 2508.0	\\
2500&   3000 &3742.0	&	 3102.7	&	 3741.1	&	 3905.2	\\
\hline
3000&    200 &2691.2	&	 970.5	&	 2690.0	&	 3006.7	\\
3000&   3000&4025.2	&	 3146.7	&	 4024.4	&	 4242.7	\\
\hline
3500&    200   &  3137.3	&	 1125.6	&	 3136.3	&	 3505.7	\\
3500&    3000  & 4336.2	&	 3198.0	&	 4335.4	&	 4609.8	\\
\hline
4000&    200   &3583.6	&	 1281.4	&	 3582.7	&	 4005.0	\\
4000&   3000   &4669.3	&	 3256.1	&	 4668.6	&	 5000.0	\\
\hline
4500&    200   & 4030.2	&	 1437.7	&	 4029.4	&	 4504.5	\\
4500&   3000   & 5020.2	&	 3320.7	&	 5019.6	&	 5408.4	\\
\hline
5000&   200    &  4476.9	&	 1594.3	&	 4476.2	&	 5004.0	\\
5000&   3000   &  5385.4	&	 3391.4	&	 5384.8	&	 5831.0	\\
\hline
\end{tabular}
\end{center}
\end{table}
\begin{table}[htp]
\caption{Branching ratios of the $Z'_{\rm S}$ with the MSSM parameters at the Reference
Point and for a few values of $m_{Z'}$, expressed in TeV. The initial 
slepton mass is fixed to 0.2 TeV,
since the decay rates are independent of $m^0_{\tilde\ell}$. \label{brs}}
\begin{center}
\small
\begin{tabular}{|c|c|ccccccccc|c|c|}
\hline
\MZprime & $\Mslep^0$   &\Brqqbar  &\Brlep         &\Brnu   & \BrWW & BR$_{Zh}$ &
\Brchinochino & \Brninonino &  BR$_{\tilde\ell^+\tilde\ell^-}$ & BR$_{\tilde q\tilde q^*}$
& $\rm{BR}_{\rm{SM}}$ & $\rm{BR}_{\rm{BSM}}$	\\
\hline
\hline
1.0 & 0.2 & 42.29 & 13.70 & 34.57 & 0.15 & 0.14 & 3.33 & 5.75 & 
0.07 & 0.00 & 90.71 & 9.29\\
1.5 & 0.2 & 41.84 & 13.54 & 34.16 & 0.15 & 0.14 & 3.51 & 6.59 &
0.07 & 0.00 & 89.68 & 10.32	\\
2.0 & 0.2 & 41.67 & 13.48 & 34.02 & 0.14 & 0.14 & 3.57 & 6.90 & 
0.08 & 0.00 & 89.32 & 10.68\\
2.5 & 0.2 & 41.56 & 13.44 & 33.91 & 0.14 & 0.14 & 3.59 & 7.03 & 
0.08 & 0.00 & 89.06 & 10.94 \\
3.0 & 0.2 & 41.25 & 13.34 & 33.66 & 0.14 & 0.14 & 3.58 & 7.06 &
0.08 & 0.00 & 88.39 & 11.61\\
3.5 & 0.2  & 40.99 & 13.26 & 33.45 & 0.14 & 0.14 & 3.57 & 7.07 & 
    0.08 & 0.00 & 87.84 & 12.16 \\
4.0 & 0.2 & 40.81 & 13.20 & 33.30 & 0.14 & 0.14 & 3.56 & 7.07 & 
0.08 & 0.00 & 87.44 & 12.56\\
4.5 & 0.2 &40.67 & 13.15 & 33.19 & 0.14 & 0.14 & 3.56 & 7.07 &
0.08 & 0.00 & 87.15 & 12.85\\
5.0 & 0.2 & 37.34 & 12.07 & 30.46 & 0.13 & 0.13 & 3.27 & 6.50 & 
0.07 & 7.97 & 80.00 & 20.00\\
\hline									       				    
\end{tabular}								       				      
\end{center}\end{table}

\subsection{Reference Point: {$\mathbf\ZCHI$}}\label{sec:ZCHI}

The \Uprime\  group corresponding to a mixing angle $\theta=-\pi/2$ and a boson $Z'_\chi$
does not lead to a meaningful sfermion scenario in the explored range of parameters, 
as the sfermion masses are unphysical after the addition of the D-term.
This feature of the $Z'_{\rm\chi}$ model, already observed in the Representative
Point parametrization (see Section 4.1 and the $m_{\tilde d_2}$ spectrum in
Fig.~\ref{thetamass}), holds even for a higher initial squark mass, such as 
$m^0_{\tilde q}=5$~TeV, as in the Reference Point.
It is nevertheless worthwhile presenting in Table~\ref{tabchi} the Standard Model branching
ratios, along with those into Higgs and vector bosons in a generic Two Higgs Doublet Model.
For any $m_{Z'}$ the rates into quark and neutrino pairs
are the dominant ones, being about 40-45\%, whereas the branching ratio into lepton states is
approximately 12\% and the other modes ($W^+W^-$, $Zh$, $HA$ and $H^+H^-$) account for the
remaining 1-3\%.  
\begin{table}[htp]
\caption{Branching ratios of the $\ZCHI$ boson as a function of the $Z'$ mass, given in TeV.
The rates into sfermion pairs are not presented, since the sfermion mass spectrum
is unphysical for the $Z'_\chi$ model in the Reference Point scenario.
\label{tabchi}}
\begin{center}
\small
\begin{tabular}{|c|ccccccc|c|c|}
\hline
\MZprime         & \Brq    &  \Brlep &        \Brnu &   \BrWW &  \BrHpHm & \BrZH &  \BrhA
& BR$_{\rm SM}$ & BR$_{\rm BSM}$	\\
\hline
\hline
1.0  & 44.35 & 12.44 & 42.29 & 0.90 & 0.00 & 0.02 &
${\cal O} (10^{-3})$ & 99.98 & 0.02\\
2.0  & 44.32 & 12.34 & 41.96 & 0.84 & 0.00 & 0.28 & 0.26 &
99.46 & 0.54\\
3.0  & 44.03 & 12.24 & 41.63 & 0.82 & 
0.24 & 0.53 & 0.52 & 98.71 & 1.29 \\
4.0  & 43.84 & 12.18 & 41.43 & 0.82 & 0.46 & 0.64 & 
      0.63 & 98.27 & 1.73\\
5.0  & 43.74 & 12.15 & 41.33 & 0.81 & 0.58 & 0.70 &
      0.69 & 98.03 & 1.97\\
\hline
\end{tabular}
\end{center}
\end{table}

\subsection{Reference Point: {$\mathbf{Z'_{\rm{SSM}}}$}}\label{sec:ZSSM}
A widely used model in the analyses of the experimental data is the
Sequential Standard Model (SSM):
in this framework the $Z'$ coupling to SM and MSSM
particles is the same as the $Z$ boson. The SSM is considered as 
a benchmark, since the production cross section is only function of
the $Z'$ mass and there is no dependence on the mixing angle $\theta$ and possible new physics
parameters, such as the MSSM ones. 

As for the supersymmetric sector, the sfermion masses get the D-term contribution associated 
with the  hyperfine 
splitting, Eq.~(\ref{d1}), but not the one due to 
further extensions of the MSSM, namely
Eq.~(\ref{dt}), proportional to $g'^2$ in the case of \Uprime.
Moreover, the $Z'_{\rm SSM}$ coupling to sfermions 
is simply given by $g_{\rm{SSM}}=g_2/(2\cos\theta_W)$, as in the SM.
Since the hyperfine-splitting D-term is quite small, the sfermion spectrum 
is physical even for low values of $m^0_{\tilde\ell}$.
Table~\ref{tabssm} reports the sfermion masses obtained at the Reference Point,
Eq.~(\ref{refpoint}), for a few values of $m_{Z'}$ and varying $m^0_{\tilde\ell}$ from 
100 GeV to $m_{Z'}/2$, the highest value kinematically allowed.   
For $m^0_{\tilde\ell}=100$~GeV, because of the D-term, 
$m_{\tilde\nu_1}$ decreases by about 25\%, $m_{\tilde\ell_1}$ and
$m_{\tilde\ell_2}$ slightly increase and $m_{\tilde\nu_2}$ is roughly unchanged.
For large values of $m^0_{\tilde\ell}$,  the
D-term is negligible and all slepton masses are approximately equal to $m^0_{\tilde\ell}$. 

\begin{figure}[htp]
\centerline{\resizebox{0.49\textwidth}{!}{\includegraphics{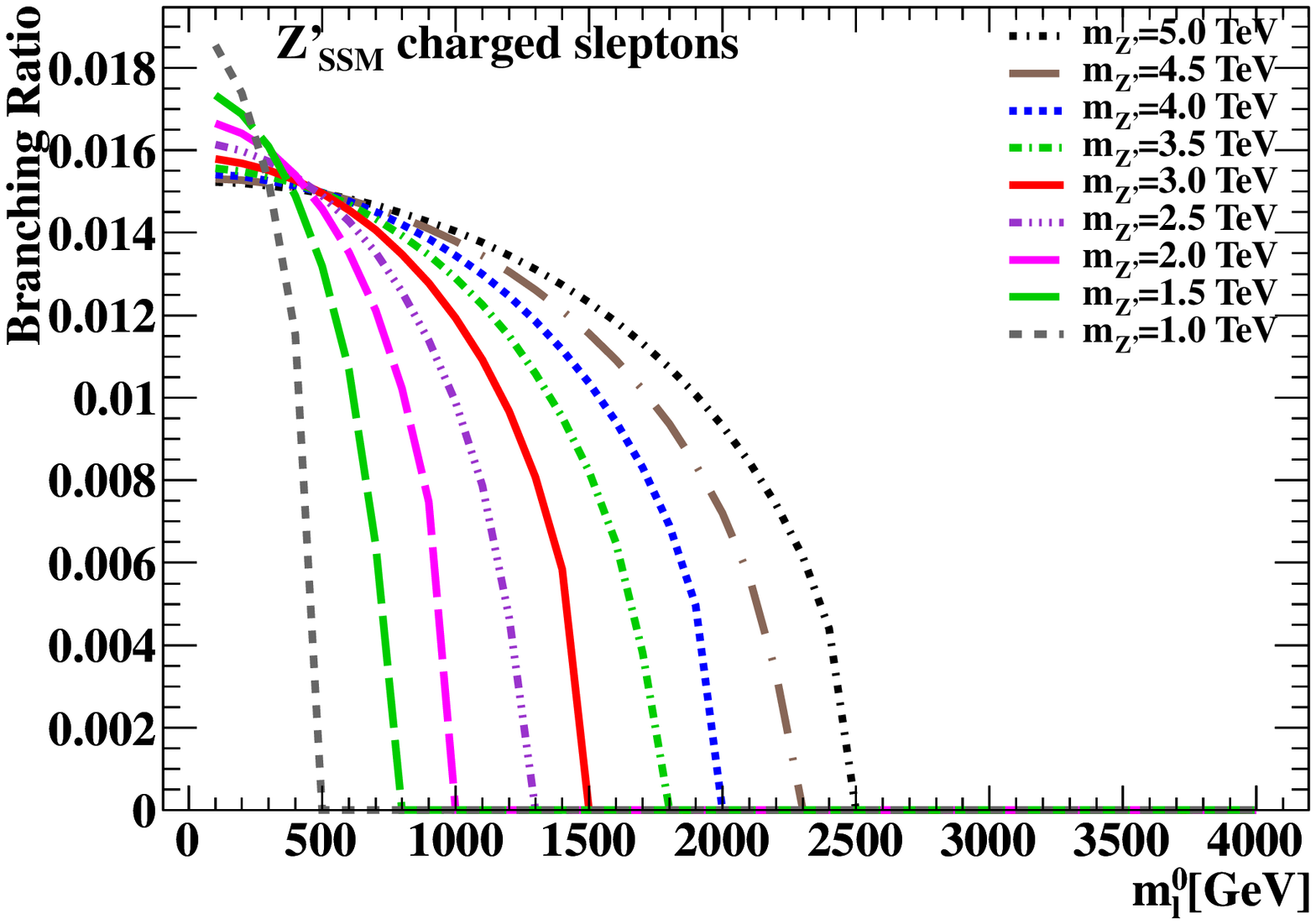}}%
\hfill%
\resizebox{0.49\textwidth}{!}{\includegraphics{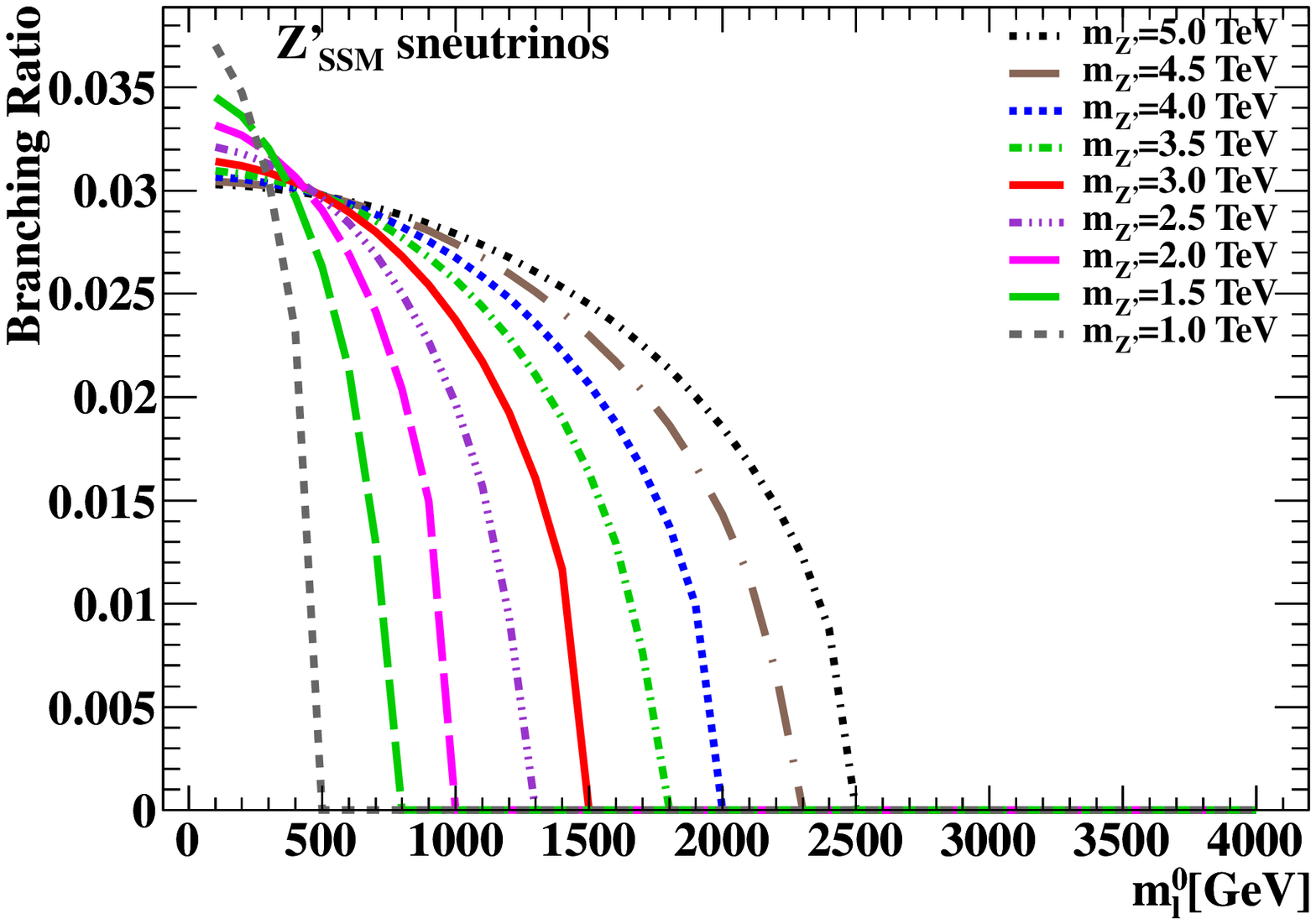}}}
\caption{Branching ratios of the $Z'_{\rm SSM}$ as a function of $m^0_{\tilde\ell}$
for several values of the $Z'$ mass. Left: branching fraction into charged sleptons.
Right: branching fraction into sneutrinos.}
\label{figssm}
\end{figure}

Tables~\ref{brsm} and \ref{brssm} present, respectively, 
the SM and BSM branching ratios of the $Z'_{\rm SSM}$ at the
Reference Point,  for the values of $Z'$ and
slepton masses listed 
in Table~\ref{tabssm}. The decays into BSM particles exhibit rates, about
60-65\%, which can be even higher than the SM ones, accounting
for the remaining 35-40\%. In fact, this turns out to be mostly due to the decays into
neutralinos, accounting for more than 30\%, and into charginos, about 16-18\%.
The branching fractions into sleptons are quite small: the one into sneutrinos 
is less than 4\% and the one into charges sleptons about 1-2\%.
The $W^+W^-$ mode contributes with a rate about 4-5\%, 
the $H^+H^-$ one is relevant only for $m_{Z'}>2.5$~TeV, with a branching ratio which
can reach 3\%, the $Zh$ and $hA$ channels are accessible for $m_{Z'}>1.5$~TeV, 
with decay fractions between 1 and 4\%.
The variation of the sneutrino and charged-slepton branching ratios as a function of the
slepton mass at the $Z'$ scale is displayed in Fig.~\ref{figssm} for 
1 TeV$<m_{Z'}<$4 TeV.
\begin{table}[htp]
\caption{Slepton masses in the $Z'_{\rm{SSM}}$ model varying $m_{Z'}$ and $m^0_{\tilde\ell}$.
All masses are quoted in GeV.
\label{tabssm} }
\begin{center}
\small
\begin{tabular}{|c|c|cccc|}
\hline
 $m_{Z'}$ &  $\Mslep^0$    &       \msluno & \msldue    &   \msnuno   &  \msndue  \\
\hline\hline                                                           
1000 & 100 & 110.6 & 109.1 & 76.6 & 100.0	\\		
1000 & 500 & 502.2 & 501.9 & 495.8 & 500 	\\
1500 & 100 &  110.6 & 109.1 & 76.6 & 100.0 \\
1500 & 750 & 751.5 & 751.3 & 747.2 & 750.0   \\
2000 & 100 & 110.6 & 109.1 & 76.6 & 100.0  \\
2000 & 1000 & 1001.1 & 1000.9 & 997.9 & 1000.0 \\
2500 & 100 & 110.6 & 109.1 & 76.6 & 100.0 \\
2500 & 1250 & 1250.9 & 1250.8 & 1248.3 & 1250.0 \\
3000 & 100 & 110.6 & 109.1 & 76.6 & 100.0 \\
3000 & 1500 & 1001.1 & 1000.9 & 997.9 & 1000.0 \\
3500 & 100 & 110.6 & 109.1 & 76.6 & 100.0  \\
3500 & 1750 & 1750.6 & 1750.6 & 1748.8 & 1750.0 \\
4000 & 100 & 110.6 & 109.1 & 76.6 & 100.0 \\
4000 & 2000 & 2000.6 & 2000.5 & 1999.0 & 2000.0 \\
4500 & 100 & 110.6 & 109.1 & 76.6 & 100.0\\
4500 & 2250 & 2250.5 & 2250.4 & 2249.1 & 2250.0 \\
5000 & 100 & 110.6 & 109.1 & 76.6 & 100.0  \\
5000 & 2500 & 2500.4 & 2500.4 & 2499.2 & 2500.0 \\
\hline
\end{tabular}
\end{center}
\end{table}
\begin{table}
\caption{Branching ratios into SM particles of 
the $Z'_{\rm SSM}$, varying $m_{Z'}$ and $m^0_{\ell}$, as debated in the text.
Slepton and $Z'$ masses are expressed in TeV.\label{brsm}}
\begin{center}
\small
\begin{tabular}{|c|c|cccc|c|}
\hline
\MZprime & $\Mslep^0$   & BR$_{q\bar q}$    & BR$_{\ell^+\ell^-}$ & BR$_{\nu\bar\nu}$ & BR$_{W^+W^-}$ & 
BR$_{\rm{SM}}$ 	\\
\hline
\hline	
1.0 & 0.10 & 29.61 & 3.87 & 7.69 & 5.56 & 46.73\\	 
1.0 & 0.50 & 31.38 & 4.10 & 8.15 & 5.90 & 49.53 \\
1.5 & 0.10 & 27.38 & 3.53 & 7.02 & 4.86 & 42.79 \\
1.5 & 0.75 & 28.89 & 3.73 & 7.41 & 5.13 & 45.15 \\
2.0 & 0.10 & 26.21 & 3.36 & 6.69 & 4.56 & 40.83\\
2.0 & 1.00 & 27.59 & 3.54 & 7.04 & 4.80 & 42.98\\
2.5 & 0.10 & 25.35 & 3.25 & 6.46 & 4.37 & 39.42 \\
2.5 & 1.25 & 26.64 & 3.41 & 6.79 & 4.59 & 41.42 \\
3.0 & 0.10 & 24.78 & 3.17 & 6.31 & 4.25 & 38.51 \\
3.0 & 1.50 & 26.01 & 1.66 & 6.62 & 4.46 & 40.42 \\
3.5 & 0.10 & 24.42 & 3.12 & 6.21 & 4.17 & 37.92 \\
3.5 & 1.75 & 25.61 & 1.40 & 6.51 & 4.37 & 39.78 \\
4.0 & 0.10 & 24.18 & 3.09 & 6.15 & 4.12 & 37.54 \\
4.0 & 2.00 & 25.35 & 1.21 & 6.44 & 4.32 & 39.35 \\
4.5 & 0.10 & 24.01 & 3.07 & 6.10 & 4.09 & 37.27 \\
4.5 & 2.25 & 25.16 & 1.07 & 6.39 & 4.28 & 39.06 \\
5.0 & 0.10 & 23.89 & 3.05 & 6.07 & 4.06 &  37.07\\
5.0 & 2.50 & 25.03 & 0.96 & 6.36 & 4.25 & 38.84 \\
\hline								
\end{tabular}							
\end{center}							
\end{table}
\begin{table}[htp]
\caption{Branching ratios into BSM particles of 
the $Z'_{\rm SSM}$ for a few values of $m_{Z'}$ and $m^0_{\tilde\ell}$, expressed in TeV.
\label{brssm}}
\begin{center}
\small
\begin{tabular}{|c|c|ccccccc|c|}
\hline
\MZprime & $\Mslep^0$   &
BR$_{H^+H^-}$ & BR$_{Zh}$ & B$_{hA}$ 
& BR$_{\tilde\chi^+\tilde\chi^-}$ & BR$_{\tilde\chi^0\tilde\chi^0}$ 
& BR$_{\tilde\ell^+\tilde\ell^-}$ & BR$_{\tilde\nu\tilde\nu^*}$ & BR$_{\rm{BSM}}$		\\
\hline
\hline	
1.0 & 0.10 & 0.00 & $\sim 10^{-6}$ &  0.00 & 18.31 & 
29.30 & 1.89 & 3.77 &  53.27\\	 
1.0 & 0.50 & 0.00 & $\sim 10^{-6}$ & 0.00 & 19.41 &  
31.06 & 0.00 & 0.00 & 50.47 \\
1.5 & 0.10 & 0.00 & 0.87 & 0.76 & 17.84 &
32.52 & 1.75 & 3.48 & 57.21 \\
1.5 & 0.75 & 0.00 & 0.92 & 0.80 & 18.82 &
34.31 & 0.00 & 0.00 & 54.55\\
2.0 & 0.10 & 0.00 & 1.93 & 1.85 & 17.37 &  
33.01 & 1.67 & 3.33 &  59.17\\
2.0 & 1.00 & 0.00 & 2.04 & 1.95 & 18.28 & 
34.75 & 0.00 & 0.00 & 57.02 \\
2.5 & 0.10 & 0.91 & 2.59 & 2.53 &  
16.93 & 32.78 & 1.62 & 3.22 & 60.58\\
2.5 & 1.25 & 0.95 & 2.72 & 2.66 & 
17.79 & 34.45 & 0.00 & 0.00 & 58.57\\
3.0 & 0.10 & 1.72 & 2.98 & 2.94 &  
16.62 & 32.51 & 1.58 & 3.15 & 61.49 \\
3.0 & 1.50 & 1.81 & 3.13 & 3.08 &  
17.44 & 34.12 & 0.00 & 0.00 & 59.58 \\
3.5 & 0.10 & 2.27 & 3.23 & 3.20 &  
16.42 & 32.30 & 1.56 & 3.10 & 62.08 \\
3.5 & 1.75 & 2.38 & 3.38 & 3.35 &  
17.22 & 33.88 & 0.00 & 0.00 & 60.22 \\
4.0 & 0.10 & 2.65 & 3.39 & 3.37 &  
16.28 & 32.16 & 1.54 & 3.07 & 62.46\\
4.0 & 2.00 & 2.78 & 3.56 & 3.53 & 
17.07 & 33.71 & 0.00 & 0.00 & 60.65 \\
4.5 & 0.10  & 2.91 & 3.51 & 3.49 &  
16.19 & 32.06 & 1.53 & 3.05 & 62.73 \\
4.5 & 2.25 & 3.05 & 3.67 & 3.65 &  
16.96 & 33.59 & 0.00 & 0.00 & 60.94 \\
5.0 & 0.10 & 3.11 & 3.59 & 3.57 &
16.12 & 31.98 & 1.52 & 3.03 & 62.93\\
5.0 & 2.50 & 3.26 & 3.76 & 3.74 & 
16.89 & 33.51 & 0.00 & 0.00 & 61.16\\
\hline
\end{tabular}
\end{center}
\end{table}
\section{Cross sections and event rates at the LHC}
In this section we present the total cross section
for $Z'$ production at the LHC according to the 
models discussed throughout this paper, i.e. Table~\ref{tab:Models}, as well as the
Sequential Standard Model. We consider $pp$
collisions at three centre-of-mass energies: 
7 TeV (the 2011 LHC run), 8 TeV (the 2012 run) and 
14 TeV, the ultimate project energy.
For each energy we shall calculate the cross section and estimate
the expected number of events with a $Z'$ boson decaying into supersymmetric
particles, for a few values of integrated luminosity.

\subsection{Leading order ${\mathbf Z'}$ production cross section}

The cross sections are computed at leading order (LO), employing 
the LO parton distribution functions CTEQ6L \cite{cteq} and setting
the factorization scale  equal to the $Z'$ mass.
Using a different LO PDF has a negligible impact on
the results. 
The cross section for Drell--Yan like processes
has been computed up to next-to-next-to leading order
(NNLO) in QCD and, in principle, the calculations carried out
in Refs.~\cite{nnlo1,nnlo2} can be easily extended to $Z'$ production processes.
However, since all $Z'$ partial widths and branching ratios 
have been evaluated at LO, for the sake of consistency, we decided to stick 
to the lowest-level approximation.

\begin{figure}
\centerline{\resizebox{0.49\textwidth}{!}{\includegraphics{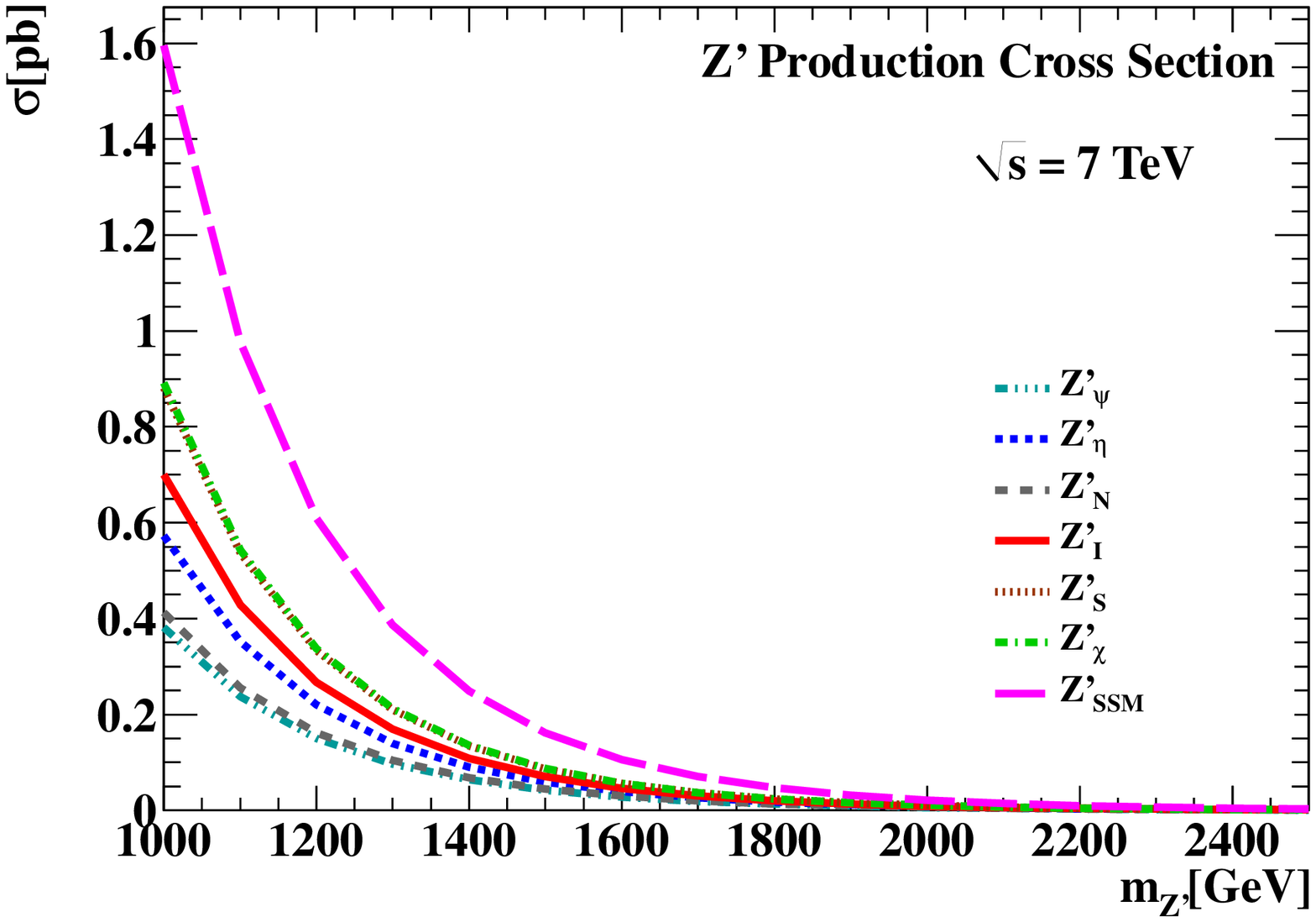}}%
\hfill%
\resizebox{0.49\textwidth}{!}{\includegraphics{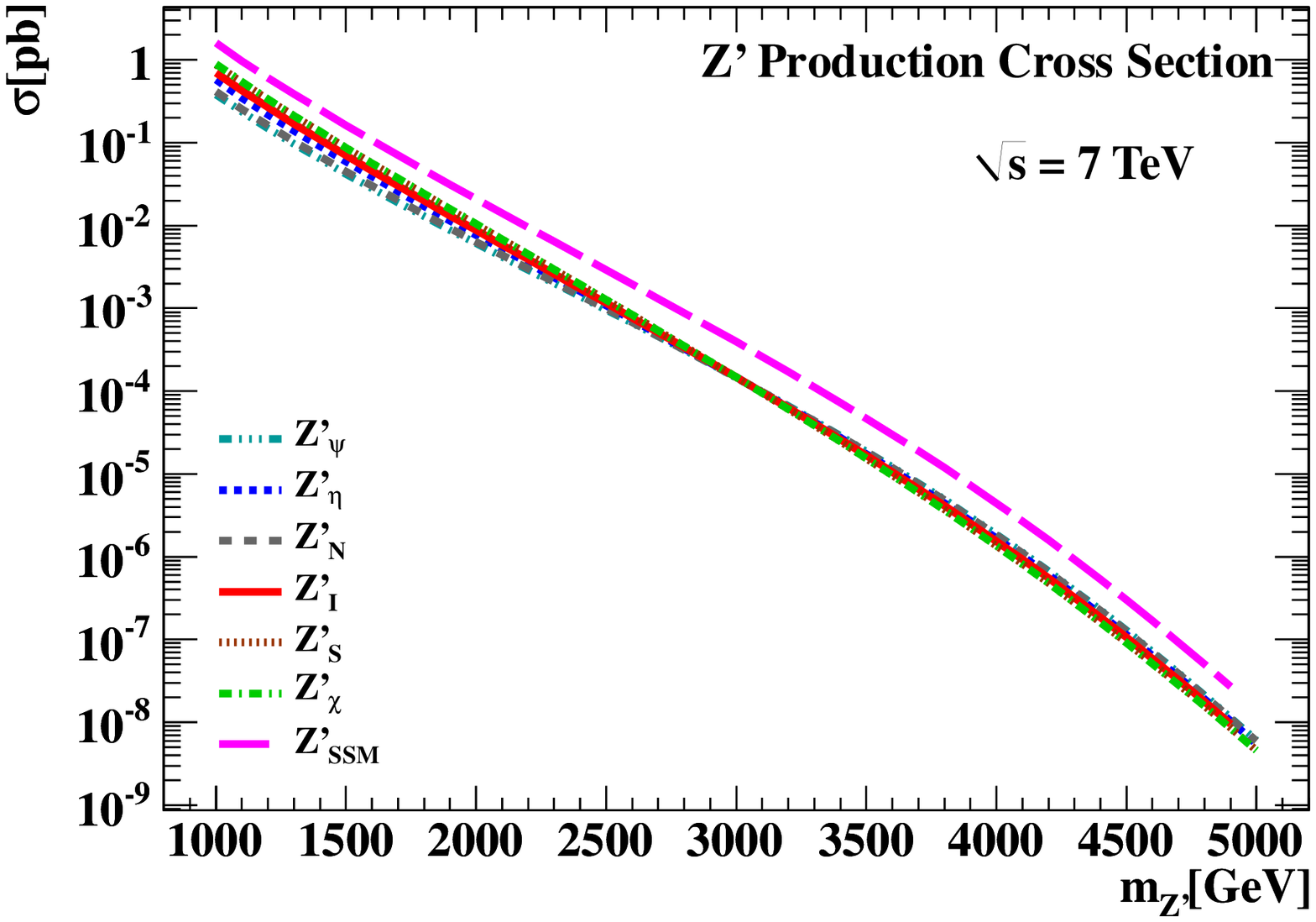}}}
\caption{Cross section of $Z'$ production in $pp$ collisions
at 7 TeV. Left: linear scale. Right: logarithmic scale. }
\label{fig7}
\end{figure}

\begin{figure}
\centerline{\resizebox{0.49\textwidth}{!}{\includegraphics{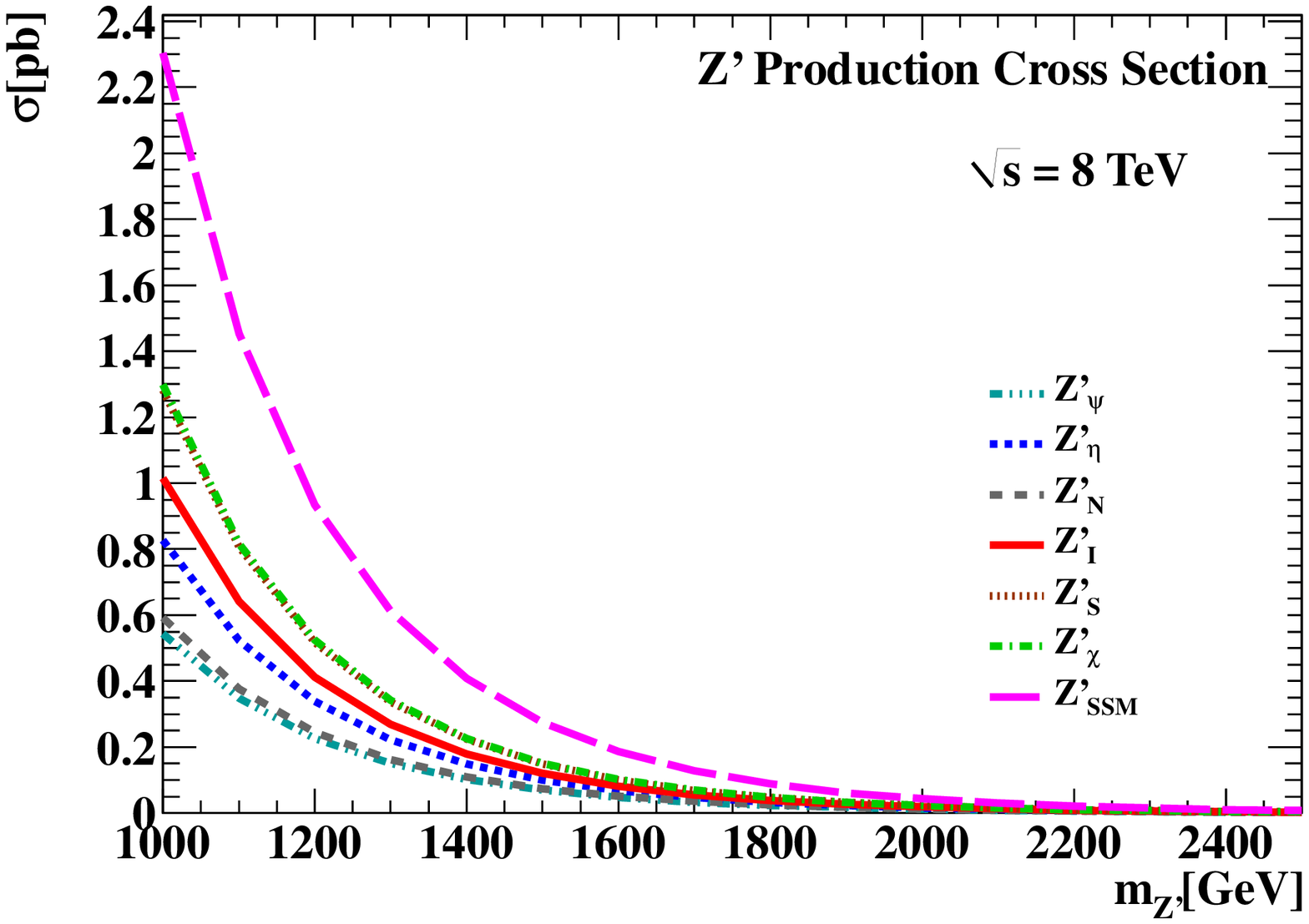}}%
\hfill%
\resizebox{0.49\textwidth}{!}{\includegraphics{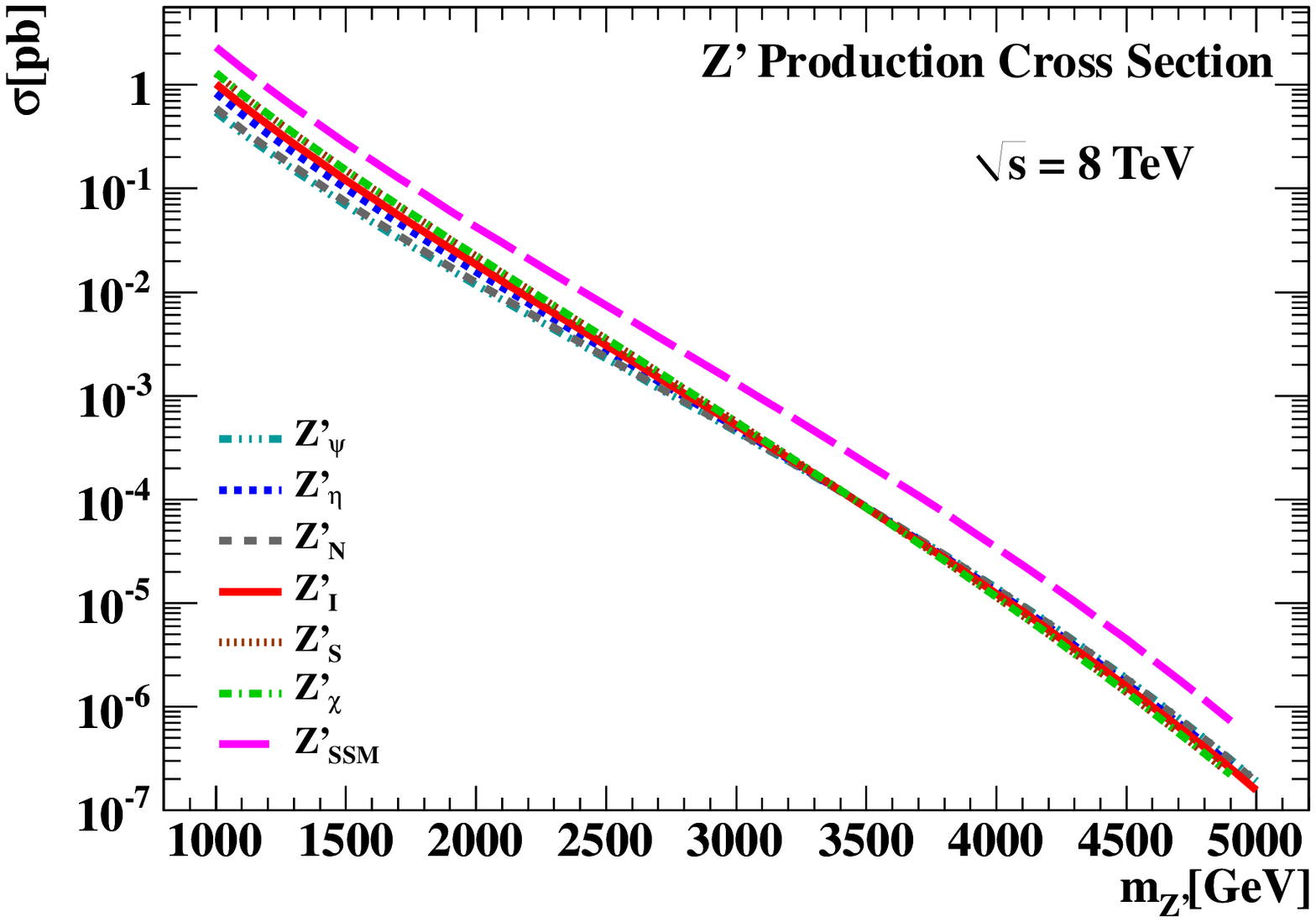}}}
\caption{Cross section for $Z'$ production in $pp$ collisions
at 8 TeV. Left: linear scale. Right: logarithmic scale. }
\label{fig8}
\end{figure}

\begin{figure}
\centerline{\resizebox{0.49\textwidth}{!}{\includegraphics{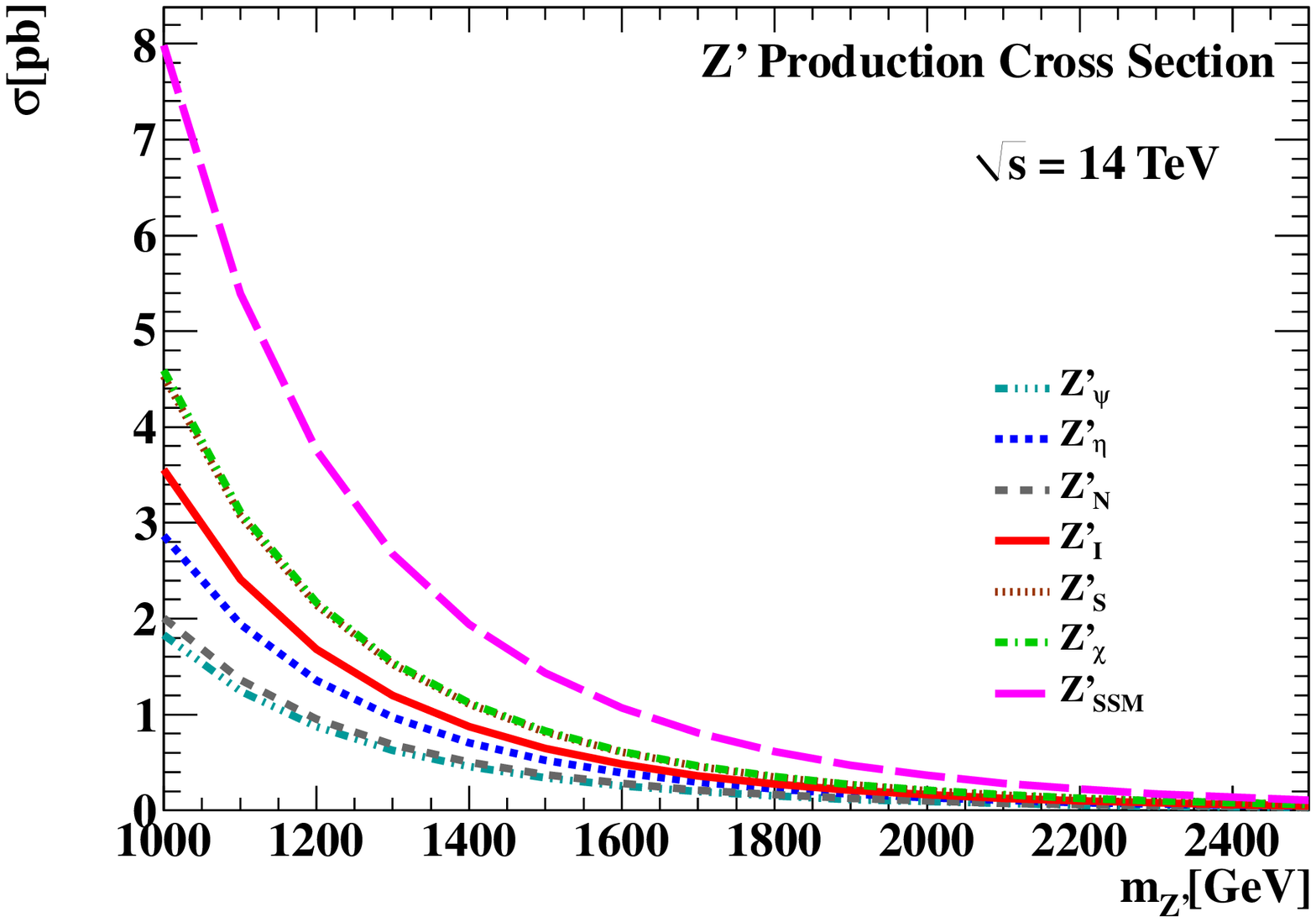}}%
\hfill%
\resizebox{0.49\textwidth}{!}{\includegraphics{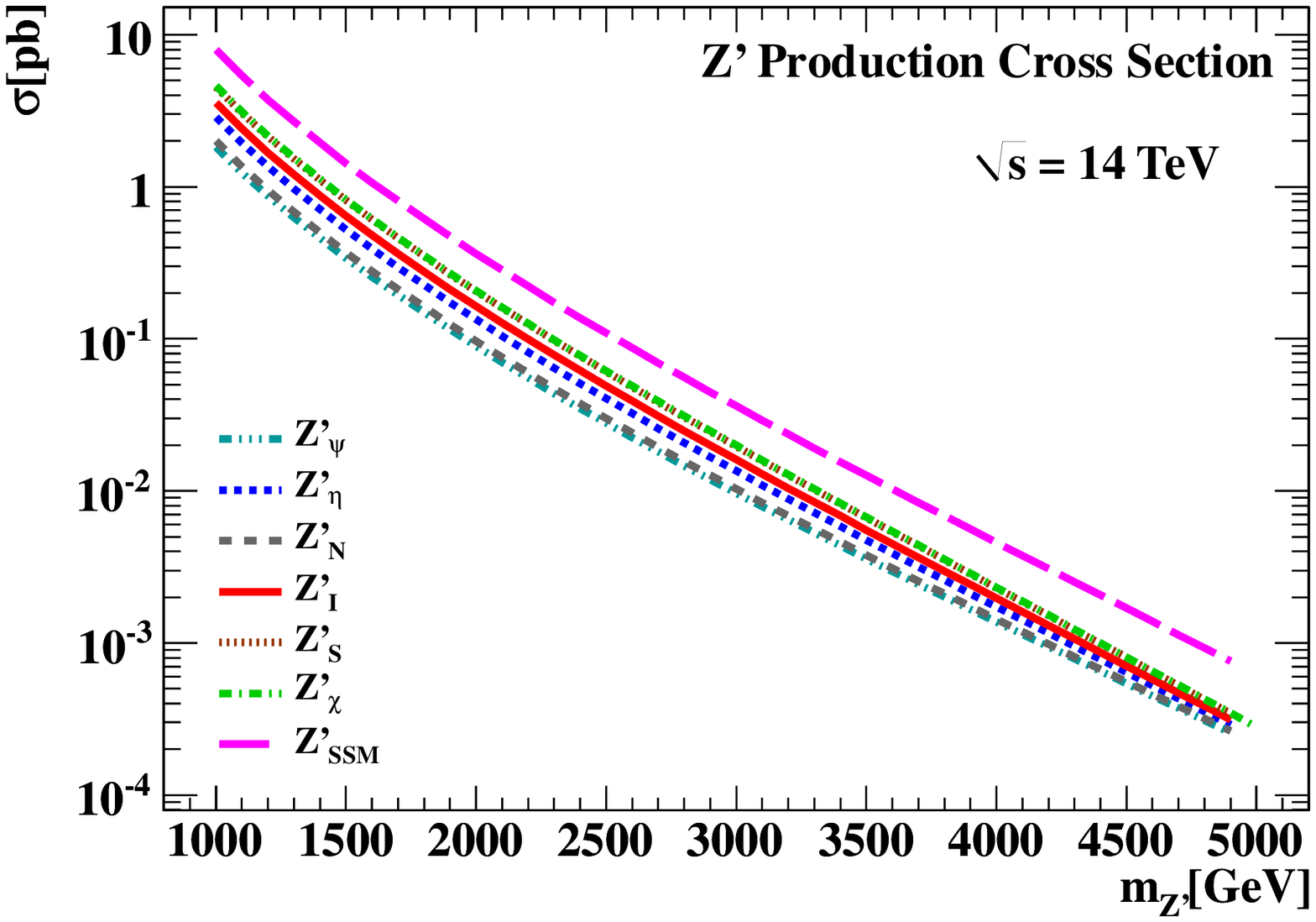}}}
\caption{Cross section for $Z'$ production in $pp$ collisions
at 14 TeV. Left: linear scale. Right: logarithmic scale. }
\label{fig14}
\end{figure}

The parton-level
process is analogous to $Z$ production, i.e. it is the
purely SM quark-antiquark annihilation  $q\bar q\to Z'$.
Since the coupling of the $Z'$ to the quarks depends on the
specific \Uprime\  scenario, 
the production rate is a function of the mixing angle $\theta$
and of the $Z'$ mass,
but is independent of the MSSM parameters. In the Sequential
Standard Model, the cross section just depends on the mass of the $Z'_{\rm SSM}$.
Figures~\ref{fig7}--\ref{fig14} present the total cross section for the
different models investigated throughout this work, 
as a function of $m_{Z'}$, at the energies 
of 7 TeV (Fig.~\ref{fig7}), 8 TeV (Fig.~\ref{fig8}) and 14 TeV (Fig.~\ref{fig14}).
For each centre-of-mass energy, 
we present the results on linear (left) and logarithmic (right) scales.
Tables~\ref{cs7}, \ref{cs8} and \ref{cs14} quote the numerical
values of the LO $Z'$ production cross section, varying $m_{Z'}$
from 1 to 5 TeV, with steps of 500 GeV, in  \Uprime\
models and in the Sequential Standard Model.

The highest production
cross section is given by the SSM, whereas the
$Z'_\psi$ model yields the lowest rate;
the predictions of the other models lie between these results and are
almost indistinguishable for large $m_{Z'}$.
Moreover, the rates decrease by
several orders of magnitude once $m_{Z'}$ increases. 
In detail, at  $\sqrt{s}=7$~TeV, the SSM cross section
runs from 1.6 pb ($m_{Z'}=1$~TeV) 
to ${\cal O}(10^{-8})$~pb ($m_{Z'}=5$~TeV). The production rate for
the \Uprime-based $Z'$ varies from ${\cal O}(10^{-1})$
to  ${\cal O}(10^{-9})$ pb in the same $m_{Z'}$ range, with very
little differences among the models. 
At the centre-of-mass energy of 8 TeV, the variation is between 2.3 pb
($Z'_{\rm SSM}$ at $m_{Z'}=1$~TeV) and ${\cal O}(10^{-9})$ pb (all other
models at $m_{Z'}=$~5 TeV).
At $\sqrt{s}=14$~TeV, for a $Z'$ mass of 1 TeV
the cross section varies from about 8 pb ($Z'_{\rm SSM}$) to 
1.8 pb ($Z'_\psi$); for $m_{Z'}=5$~TeV,
all models yield a rate around ${\cal O}(10^{-4})$ pb.

\begin{table}
\caption{LO $Z'$ production cross section in pb at the LHC for $pp$ collisions
at $\sqrt{s}=7$~TeV for the various models in Table~\ref{tab:Models} and
$m_{Z'}$ varying from 1 to 5 TeV, with steps of 500 GeV.
The CTEQ6L LO parton distribution functions
are employed.}
\label{cs7}
\begin{center}
\small
\begin{tabular}{|c|c|c|c|c|c|c|c|}
\hline
$m_{Z'}$ & $\sigma(Z'_\eta)$ & $\sigma(Z'_\psi)$ & $\sigma(Z'_{\rm N})$ & 
$\sigma(Z'_{\rm I})$
& $\sigma(Z'_{\rm S})$ & $\sigma(Z'_\chi)$ & $\sigma(Z'_{\rm SSM})$ \\ 
\hline\hline
1.0 & 0.57 & 0.38 & 0.41 & 0.70 & 0.88 & 0.89 & 1.59 \\
\hline 
1.5 & $6.3\times 10^{-2}$  & $4.2\times 10^{-2}$ & 
$4.5\times 10^{-2}$ & $7.0\times 10^{-2}$ & $8.6\times 10^{-2}$  
&  $8.7\times 10^{-2}$ & 0.16  \\
\hline 
2.0 & $7.7\times 10^{-3}$  & $6.1\times 10^{-3}$ & 
$6.4\times 10^{-3}$ & $8.8\times 10^{-3}$ & $1.0\times 10^{-2}$  
&  $1.0\times 10^{-2}$ & $2.1\times 10^{-2}$  \\
\hline 
2.5 & $1.0\times 10^{-3}$  & $9.6\times 10^{-4}$ & 
$9.8\times 10^{-4}$ & $1.2\times 10^{-3}$ & $1.3\times 10^{-3}$  
&  $1.3\times 10^{-3}$ & $2.9\times 10^{-2}$  \\
\hline 
3.0 & $ 1.5\times 10^{-4}$  & $1.4\times 10^{-4}$ & 
$1.4\times 10^{-4}$ & $1.5\times 10^{-4}$ & $1.5\times 10^{-4}$  
&  $1.5\times 10^{-4}$ & $3.9\times 10^{-4}$  \\
\hline 
3.5 & $1.7\times 10^{-5}$  & $1.9\times 10^{-5}$ & 
$1.8\times 10^{-5}$ & $1.7\times 10^{-5}$ & $1.5\times 10^{-5}$  
&  $1.5\times 10^{-5}$ & $4.7\times 10^{-5}$  \\
\hline 
4.0 & $1.7\times 10^{-6}$  & $1.9\times 10^{-6}$ & 
$1.8\times 10^{-6}$ & $1.5\times 10^{-6}$ & $1.4\times 10^{-6}$  
&  $1.3\times 10^{-6}$ & $4.4\times 10^{-6}$  \\
\hline 
4.5 & $1.1\times 10^{-7}$  & $1.3\times 10^{-7}$ & 
$1.2\times 10^{-7}$ & $1.0\times 10^{-7}$ & $9.2\times 10^{-8}$  
&  $9.1\times 10^{-8}$ & $3.0\times 10^{-7}$  \\
\hline 
5.0 & $5.5\times 10^{-9}$  & $6.0\times 10^{-9}$ & 
$5.9\times 10^{-9}$ & $5.1\times 10^{-9}$ & $4.6\times 10^{-9}$  
&  $4.5\times 10^{-9}$ & $1.4\times 10^{-8}$  \\
\hline 
\end{tabular}
\end{center}
\end{table}
\begin{table}[htp]
\caption{As in Table~\ref{cs7}, but at the centre-of-mass energy
$\sqrt{s}=8$~TeV.}
\label{cs8}
\begin{center}
\small
\begin{tabular}{|c|c|c|c|c|c|c|c|}
\hline
$m_{Z'}$ & $\sigma(Z'_\eta)$ & $\sigma(Z'_\psi)$ & $\sigma(Z'_{\rm N})$ & 
$\sigma(Z'_{\rm I})$ & $\sigma(Z'_{\rm S})$ & $\sigma(Z'_\chi)$ & $\sigma(Z'_{\rm SSM})$ \\ 
\hline\hline
1.0 & 0.83 & 0.54 & 1.01 & 1.28 & 1.28 & 1.30 & 2.30 \\
\hline 
1.5 & $ 0.10 $  & $6.9\times 10^{-2}$ & 
$7.4\times 10^{-2}$ & $0.12$ & $0.15$  
&  $0.15$ & 0.27  \\
\hline 
2.0 & $1.6\times 10^{-2}$  & $1.2\times 10^{-2}$ & 
$1.2\times 10^{-2}$ & $1.8\times 10^{-2}$ & $2.2\times 10^{-2}$  
&  $2.2\times 10^{-2}$ & $4.3\times 10^{-2}$  \\
\hline 
2.5 & $2.8\times 10^{-3}$  & $2.3\times 10^{-3}$ & 
$2.4\times 10^{-3}$ & $3.1\times 10^{-3}$ & $3.5\times 10^{-3}$  
&  $3.5\times 10^{-3}$ & $7.4\times 10^{-3}$  \\
\hline 
3.0 & $ 4.9\times 10^{-4}$  & $4.5\times 10^{-4}$ & 
$4.6\times 10^{-4}$ & $5.2\times 10^{-4}$ & $5.5\times 10^{-4}$  
&  $5.6\times 10^{-4}$ & $1.3\times 10^{-3}$  \\
\hline 
3.5 & $8.4\times 10^{-5}$  & $8.5\times 10^{-5}$ & 
$8.5\times 10^{-5}$ & $8.4\times 10^{-5}$ & $8.3\times 10^{-5}$  
&  $8.3\times 10^{-5}$ & $2.2\times 10^{-4}$  \\
\hline 
4.0 & $1.3\times 10^{-5}$  & $1.4\times 10^{-5}$ & 
$1.4\times 10^{-5}$ & $1.2\times 10^{-5}$ & $1.1\times 10^{-5}$  
&  $1.1\times 10^{-5}$ & $3.4\times 10^{-5}$  \\
\hline 
4.5 & $1.7\times 10^{-6}$  & $1.9\times 10^{-6}$ & 
$1.8\times 10^{-6}$ & $1.6\times 10^{-6}$ & $1.4\times 10^{-6}$  
&  $1.4\times 10^{-6}$ & $4.4\times 10^{-6}$  \\
\hline 
5.0 & $1.7\times 10^{-7}$  & $1.9\times 10^{-7}$ & 
$1.9\times 10^{-7}$ & $1.6\times 10^{-7}$ & $1.4\times 10^{-7}$  
&  $1.4\times 10^{-7}$ & $4.5\times 10^{-7}$  \\
\hline 
\end{tabular}
\end{center}
\end{table}
\begin{table}[htp]
\caption{As in Tables~\ref{cs7} and \ref{cs8}, but at the centre-of-mass energy
$\sqrt{s}=14$~TeV.}
\label{cs14}
\begin{center}
\small
\begin{tabular}{|c|c|c|c|c|c|c|c|}
\hline
$m_{Z'}$ & $\sigma(Z'_\eta)$ & $\sigma(Z'_\psi)$ & $\sigma(Z'_{\rm N})$ & 
$\sigma(Z'_{\rm I})$ & $\sigma(Z'_{\rm S})$ & $\sigma(Z'_\chi)$ & $\sigma(Z'_{\rm SSM})$ \\ 
\hline\hline
1.0 & 2.87 & 1.83 & 2.00 & 3.56 & 4.53 & 4.59 & 7.98 \\
\hline 
1.5 & 0.52  & 0.34 & 0.37 & 0.64 & 0.82  
& 0.83 & 1.43  \\
\hline 
2.0 & 0.13  & $9.0\times 10^{-2}$ & 
$9.7\times 10^{-2}$ & 0.16 & 0.20  
&  0.21 & 0.36  \\
\hline 
2.5 & $4.0\times 10^{-2}$  & $2.8\times 10^{-2}$ & 
$3.0\times 10^{-2}$ & $4.9\times 10^{-2}$ & $6.1\times 10^{-2}$  
&  $6.2\times 10^{-2}$ & 0.11  \\
\hline 
3.0 & $ 1.3\times 10^{-2}$  & $9.7\times 10^{-3}$ & 
$1.0\times 10^{-2}$ & $1.6\times 10^{-2}$ & $2.0\times 10^{-2}$  
&  $2.0\times 10^{-2}$ & $3.6\times 10^{-2}$  \\
\hline 
3.5 & $4.8\times 10^{-3}$  & $3.6\times 10^{-3}$ & 
$3.8\times 10^{-3}$ & $5.5\times 10^{-3}$ & $6.6\times 10^{-3}$  
&  $6.7\times 10^{-3}$ & $1.3\times 10^{-2}$  \\
\hline 
4.0 & $1.7\times 10^{-3}$  & $1.4\times 10^{-3}$ & 
$1.4\times 10^{-3}$ & $2.0\times 10^{-3}$ & $2.3\times 10^{-3}$  
&  $2.3\times 10^{-3}$ & $4.6\times 10^{-3}$  \\
\hline 
4.5 & $6.4\times 10^{-4}$  & $5.4\times 10^{-4}$ & 
$5.6\times 10^{-4}$ & $7.1\times 10^{-4}$ & $8.0\times 10^{-4}$  
&  $8.1\times 10^{-4}$ & $1.7\times 10^{-3}$  \\
\hline 
5.0 & $2.4\times 10^{-4}$  & $2.1\times 10^{-4}$ & 
$2.2\times 10^{-4}$ & $2.5\times 10^{-4}$ & $2.8\times 10^{-4}$  
&  $2.8\times 10^{-4}$ & $6.2\times 10^{-4}$  \\
\hline 
\end{tabular}
\end{center}
\end{table}

\subsection{Event rates with sparticle production in {$\mathbf Z'$} decays at the LHC}

In the following, we wish  to investigate the domain where possible $Z'$ decays
into supersymmetric particles could be detectable.
For this purpose, we consider two scenarios: \rts =8 \TeV, 
with an integrated luminosity, \IL=20 \ifb, as expected in the 2012 LHC data taking, and,
in future perspective,  \rts =14  \TeV \ with \IL=100 \ifb.
In the narrow-width approximation, the foreseen number of events
in $Z'$ decays is simply given by the
product of integrated luminosity, production cross section and relevant branching ratio.

The expected event rates in the two considered scenarios are summarized in 
Tables~\ref{tabnum1} and \ref{tabnum2}, for \MZprime =1.5 and 2 TeV and setting
the slepton mass $m^0_{\tilde\ell}$ to the values which in 
Tables~\ref{tabeta}--\ref{brssm} maximize the slepton rate.
We discarded the $Z'_\chi$ model as it does not yield a sfermion spectrum
after the addition of the D-term to squark and slepton masses.
As discussed in  Section \ref{sec:Reference Points}, leptonic final states in
supersymmetric events can be yielded
by direct decays $Z'\to \tilde\ell^+\tilde\ell^-$ (Fig.~\ref{zsl})
or by a cascade originated from
primary decays into sneutrino, chargino or neutralino pairs 
(see Figs.~\ref{zch}, \ref{zneu} and \ref{zsneu}).
By adding up such rates, one obtains the so-called {\it cascade} branching ratio:
\begin{equation}
{\rm BR}_{\rm casc}={\rm BR}_{\tilde\nu\tilde\nu^*}+{\rm BR}_{\tilde\chi^+\tilde\chi^-}+
{\rm BR}_{\tilde\chi^0\tilde\chi^0}.
\end{equation}

In Tables~\ref{tabnum1} and \ref{tabnum2} 
N$_{\rm slep}$ and N$_{\rm casc}$ are the number of events with
a $Z'$ decaying into a primary charged-slepton pairs or into a supersymmetric cascade, respectively.
In both luminosity (energy) regimes, due to the large cross section,
the Sequential Standard Model
is the one yielding the highest production of
supersymmetric particles in $Z'$ decays, up to ${\cal O}(10^4)$-${\cal O}(10^5)$ 
for cascade events at $\sqrt{s}=14$~TeV and \IL=100~fb$^{-1}$ and a $Z'$ mass
$m_{Z'}=1.5$~TeV. 
As discussed in Sections 
\ref{sec:ZETA} and \ref{sec:ZI}, in  the $Z'_\eta$  and $Z'_{\rm I}$
models direct decays into charged sleptons are prevented, but
sneutrino, neutralino and chargino productions are accessible,
with expected number of
events from 50 to ${\cal O}(10^4)$ according to
$m_{Z'}$, energy and integrated luminosity.
As for the $Z'_N$ model, in the high-luminosity phase,  
a few hundreds of direct sleptons and up to $10^4$
cascade particles can be produced 
for $m_{Z'}=1.5$~TeV. For \IL=20~fb$^{-1}$
and $\sqrt{s}=8$~TeV, direct slepton decays are negligible, but
about 400 and 70 cascade events can be expected 
for a $Z'$ mass of 1.5 and 2 TeV, respectively.
The $Z'_S$ boson leads to many cascade particles in the 
high-luminosity regime, between $10^3$ and $10^4$, and a few tenths of direct leptons.
For the lower-luminosity case, there are no directly produced
charged sleptons, whereas the cascade sparticles are about 30 
($m_{Z'}$=1.5~TeV) and  46 ($m_{Z'}=2$~TeV).

Before concluding this subsection, we point out that, although the numbers
in Tables \ref{tabnum1} and \ref{tabnum2} encourage optimistic predictions on
\Zprime \ decays  into sparticles, especially in the high-luminosity phase,
before drawing a conclusive statement on this issue, it will be
necessary carrying out a careful study accounting for
detector acceptance and resolution, triggering efficiency and
cuts on final-state jets and leptons.
Hence, the results presented in this paper should be seen a first step
towards a more thorough investigation, which requires, above all,
the implementation of the models herein discussed into a Monte Carlo
event generator. In this perspective, one should compare
the Monte Carlo predictions with 
the experimental data following, e.g., the approach proposed in \cite{rojas} or
investigating the observables suggested in \cite{han} to search for new
physics in Drell--Yan like events mediated by a new heavy resonance.
The same analysis should be also performed for the Standard Model backgrounds:
as discussed in the Introduction, we shall defer the implementation of
the modelling for $Z'$ production and decays, as well as the comparison with the
simulation of the backgrounds, to future work.  
\begin{table}[htp]
\caption{Number of supersymmetric particles at the LHC,
for $Z'$ production \Uprime\  models and in 
the Sequential Standard Model at
$\sqrt{s}$=8 TeV and \IL=20 fb$^{-1}$, as a function of $m_{Z'}$ expressed in TeV.}
\label{tabnum1}
\begin{center}
\small
\begin{tabular}{|c|c|c|c|}
\hline
Model & $m_{Z'}$ & N$_{\rm casc}$ & N$_{\rm slep}$ \\ 
\hline\hline
$Z'_\eta$ & 1.5  & 523  & -- \\
\hline
$Z'_\eta$ & 2 &  55 & -- \\
\hline
$Z'_\psi$ & 1.5 &  599 & 36 \\
\hline
$Z'_\psi$ & 2 & 73 & 4 \\
\hline
$Z'_{\rm N}$ & 1.5 &  400 & 17 \\
\hline
$Z'_{\rm N}$ & 2 &  70 & 3 \\
\hline
$Z'_{\rm I}$ & 1.5  &  317 & -- \\
\hline
$Z'_{\rm I}$ & 2  & 50 & -- \\
\hline
$Z'_{\rm S}$ & 1.5  & 30 & -- \\
\hline
$Z'_{\rm S}$ & 2  & 46  & -- \\
\hline
$Z'_{\rm SSM}$ & 1.5  &  2968  & 95 \\
\hline
$Z'_{\rm SSM}$ & 2  & 462 & 14 \\
\hline
\end{tabular}
\end{center}
\end{table}
\begin{table}[htp]
\caption{As in Table~\ref{tabnum1}, but for $\sqrt{s}=14$~TeV and  
\IL=100~fb$^{-1}$.}
\label{tabnum2}
\begin{center}
\small
\begin{tabular}{|c|c|c|c|}
\hline
Model & $m_{Z'}$ & N$_{\rm casc}$ & N$_{\rm slep}$ \\ 
\hline\hline
$Z'_\eta$ & 1.5  &  13650 & -- \\
\hline
$Z'_\eta$ & 2.0 &  2344 & -- \\
\hline
$Z'_\psi$ & 1.5 &  10241 & 622 \\
\hline
$Z'_\psi$ & 2.0 & 2784 & 162 \\
\hline
$Z'_{\rm N}$ & 1.5 &  9979 & 414 \\
\hline
$Z'_{\rm N}$ & 2.0 &  2705 & 104 \\
\hline
$Z'_{\rm I}$ & 1.5 &  8507 & -- \\
\hline
$Z'_{\rm I}$ & 2.0 & 2230 & -- \\
\hline
$Z'_{\rm S}$ & 1.5 & 8242 & 65 \\
\hline
$Z'_{\rm S}$ & 2.0 & 2146  & 16 \\
\hline
$Z'_{\rm SSM}$ & 1.5 &  775715  & 24774 \\
\hline
$Z'_{\rm SSM}$ & 2 & 19570 & 606 \\
\hline
\end{tabular}
\end{center}
\end{table}

\section{Conclusions}

In this paper, we discussed production and decay of new neutral $Z'$ bosons, according to
new physics models based on a \Uprime\  gauge group and in the
Sequential Standard Model. Unlike most analyses undertaken so far,
based on SM decays, 
we also included $Z'$ supersymmetric decay modes, as predicted by the Minimal Supersymmetric 
Standard Model: in this perspective, the current $Z'$ mass limits
may have to be revisited.
Extending the MSSM with  the \Uprime\  symmetry implies 
new features, such as an extra scalar neutral Higgs boson, two novel
neutralinos and a modification of the sfermion masses due to an additional
contribution to the so-called D-term. 
The particle mass spectra were studied in terms of the parameters characterizing
the \Uprime\  group and the MSSM;
in particular, we discarded scenarios wherein, for fixed values of
the sfermion soft mass,
squarks or sleptons are not physical after the addition of the D-term. 
The same study has been performed for the purpose of the 
$Z'$ partial widths and branching ratios, paying special attention to final states
with charged leptons and missing energy.
In fact, these configurations are favourable for an experimental detection 
at hadron colliders and
can be yielded by intermediate charged sleptons or a 
supersymmetric cascade through neutralinos, chargino and sneutrinos.
The branching ratios of these $Z'$ decays have been investigated
in all the models, as a function of the slepton mass.

We finally computed the $Z'$ production LO cross section in all scenarios 
and gave an estimate of the number of supersymmetric events
in $Z'$ decays, in the narrow-width approximation and 
for few values of centre-of-mass energy and integrated luminosity. 
The outcome of this study is that, for some models and parametrizations,
one can even have up to $10^4$-$10^5$ events with sparticle production
in $Z'$  decays.
As an additional remark, we wish to point out that 
the $\Zprime \to \slepton^+\slepton^-$ decay presents 
two interesting features. First, the $Z'$ mass will set an additional
constrain on the slepton invariant mass; second, it allows the exploration
of corners of the phase space which would be instead unaccessible through other 
processes, e.g. Drell--Yan like events.

In summary, we  consider our investigation a useful starting
point to study $Z'$ production and decay beyond the Standard Model,
such as within supersymmetric theories, drawing guidelines
for future experimental analyses.
In future perspective, it will be very interesting performing a study
including parton showers, finite-width and hadronization corrections, 
as well as experimental effects, like the detector simulation
and the acceptance cuts.
In this way, one will eventually be able to draw a statement on the
$Z'$ mass limits within supersymmetry.
To reach these objectives,
the models for $Z'$ production and decay, 
examined throughout this paper, will have to be implemented
in Monte Carlo programs, such as HERWIG or PYTHIA, and the
supersymmetry signals  compared with the Standard Model
backgrounds simulated, e.g., by means of the ALPGEN code \cite{alpgen}.
In the framework of an event generator, it will also be possible,
in the same manner as the experimental analyses do, rescaling 
the total cross section in such a way to include higher-order
QCD corrections.  For this purpose,
the use of the FEWZ code \cite{fewz}, which simulates vector boson
production at hadron colliders at NNLO, with fully exclusive final states,
is planned.
Other possible extensions of our analysis consist
in investigating more thoroughly the unconventional 
assignment of the SM and exotic fields to the SU(10) 
representations, as well as scenarios wherein the exotic leptons
(sleptons) and quarks (squarks), 
predicted by the grand-unified group E$_6$, but discarded in the present work,
are lighter
than the $Z'$ and therefore capable of 
contributing to its decay
width. This is in progress.

\section{Acknowledgements}
We are especially 
indebted to T.~Gherghetta for a very useful correspondence
aiming at understanding the results and the modelling of 
Ref.~\cite{Gherghetta:1996yr}.
We acknowledge R.~Barbieri, M.L.~Mangano, B.~Mele, E.~Nardi 
and M.H.~Seymour for discussions on these and related topics.
We are grateful to V.~Sanz, who pointed out a mistake in 
a Feynman diagram presented in the previous versions of this paper.

\appendix

\section{${\mathbf Z'}$ decay rates into standard and supersymmetric 
particles}

The Lagrangian term describing the interaction of the
$Z'$ with fermions is given by:
\begin{equation}
{\cal L}_f=g'\bar f\gamma^\mu(v_f-a_f\gamma_5)f Z'_\mu,
\label{lzf}
\end{equation}
with
\begin{equation}
f=
\begin{pmatrix}
f_L\\
f_R^c
\end{pmatrix}.
\end{equation}
Setting $Q'(f_R)=-Q'(f_L^c)$, the vector and axial-vector 
couplings read:
\begin{equation} 
v_f=\frac{1}{2}\left[Q'(f_L)+Q'(f_R)\right]\ ,\ 
a_f=\frac{1}{2}\left[Q'(f_L)-Q'(f_R)\right],
\label{coup}
\end{equation}
where the \Uprime\  charges of left- and right-handed fermions
can be obtained by using Eq.~(\ref{qphi}) and
Table~\ref{tabq}. 
In terms of the mixing angle $\theta$, such couplings read:
\begin{eqnarray} 
v_f&=&\frac{1}{2}\left[(Q'_\psi(f_L)+Q'_\psi(f_R))\cos\theta-
(Q_\chi'(f_L)+Q_\chi'(f_R))\sin\theta
\right]\ ,\ \nonumber\\
a_f&=&\frac{1}{2}\left[(Q'_\psi(f_L)-Q'_\psi(f_R))\cos\theta-
(Q_\chi'(f_L)-Q_\chi'(f_R))\sin\theta
\right].\label{vfaf}
\end{eqnarray}
One can thus write the $Z'$ width into fermion pairs as:
\begin{equation}\label{gamf}
\Gamma(Z'\to f\bar f) = C_f\frac{g'^2}{12\pi}m_{Z'}
\left[v_f^2\left(1+2\frac{m_f^2}{m^2_{Z'}}\right)+a_f^2\left(
1-4\frac{m_f^2}{m^2_{Z'}}\right)\right]\left(1-4\frac{m_f^2}
{m^2_{Z'}}\right)^{1/2},\end{equation}
where the colour factor is $C_f=3$ for quarks and
$C_f=1$ for leptons.  
With the charges listed in Table~\ref{tabq} and employing
Eq.~(\ref{vfaf}), one can
show that, in the $Z'_{\rm I}$ model, namely
$\theta=\arccos\sqrt{5/8}-\pi/2$, the vector and
vector-axial couplings of the $Z'$ with up-type quarks
vanish, i.e. $v_u=a_u=0$. In fact, when discussing
$Z'_{\rm I}$ phenomenology at the Representative
Point (Section 4), it was pointed out that its
branching ratio into $u\bar u$ pairs is null.

Likewise, the interaction Lagrangian of the sfermions with the $Z'$ reads:
\begin{equation}
{\cal L}_{\tilde f}=g'(v_f\pm a_f)[\tilde f^*_{L,R} 
(\partial_\mu\tilde f_{L,R})-(\partial_\mu \tilde f^*_{L,R})\tilde f_{L,R}]  Z'^\mu.
\label{lzsf}
\end{equation}
The width into left- or right-handed sfermions is given by:
\begin{equation}
\Gamma(Z'\to \tilde f_{L,R}\tilde f^*_{L,R})=
C_f\frac{g'^2}{48\pi}m_{Z'}(v_f\pm a_f)^2
\left(1-4\frac{m_{\tilde f}^2}
{m^2_{Z'}}\right)^{1/2},\label{gsf}
\end{equation}
where the $\pm$ sign refers to left- and right-handed sfermions,
respectively.
Eq.~(\ref{gsf}) is expressed in terms of weak eigenstates $\tilde f_{L,R}$; 
its generalization to the mass eigenstates $\tilde f_{1,2}$
is straightforward and discussed in \cite{Gherghetta:1996yr}. 
However, for the parametrizations used throughout this paper, sfermion mixing
is always negligible and Eq.~(\ref{gsf}) can be safely used even
to calculate the branching ratios into $\tilde f_1\tilde f^*_1$ and
$\tilde f_2\tilde f^*_2$ final states. 

From Eq.~(\ref{gsf}) one can learn that
the $Z'$ rate into left- and right-handed sfermions vanishes for
$v_f=-a_f$ and $v_f=a_f$, respectively.
In fact, for $v_f=a_f$, according to Eq.~(\ref{lzf}), 
the $Z'$ only couples to left-handed fermions and therefore
in the MSSM, in absence of left-right mixing, there is no coupling with right-handed 
sfermions. Likewise, for $v_f=-a_f$, 
the $Z'$ only couples to right-handed fermions and sfermions and
the rate into $\tilde f_L\tilde f^*_L$ pairs is null.
For example, in the  $Z'_{\rm N}$ model, it is 
$v_{\tilde\nu}=a_{\tilde\nu}$, whereas, in the $Z'_{\rm I}$ model, 
$v_{\tilde\ell}=a_{\tilde\ell}$. Therefore, as remarked in Sections
5.3 and 5.4, the $Z'_{\rm N}\to \tilde\nu_2\tilde\nu_2^*$ 
and  $Z'_{\rm I}\to \tilde\ell_2\tilde\ell_2$ are suppressed,
although they are kinematically permitted at the Reference Point.

As for the Higgs sector, defining $Q'_1$, $Q'_2$ and $Q'_3$ the \Uprime\ charges as in
Eq.~(\ref{q123}) and $\beta=\arctan(v_2/v_1)$,
one can obtain the \Zprime\ rate for decays into charged-Higgs pairs \cite{trampe}
\begin{equation}
\Gamma(Z'\to H^+H^-)=\frac{g'^2}{48\pi}
\left(Q'_1\sin^2\beta-Q'_2\cos^2\beta\right)^2 m_{Z'}
\left(1-4\frac{m_{H^\pm}^2}
{m^2_{Z'}}\right)^{3/2}
\end{equation}
and associated production of a $W$ boson with a charged Higgs
\footnote{Eq.~(\ref{wh}) corrects a typing mistake present 
in Ref.~\cite{Gherghetta:1996yr},
wherein the decay width $Z'\to W^\pm H^\mp$ 
is instead 4 times smaller than in
(\ref{wh}).}
\begin{eqnarray}
\Gamma(Z'\to W^\pm H^\mp) &=&   
\frac{g'^2}{48\pi} (Q'_1+Q'_2)^2m_{Z'}\sin^2\beta\cos^2\beta 
\left[1+2\frac{5m_W^2-m^2_{H^\pm}}{m^2_{Z'}}+
\frac{(m_W^2-m^2_{H^\pm})^2}{m^4_{Z'}}
\right]\nonumber\\
&\times & \sqrt{1-2\frac{m^2_W+m^2_{H^\pm}}{m^2_{Z'}}+
\frac{(m_W^2-m^2_{H^\pm})^2}{m^4_{Z'}}}.\label{wh}
\end{eqnarray}
As the $Z'$ has no direct coupling with $W$'s, the
decay into $W^+W^-$ pairs occurs by means of the $Z$-$Z'$ mixture.
For small values of the $Z$-$Z'$ mixing angle, this width reads \cite{trampe}:
\begin{equation}
\Gamma(Z'\to W^+W^-)=\frac{g'^2}{48\pi}
\left(Q'_1\cos^2\beta-Q'_2\sin^2\beta \right)^2 m_{Z'}.
\end{equation}
In order to obtain the widths into $Z$-Higgs pairs,
i.e. $Zh$, $ZH$ or $ZH'$ final states, or into
scalar-pseudoscalar neutral-Higgs pairs, such as $hA$,
$HA$ or $H'A$, one first needs to diagonalize the 
neutral Higgs mass matrix (see  \cite{Gherghetta:1996yr}). 
The $Z$-Higgs rate can be written in compact form as:
\begin{eqnarray}
\Gamma(Z'\to Z h_i) &=&   
\frac{g'^2}{48\pi} (Q'_1\cos\beta U_{1i}-Q'_2\sin\beta U_{2i})^2
m_{Z'}
\left[1+2\frac{5m_Z^2-m^2_{h_i}}{m^2_{Z'}}+
\frac{(m_W^2-m^2_{h_i})^2}{m^4_{Z'}}
\right]\nonumber\\
&\times & \sqrt{1-2\frac{m^2_Z+m^2_{h_i}}{m^2_{Z'}}+
\frac{(m_W^2-m^2_{h_i})^2}{m^4_{Z'}}},
\label{zhh}
\end{eqnarray}
where $i=1,2,3$ for final states $Zh$, $ZH$ and $ZH'$, respectively,
and $U_{ij}$ is the matrix which diagonalizes the
Higgs mass matrix in the $(h\  \ H\ \  H')$ basis.
Likewise, using the same notation as in Eq.~(\ref{zhh}),
the scalar-pseudoscalar Higgs width reads:
\begin{eqnarray}
\Gamma(Z'\to h_iA) &=&   
\frac{g'^2}{48\pi} \frac{v^2}{N^2}
(v_3Q'_1\sin\beta U_{1i}+v_3Q'_2\cos\beta U_{2i}+
v Q'_3\sin\beta\cos\beta U_{3i})^2m_{Z'}
\nonumber\\
&\times & \left[1-2\frac{m^2_{h_i}+m^2_A}{m^2_{Z'}}+
\frac{(m_{h_i}^2-m^2_{A})^2}{m^4_{Z'}}\right]^{3/2}.
\label{gha}
\end{eqnarray}
In Eq.~(\ref{gha}), following \cite{Gherghetta:1996yr},
we defined $N=\sqrt{v_1^2v_2^2+v_1^2v_3^2+v_2^2v_3^2}$
and $v=\sqrt{v_1^2+v_2^2}$.

Finally, one can derive the decay widths into gauginos.
As for neutralinos, after diagonalizing the mass matrix (\ref{neumass}),
the interaction Lagrangian reads:
\begin{equation}
{\cal L}_{\tilde\chi^0}=\sum_{i,j}g_{ij}\overline{\tilde\chi}^0_i
\gamma^\mu\gamma_5\tilde\chi^0_j Z'_\mu,
\end{equation}
where $g_{ij}$ is a generalized coupling depending on 
the diagonalizing-matrix elements and has been calculated numerically. 
The partial rate into neutralino pairs ($\tilde\chi^0_i\tilde\chi^0_j$)
with masses $m_i$ and $m_j$ is thus given by:
\begin{eqnarray}
\Gamma(Z'\to \tilde\chi^0_i\tilde\chi^0_j)&=&
\frac{g_{ij}^2}{12\pi}m_{Z'}\left[1-\frac{m_i^2+m_j^2}{2m_{Z'}^2}
-\frac{(m_i^2-m_j^2)^2}{2m_{Z'}^4}-3\frac{m_im_j}{m^2_{Z'}}\right]
\nonumber\\
& \times & \sqrt{\left[1-\frac{(m_i+m_j)^2}{m_{Z'}^2}\right]
\left[1-\frac{(m_i-m_j)^2}{m^2_{Z'}}\right]}.
\end{eqnarray}
Finally, the Lagrangian term corresponding to the coupling of the
$Z'$ with charginos is given by:
\begin{equation}
{\cal L}_{\tilde\chi^\pm}=\frac{g'}{2}\sum_{i,j}\overline{\tilde\chi}^\pm_i
\gamma^\mu(v_{ij}+a_{ij}\gamma_5)\tilde\chi^\pm_jZ'_\mu.
\end{equation}
The generalized vector and vector-axial couplings can be expressed
in terms of $\phi_{\pm}$, the angles of the unitary transformation  
diagonalizing the
chargino mass matrix \cite{Martin:1997ns},  
and the Higgs \Uprime\ charges as follows \cite{Gherghetta:1996yr}:
\begin{eqnarray}
v_{11} &=& Q'_1\sin^2\phi_--Q'_2\sin^2\phi_+,\nonumber\\
a_{11} &=& Q'_1\sin^2\phi_-+Q'_2\sin^2\phi_+,\nonumber\\
v_{12}=v_{21} &=& Q'_1\sin^2\phi_-\cos\phi_--\delta
Q'_2\sin\phi_++\cos\phi_+,\nonumber\\
a_{12}=a_{21} &=& Q'_1\sin^2\phi_-\cos\phi_++\delta
Q'_2\sin\phi_++\cos\phi_+,\nonumber\\
v_{22} &=& Q'_1\cos^2\phi_--Q'_2\cos^2\phi_+,
\nonumber\\
a_{22} &=& Q'_1\cos^2\phi_-+Q'_2\cos^2\phi_+.
\end{eqnarray}
In the above equations, $\delta=\mathrm{sgn}(m_{\tilde\chi^\pm_1})
\mathrm{sgn}(m_{\tilde\chi^\pm_2})$. The analytical expressions for
$\phi_\pm$ can be found in \cite{Martin:1997ns} and are not reported here
for brevity.
The rate into chargino pairs is finally given by:
\begin{eqnarray}
\Gamma(Z'\to \tilde\chi^\pm_i\tilde\chi^\mp_j)&=&
\frac{g'^2}{48\pi}m_{Z'}\left\{(v_{ij}^2+a_{ij}^2)
\left[1-\frac{m_i^2+m_j^2}{2m_{Z'}^2}-
\frac{(m_i^2-m_j^2)^2}{2m_{Z'}^4}\right]-3(v_{ij}-a_{ij})^2\
\frac{m_im_j}{m^2_{Z'}}\right\}
\nonumber\\
&\times & \sqrt{\left[1-\frac{(m_i+m_j)^2}{m_{Z'}^2}\right]
\left[1-\frac{(m_i-m_j)^2}{m^2_{Z'}}\right]}.
\end{eqnarray}

\end{document}